\documentclass[%
 reprint,
superscriptaddress,
nofootinbib,
 amsmath,amssymb,
 aps,
]{revtex4-2}

\usepackage{graphicx}
\usepackage{dcolumn}
\usepackage{bm}
\usepackage{color}
\usepackage{multirow}
\usepackage{booktabs}
\usepackage{hyperref}

\newcommand{\tr}{\mathrm{Tr}}

\DeclareMathOperator*{\mean}{\scalebox{2}{$\overline{\Sigma}$}}

\newcommand{\Bens}{\texttt{cB211.72.64}}
\newcommand{\Cens}{\texttt{cC211.60.80}}
\newcommand{\Dens}{\texttt{cD211.54.96}}

\begin{document}

\preprint{APS/123-QED}

\newcommand{\UCY}{Department of Physics, University of Cyprus, P.O. Box 20537, 1678 Nicosia, Cyprus}
\newcommand{\CYI}{Computation-based Science and Technology Research Center, The Cyprus Institute, Nicosia, Cyprus}
\newcommand{\TU}{Department of Physics, Temple University, Philadelphia, PA 19122 - 1801, USA }
\newcommand{\Wuppertal}{University of Wuppertal, Wuppertal, Germany}
\newcommand{\Romadue}{Dipartimento di Fisica and INFN, Universit\`a di Roma ``Tor Vergata", Via della Ricerca Scientifica 1, I-00133 Roma, Italy}
\newcommand{\hiskp}{HISKP (Theory), Rheinische Friedrich-Wilhelms-Universit\"at Bonn, Nussallee 14-16, 53115 Bonn, Germany}
\newcommand{\hpca}{High Performance Computing and Analytics Lab, Rheinische Friedrich-Wilhelms-Universit\"at Bonn, Friedrich-Hirzebruch-Allee 8, 53115 Bonn, Germany}

\title{Nucleon axial and pseudoscalar form factors using twisted-mass fermion ensembles at the physical point}

\author{Constantia Alexandrou}
\affiliation{\UCY}
\affiliation{\CYI}
\author{Simone Bacchio}
\affiliation{\CYI}
\author{Martha Constantinou}
\affiliation{\TU}
\author{Jacob Finkenrath}
\affiliation{\Wuppertal}
\affiliation{\CYI}
\author{Roberto Frezzotti}
\affiliation{\Romadue}
\author{Bartosz Kostrzewa}
\affiliation{\hpca}
\author{Giannis Koutsou}
\affiliation{\CYI}
\author{Gregoris Spanoudes}
\affiliation{\UCY}
\author{Carsten Urbach}
\affiliation{\hiskp}

\collaboration{Extended Twisted Mass Collaboration}

\date{\today}

\begin{abstract}
We compute the nucleon axial and pseudoscalar form factors using three $N_f=2+1+1$ twisted mass fermion ensembles with all quark masses tuned to approximately their physical values.  The values of the lattice spacings of these three physical point ensembles are  $0.080$~fm, $0.068$~fm and $0.057$~fm, and spatial sizes 5.1 fm, 5.44 fm, and 5.47 fm, respectively, yielding $m_\pi L>3.6$.  Convergence to the ground state matrix elements is assessed using multi-state fits. We study the momentum dependence of the three form factors and check the partially conserved axial-vector current (PCAC) hypothesis and the pion pole dominance (PPD). We show that in the continuum limit, the PCAC and PPD relations are satisfied. We also show that the Goldberger-Treimann relation is approximately fulfilled and determine the Goldberger-Treiman discrepancy.  We find for the nucleon axial charge  $g_A=1.245(28)(14)$, for the axial radius $\langle r^2_A \rangle=0.339(48)(06)~{\rm fm}^2$, for the pion-nucleon coupling constant $g_{\pi NN} \equiv \lim_{Q^2 \rightarrow -m_\pi^2} G_{\pi NN}(Q^2)=13.25(67)(69)$ and for $G_P(0.88m_{\mu}^2)\equiv g_P^*=8.99(39)(49)$.
\end{abstract}

\maketitle

\section{Introduction}
The nucleon axial form factors are important quantities for
weak interactions, neutrino scattering,  and parity violation experiments.
There are currently a number of neutrino scattering experiments that require knowledge of the axial form factors. At Fermi Lab, the two neutrino experiments, NO$\nu$A and MINER$\nu$A~\cite{MINERvA:2023avz}, share the same neutrino beam. The former is designed to study neutrino oscillations and the latter to perform high-precision measurements of neutrino interactions on a wide variety of materials, including helium, carbon, iron, and lead. The MicroBooNE experiment, also at Fermi Lab, aims at measuring low-energy neutrino cross sections, investigating the low-energy excess events observed by the MiniBooNE experiment, and studying neutrinos produced in supernovae. The  T2K experiment at KEK  in Japan and the CNGS experiment in Europe investigate neutrino flavor changes. The upcoming experiment DUNE will be the next-generation flagship experiment on neutrino physics.

These experimental efforts need to be matched by theoretical investigations. Computing reliably the nucleon axial form factors provides crucial input for these experiments.
However, the theoretical extraction of these form factors is difficult due to their non-perturbative nature.  Phenomenological approaches include chiral perturbation theory that provides a non-perturbative framework suitable for low values of  $Q^2$ up to about $0.4$~GeV$^2$~\cite{Schindler:2006jq,Schindler:2006it,Fuchs:2003vw}.  Other  models used include  the perturbative chiral quark model~\cite{Khosonthongkee:2004qm}, the chiral
constituent quark model~\cite{Glozman:2001zc} and light-cone sum rules~\cite{Anikin:2016teg}. 
 Lattice QCD  provides the {\it ab initio} non-perturbative framework for computing such quantities using directly the QCD Lagrangian.   Early studies of nucleon axial form factors were carried out within the quenched approximation~\cite{Liu:1991nk,Liu:1992ab}, as well as, using dynamical fermion simulations at heavier than physical pion masses~\cite{Alexandrou:2009vqd}. Only recently,  several groups are computing the axial form factors including simulations generated directly at the physical value of the pion mass~\cite{Alexandrou:2017hac,Jang:2019vkm,Rajan:2017lxk,Chang:2018uxx,Bali:2018qus,Shintani:2018ozy,Ishikawa:2018rew,RQCD:2019jai,Alexandrou:2020okk,Alexandrou:2020sml,Alexandrou:2021wzv,Djukanovic:2022wru,Jang:2023zts}. This work is the first to use solely simulations performed at physical values of the pion mass to take the continuum limit, avoiding a chiral extrapolation.

The nucleon matrix element of the isovector axial-vector current $A_\mu $  is written in terms of two form factors,  the axial, $G_A(Q^2)$, and the induced pseudoscalar,  $G_P(Q^2)$. The axial form factor, $G_A(Q^2)$, is experimentally determined from elastic scattering of neutrinos with protons, $\nu_\mu + p \rightarrow \mu^+ + n$~\cite{Ahrens:1988rr,Meyer:2016oeg,Bodek:2007vi}, while   $G_P(Q^2)$ from the longitudinal cross-section in pion electro-production~\cite{Choi:1993vt,Bernard:1994pk,Fuchs:2003vw}. At zero momentum transfer the axial form factor gives the axial charge $g_A \equiv G_A(0)$, which is measured in high precision from $\beta$-decay experiments~\cite{Brown:2017mhw,Darius:2017arh,Mendenhall:2012tz,Mund:2012fq}. The induced pseudoscalar coupling $g_P^*$  can be determined via the ordinary muon capture process $\mu^- + p \rightarrow n + \nu_\mu$ from the singlet state of the muonic hydrogen atom at the muon capture point, which corresponds to momentum transfer squared of $Q^2=0.88 m_\mu^2$~\cite{Castro:1977ep,Bernard:1998rs,Bernard:2000et,Andreev:2012fj,Andreev:2007wg}, where $m_\mu$ is the muon mass. We also study the nucleon matrix element of the isovector pseudoscalar current that determines the pseudoscalar form factor $G_5(Q^2)$ and from it the pion-nucleon coupling constant $g_{\pi NN}$.

 In this work, we use three ensembles generated at physical quark masses of the light, strange, and charm quarks and at three values of the lattice spacing, namely $a=0.080$~fm,  $a=0.068$~fm, and $a=0.057$~fm. This same setup has been used in the calculation of the electromagnetic form factors~\cite{Alexandrou:2023qor} and transversity form factors~\cite{Alexandrou:2022dtc}.  This allows us to directly take the continuum limit of the axial and pseudoscalar form factors using, for the first time, only simulations performed at the physical pion mass. This is a  major achievement since it avoids chiral extrapolation which, for the baryon sector, may introduce an uncontrolled systematic error. 
 Such simulations at the physical pion mass can be used to check important relations, such as the partially conserved axial-vector current (PCAC) relation that at form factor level connects $G_A(Q^2)$ and $G_P(Q^2)$  with  $G_5(Q^2)$. 
 At low $Q^2$ and assuming pion pole dominance (PPD) one can further relate $G_A(Q^2)$ to  $G_P(Q^2)$ and derive the Goldberger-Treiman relation. These relations have been studied within lattice QCD and will be discussed in detail in this paper.

 The remainder of this paper is organized as follows: In section~\ref{sec:AP_ME} we discuss the decomposition of the nucleon matrix elements of the axial-vector and pseudoscalar operators in terms of form factors and the  PCAC and Goldberger-Treiman relations and the pion pole dominance.  In section~\ref{sec:ensembles} we give the details on the parameters of the twisted mass fermion ensembles analyzed and in section~\ref{sec:matrix_elements} we discuss the extraction of the form factors from the two- and three-point correlators including the renormalization procedure. In section~\ref{sec:fit_strategy} we present the methods we employ for the identification of excited states and the extraction of the ground state matrix element, as well as the various fits we perform and the model averaging procedure. In section~\ref{sec:Q2Fit}, we discuss our procedure of fitting the $q^2$-dependence of the form factors and taking the continuum limit, and in section~\ref{sec:GA}, we give the results on the axial form factor, $G_A(Q^2)$, in the continuum limit. In section~\ref{sec:GPandG5} we present the analogous analysis for the induced pseudoscalar, $G_P(Q^2)$, and pseudoscalar, $G_5(Q^2)$,  form factors. We also investigate the PCAC and Goldberger-Treiman (GT) relations and evaluate the GT discrepancy. In section~\ref{sec:comparison} we compare with other recent lattice QCD results and in section~\ref{sec:conclusions} we summarize and provide our conclusions. In the appendix~\ref{sec:appendix}, we provide values and parameterization of form factors at the continuum limit.

\section{Decomposition of the  nucleon axial-vector and pseudoscalar matrix elements }\label{sec:AP_ME}

In this work, we consider only isovector quantities and neglect isospin-breaking effects due to QED interactions and $u$--$d$ quark mass difference. Any corrections arising from such isospin-breaking effects are in fact immaterial as compared to our present accuracy and are expected to become relevant only at better than one percent precision.  We summarize here for completeness the various relations using the same notation as that used in our previous work~\cite{Alexandrou:2020okk}. The isovector axial-vector operator is given by
\begin{equation}
    A_\mu = \bar{u} \gamma_\mu \gamma_5  u -\bar{d}\gamma_\mu \gamma_5  d
    \label{Eq:AVcurrent}
\end{equation}
where $u$ and $d$ are the up and down quark fields respectively. In the chiral limit, where the pion mass $m_\pi = 0$, the axial-vector current is conserved, namely $\partial^\mu A_\mu = 0$. 
For a non-zero pion mass, the spontaneous breaking of chiral symmetry relates the axial-vector current to the pion field $\psi_\pi$, through the relation
\begin{equation}
    \partial^\mu A_\mu = F_\pi m_\pi^2 \psi_\pi.
\end{equation}
We use the convention $F_\pi = 92.9$~MeV for the pion decay constant. In QCD, the axial Ward-Takahashi identity leads to the  partial conservation of the axial-vector current (PCAC) 
\begin{equation}
  \partial^\mu A_\mu= 2 m_q P,
  \label{Eq:PCAC}
\end{equation}
where $P$ is the pseudoscalar operator and $m_q=m_u=m_d$ is the light quark mass for degenerate up and down quarks. Using the PCAC relation, it then follows that  the pion field can be expressed as
\begin{equation}
    \psi_\pi = \frac{2 m_q P}{F_\pi m_\pi^2}.
    \label{Eq:piToP}
\end{equation}

The nucleon matrix element of the axial-vector current of Eq.~(\ref{Eq:AVcurrent}) can be written in terms of the axial, $G_A(Q^2)$, and induced pseudoscalar, $G_P(Q^2)$, form factors as
\begin{eqnarray}
    &&\langle N(p',s') \vert A_\mu\vert N(p,s) \rangle = \bar{u}_N(p',s')  \nonumber \\
    && \bigg[\gamma_\mu G_A(Q^2) - \frac{Q_\mu}{2 m_N} G_P(Q^2)\bigg] \gamma_5 u_N(p,s),
    \label{Eq:DecompA}
\end{eqnarray}
where $u_N$ is the nucleon spinor with initial (final) 4-momentum $p$ ($p'$) and spin $s$ ($s'$), $q=p'-p$ the momentum transfer, $q^2=-Q^2$ and $m_N$ the nucleon mass. The axial form factor is commonly parameterized as
\begin{equation}\label{Eq:GA}
    G_A(Q^2)=g_A\left(1-\frac{\langle r_A^2\rangle}{6}Q^2\right) + \mathcal{O}(Q^4),
\end{equation}
where
\begin{align}
    g_A&\equiv G_A(0)\label{Eq:gA}\\
    \langle r_A^2\rangle &\equiv - \frac{6}{ g_A} \frac{\partial G_A(Q^2)}{\partial Q^2} \bigg \vert_{Q^2 \rightarrow 0}\,
\label{Eq:radius}
\end{align}
are the axial charge and radius, respectively. A quantity of interest for the induced pseudoscalar form factor is the induced pseudoscalar coupling determined at the muon capture point~\cite{Egger:2016hcg},  namely
 \begin{equation}
     g_P^* \equiv \frac{m_\mu}{2 m_N} G_P(0.88\,m_\mu^2)
     \label{Eq:gP*}
 \end{equation}
with $m_\mu=105.6$~MeV the muon mass. It was computed in chiral perturbation theory in Ref.~\cite{Bernard:1994wn}.

The nucleon pseudoscalar matrix element is given by
\begin{equation}
    \langle N(p',s') \vert P \vert N(p,s) \rangle =  G_5(Q^2) \bar{u}_N(p',s')\gamma_5 u_N(p,s),
    \label{Eq:Decomp5}
\end{equation}
where $P=\bar{u} \gamma_5  u - \bar{d}  \gamma_5 d$ is the isovector pseudoscalar current.
The PCAC relation at the form factors level relates the axial and induced pseudoscalar form factors to the pseudoscalar form factor via the relation
\begin{equation}
    G_A(Q^2) - \frac{Q^2}{4 m_N^2} G_P(Q^2) = \frac{m_q}{m_N} G_5(Q^2).
    \label{Eq:PCAC_FFs}
\end{equation}
Making use of Eq.~(\ref{Eq:piToP}), one can connect the pseudoscalar form factor to the pion-nucleon form factor $G_{\pi NN}(Q^2)$ as follows
\begin{equation}
    m_qG_5(Q^2) = \frac{F_\pi m_\pi^2}{m_\pi^2 + Q^2} G_{\pi NN}(Q^2).
    \label{Eq:PtoPiNN}
\end{equation}
Eq.~(\ref{Eq:PtoPiNN}) is written so that it illustrates the pole structure of $G_5(Q^2)$ and the preferred usage of $m_qG_5(Q^2)$, which is a scale-independent quantity unlike $G_5(Q^2)$.  Substituting $m_qG_5(Q^2)$ in Eq.~(\ref{Eq:PCAC_FFs}),  one obtains the Goldberger-Treiman relation~\cite{Alexandrou:2007hr,Alexandrou:2009vqd}
\begin{equation}
G_A(Q^2)-\frac{Q^2}{4m_N^2} G_P(Q^2)=\frac{F_\pi m_\pi^2}{m_N(m^2_\pi+Q^2)}G_{\pi NN}(Q^2).
  \label{GT}    
\end{equation}
The pion-nucleon form factor $G_{\pi NN}(Q^2)$ at the pion pole gives the pion-nucleon coupling 
\begin{equation}
 g_{\pi NN} \equiv \lim_{Q^2 \rightarrow -m_\pi^2} G_{\pi NN}(Q^2)\,,
\end{equation}
which can be computed  using Eq.~\eqref{Eq:PtoPiNN} to obtain
\begin{equation}
 \lim_{Q^2 \rightarrow -m_\pi^2} (Q^2+m_\pi^2) m_qG_5(Q^2) = F_\pi m_\pi^2 g_{\pi NN}.
    \label{Eq:gpiNN_G5}
\end{equation}
Equivalently, $g_{\pi NN}$ can be computed using Eq.~(\ref{GT}), where the pole on the right-hand side of Eq.~(\ref{GT}) must be compensated by a similar pole in $G_P(Q^2)$, since $G_A(-m_\pi^2)$ is finite, thus obtaining
\begin{equation}
    \lim_{Q^2 \rightarrow -m_\pi^2} (Q^2+m_\pi^2) G_P(Q^2) = 4 m_N F_\pi g_{\pi NN}.
    \label{Eq:gpiNN_GP}
\end{equation}
 Additionally, close to the pole, the following relation holds 
\begin{equation}
G_P(Q^2)=\frac{4m_N F_\pi }{m^2_\pi+Q^2}G_{\pi NN}(Q^2)\bigg|_{Q^2\rightarrow -m^2_\pi}
\end{equation}
due to pion pole dominance (PPD). Inserting it in Eq.~\eqref{Eq:PtoPiNN} we obtain the relation 
\begin{equation}\label{Eq:G5_GP}
    G_P(Q^2)=\frac{4m_N}{m_\pi^2} m_qG_5(Q^2)\bigg|_{Q^2\rightarrow -m^2_\pi},
\end{equation}
which relates $G_P(Q^2)$ to $G_5(Q^2)$.
Substituting $G_P(Q^2)$ in Eq.~(\ref{GT}) we obtain 
the well-known relation~\cite{Goldberger:1958vp}
\begin{equation}
m_N G_A(Q^2)=F_\pi G_{\pi NN}(Q^2)\Big|_{Q^2\rightarrow -m^2_\pi},
   \label{Eq:GTRQ2}    
\end{equation}
 which means that $G_P(Q^2)$ can be expressed as~\cite{Scadron:1991ep}
\begin{equation}
    G_P(Q^2) = \frac{4 m_N^2}{Q^2+m_\pi^2} G_A(Q^2)\bigg|_{Q^2\rightarrow -m^2_\pi},
    \label{Eq:PPD}
\end{equation}
close to the pion pole.

From Eq.~(\ref{Eq:GTRQ2}), the pion-nucleon coupling can be expressed as \begin{equation}
    g_{\pi NN} = \frac{m_N}{F_\pi} G_A(-m_\pi^2) = \frac{m_N}{F_\pi} g_A\bigg|_{m_\pi\rightarrow 0},
    \label{Eq:GTR}
\end{equation}
where the latter holds in the chiral limit, $m_\pi=0$.
The deviation from Eq.~(\ref{Eq:GTR}) due to the finite pion mass is known as the Goldberger-Treiman discrepancy, namely
\begin{equation}
    \Delta_{\rm GT} = 1 - \frac{g_A m_N}{g_{\pi NN} F_\pi}
    \label{Eq:GTD}
\end{equation}
and it is estimated to be at the 2\% level~\cite{Nagy:2004tp} in chiral perturbation theory. The Goldberger-Treiman discrepancy is related to the low-energy constant $\bar{d}_{18}$~\cite{Fettes:1998ud,Bernard:2001rs} via
\begin{equation}
    \Delta_{\rm GT} = - \frac{2 \bar{d}_{18}m_\pi^2}{g_A}\,.
    \label{Eq:d18}
\end{equation}
Given the above relations, we define the following ratios to test whether our lattice results satisfy these relations.
\begin{align}
r_{\rm PCAC}(Q^2) &= \frac{\frac{m_q}{m_N} G_5(Q^2) + \frac{Q^2}{4 m_N^2} G_P(Q^2) }{G_A(Q^2)} \,,
\label{Eq:rPCAC}\\
r_{\rm PPD,1}(Q^2) &= \frac{m_\pi^2 + Q^2}{4 m_N^2}\frac{G_P(Q^2)}{ G_A(Q^2) }\,,
\label{Eq:rPPD1}\\
r_{\rm PPD,2}(Q^2) &= \frac{4m_N}{m_\pi^2}\frac{m_qG_5(Q^2)}{ G_P(Q^2) }\,.
\label{Eq:rPPD2}
\end{align}
The first is based on the PCAC relation in Eq.~\eqref{Eq:PCAC_FFs}.  Since PCAC is an exact operator relation, it provides a stringent test of our analysis on the form factor level. The second and third relations assume pion pole dominance and use  Eqs.~\eqref{Eq:PPD}  and \eqref{Eq:G5_GP}, respectively, and they are only expected to be unity near the pion pole.
We note that we can use the PCAC relation in Eq.~\eqref{Eq:PCAC_FFs} to write
\begin{equation}\label{Eq:rPPD2_AP}
    r_{\rm PPD,2}(Q^2)=  \frac{4m_N^2}{m_\pi^2} \frac{G_A(Q^2)}{ G_P(Q^2) } - \frac{Q^2}{m_\pi^2}\,.
\end{equation}
Using the parameterization of $G_A(Q^2)$ in Eq.~\eqref{Eq:GA} to evaluate $G_A(-m_\pi^2)$ we obtain that near  the pion pole the ratio
\begin{align*}
    \frac{4m_N^2}{m_\pi^2} \frac{G_A(Q^2)}{ G_P(Q^2) } &= \frac{g_A m_N}{g_{\pi NN} F_\pi}\left(1+\frac{\langle r_A^2\rangle m_\pi^2}{6}\right)\left(1+\frac{Q^2}{m_\pi^2}\right)\nonumber\\
    &=\left(1-\Delta_{\rm GT}+\frac{\langle r_A^2\rangle m_\pi^2}{6}\right)\left(1+\frac{Q^2}{m_\pi^2}\right),
\end{align*}
 at leading order in $m_\pi^2$, $\Delta_{\rm GT}$ and $Q^2$. Using the latter in Eq.~\eqref{Eq:rPPD2_AP} we obtain~\cite{Park:2021ypf}
\begin{equation}\label{Eq:rPPD2_GT}
    r_{\rm PPD,2}(Q^2)=  1+\left(\frac{\langle r_A^2\rangle m_\pi^2}{6}-\Delta_{\rm GT}\right)\left(1+\frac{Q^2}{m_\pi^2}\right)\,\,
\end{equation}
and therefore a deviation from unity in $r_{\rm PPD,2}(Q^2)$ can be related to the Goldberger-Treiman discrepancy.

\section{Gauge ensembles and statistics}\label{sec:ensembles}
\noindent 
We employ the twisted-mass fermion discretization scheme~\cite{Frezzotti:2003ni,Frezzotti:2000nk}, which provides automatic ${\cal O}(a)$-improvement~\cite{ETM:2010iwh}.
The bare light quark parameter $\mu_l$ is tuned to reproduce the isosymmetric pion mass $m_\pi=135$~MeV~\cite{Alexandrou:2018egz, Finkenrath:2022eon}, while the heavy quark parameters, $\mu_s$ and $\mu_c$ are tuned using the kaon mass and an appropriately defined ratio between the kaon and D-meson masses as well as the D-meson mass, following the procedure of Refs.~\cite{Finkenrath:2022eon,Alexandrou:2018egz}. The action also includes a clover term that reduces isospin-breaking effects due to the twisted-mass fermion discretization.
The values of the parameters of the ensembles analyzed in this work can be found in Table~\ref{tab:ens}. The lattice spacings and pion masses are taken from Ref.~\cite{Alexandrou:2022amy}. The values of the lattice spacing are determined both in the meson and nucleon sectors. We quote the ones from the meson sector which are compatible with the values determined from the nucleon mass in Ref.~\cite{ExtendedTwistedMass:2021gbo}. The resulting tuned pion masses, given in Table~\ref{tab:ens}, deviate by up to 4\% from the isosymmetric pion mass. This deviation is comparable with the mass difference between charged and neutral pion. Thus, we expect any correction on the form factors arising from such a  deviation to be of the same order of magnitude as isospin-breaking effects and, thus, immaterial as compared to our present accuracy.

\begin{table}[t!]
    \centering
    \begin{tabular}{c|c|c|c|c|c}
    \hline\hline
       Ensemble  & $V/a^4$ & $\beta$ & $a$ [fm] & $m_\pi$ [MeV]  & $m_\pi L$ \\
       \hline 
        \texttt{cB211.072.64} & $64^3 \times 128$ & 1.778 &  0.07957(13) & 140.2(2) & 3.62 \\
        \texttt{cC211.060.80} & $80^3 \times 160$ & 1.836 &  0.06821(13) & 136.7(2) & 3.78 \\
        \texttt{cD211.054.96} & $96^3 \times 192$ & 1.900 &  0.05692(12) & 140.8(2) & 3.90 \\
        \hline
    \end{tabular}
    \caption{Parameters for the $N_f=2+1+1$  ensembles analyzed in this work. In the first column, we give the name of the ensemble, in the second the lattice volume, in the third $\beta=6/g^2$ with $g$  the bare coupling constant, in the fourth the lattice spacing, in the fifth the pion mass, and in the sixth the value of $m_\pi L$. Lattice spacings and pion masses are taken from Ref.~\cite{Alexandrou:2022amy}.} 
    \label{tab:ens}
\end{table}

 \begin{table}[t!]
   \begin{minipage}[t]{0.32\linewidth}
     \begin{tabular}{|r|r|r|}
       \hline
       \multicolumn{3}{|c|}{\texttt{cB211.072.64}} \\
       \hline
       \multicolumn{3}{|c|}{750 configurations} \\
       \hline
       $t_s/a$ & $t_s$[fm] & $n_{src}$ \\
       \hline
       8 & 0.64 &  1 \\
       10 & 0.80 &  2 \\
       12 & 0.96 &  5 \\
       14 & 1.12 &  10 \\
     16 & 1.28 & 32 \\
     18 & 1.44 & 112 \\
     20 & 1.60 & 128 \\
    \hline
     \multicolumn{2}{|c|}{Nucleon 2pt} & 477 \\
    \hline
    \multicolumn{3}{c}{} \\
    \multicolumn{3}{c}{} \\
    \multicolumn{3}{c}{} \\
    \end{tabular}
   \end{minipage}\hfill
   \begin{minipage}[t]{0.32\linewidth}
    \begin{tabular}{|r|r|r|}
    \hline
      \multicolumn{3}{|c|}{\texttt{cC211.060.80}} \\
    \hline
      \multicolumn{3}{|c|}{400 configurations} \\
    \hline
      $t_s/a$ & $t_s$[fm] & $n_{src}$ \\
    \hline
      6 & 0.41 &   1 \\
      8 & 0.55 &   2 \\
     10 & 0.69 &   4 \\
     12 & 0.82 &  10 \\
     14 & 0.96 &  22 \\
     16 & 1.10 &  48 \\
     18 & 1.24 &  45 \\
     20 & 1.37 & 116 \\
     22 & 1.51 & 246 \\
    \hline
     \multicolumn{2}{|c|}{Nucleon 2pt} & 650 \\
    \hline 
    \multicolumn{3}{c}{} \\
    \end{tabular}
  \end{minipage}\hfill
  \begin{minipage}[t]{0.32\linewidth}
    \begin{tabular}{|r|r|r|}
    \hline
      \multicolumn{3}{|c|}{\texttt{cD211.054.96}} \\
    \hline
      \multicolumn{3}{|c|}{500 configurations} \\
    \hline
      $t_s/a$ & $t_s$[fm] & $n_{src}$ \\
    \hline
      8 & 0.46 &   1 \\
     10 & 0.57 &   2 \\
     12 & 0.68 &   4 \\
     14 & 0.80 &   8 \\
     16 & 0.91 &  16 \\
     18 & 1.03 &  32 \\
     20 & 1.14 &  64 \\
     22 & 1.25 &  16 \\
     24 & 1.37 &  32 \\
     26 & 1.48 &  64 \\
    \hline
     \multicolumn{2}{|c|}{Nucleon 2pt} & 480 \\
    \hline 
    \end{tabular}
  \end{minipage}
      \caption{Statistics used in the computation of the isovector matrix elements for the cB211.072.64 (left table) the cC211.060.80 (middle table) and the cD211.054.96 (right table) ensemble. In each table, we provide the sink-source separations used in lattice units (first column) and physical units (second column) and the number of source positions per configuration (third column). For each ensemble, the bottom row indicates the number of source positions used for the two-point functions.}
      \label{tab:statistics}
  \end{table}

The nucleon matrix elements of the axial-vector and pseudoscalar operators are obtained via appropriate combinations of three- and two-point nucleon correlation functions, as will be explained in more detail in the following section. In Table~\ref{tab:statistics}, we give the statistics used for computing the two- and three-point functions in terms of the number of configurations analyzed and the number of point sources employed per configuration. The statistics of the three-point functions are increased at increasing source-sink separation such that the errors are kept approximately constant among all the time separations.
For the twisted mass formulation employed here, the disconnected quark loop contributions are of order $a^2$  and, thus, vanish in the continuum limit~\cite{Frezzotti:2003ni}. 
For this reason, we can safely neglect them in the present work.

\section{Extraction of nucleon matrix elements}\label{sec:matrix_elements}

To evaluate the nucleon matrix elements of the operators given in Eqs.~\eqref{Eq:DecompA} and \eqref{Eq:Decomp5}, we compute three- and two-point correlation functions. The two-point function is given by
\begin{eqnarray}
C(\Gamma_0,\vec{p};t_s,t_0) &&{=}  \sum_{\vec{x}_s} \hspace{-0.1cm} e^{{-}i (\vec{x}_s{-}\vec{x}_0) \cdot \vec{p}} \times \nonumber \\
&&\tr \left[ \Gamma_0 {\langle}{\cal J}_N(t_s,\vec{x}_s) \bar{\cal J}_N(t_0,\vec{x}_0) {\rangle} \right],
\label{Eq:2pf}
\end{eqnarray}
where $x_0$ is the source, $x_s$ is the sink positions on the lattice, and $\Gamma_0$ is the unpolarized positive parity projector $\Gamma_0 = \frac{1}{2}(1+\gamma_0)$. States with the quantum numbers of the nucleon are created and destroyed by the interpolating field 
\begin{equation}
    {\cal J}_N(t,\vec{x}) = \epsilon^{abc} u^a(x) \left[ u^{b T}(x) \mathcal{C} \gamma_5 d^{c} (x) \right],
    \label{Eq:IntF}
\end{equation}
where $\mathcal{C}$ is the charge conjugation matrix. By inserting the unit operator in Eq.~(\ref{Eq:2pf}) in the form of a sum over states of the QCD Hamiltonian only states with the quantum numbers of the nucleon survive. The overlaps between the interpolating field and the nucleon state $\vert N \rangle$, such as $\langle \Omega \vert {\cal J}_N \vert N \rangle$,  need to be canceled to access the matrix element. It is desirable to increase the overlap with the nucleon state and reduce it with excited states so that the ground state dominates for as small as possible Euclidean time separations. This is because the signal-to-noise ratio decays exponentially with the Euclidean time evolution. To accomplish ground state dominance, we apply Gaussian smearing~\cite{Alexandrou:1992ti,Gusken:1989qx} to the quark fields entering the interpolating field 
\begin{equation}
    \tilde{q}(\vec{x}, t) = \sum_{\vec{y}} [\mathbf{1}  + \alpha_G H(\vec{x}, \vec{y}; U(t))]^{N_G} q(\vec{y},t),
\end{equation}
where the hopping matrix is given by
\begin{eqnarray}
H(\vec{x},\vec{y};U(t)) = \sum_{i=1}^3 \left[ U_i(x) \delta_{x,y-\hat{i}} + U_i^\dag(x-\hat{i}) \delta_{x,y+\hat{i}}  \right].
\end{eqnarray}
The parameters $\alpha_G$ and $N_G$ are tuned~\cite{Alexandrou:2018sjm,Alexandrou:2019ali} in order to approximately give a smearing radius for the nucleon of $ 0.5$~fm. For the links entering the hopping matrix, we apply APE smearing~\cite{Albanese:1987ds} to reduce statistical errors due to ultraviolet fluctuations. 
In Table~\ref{tab:smearing params}, we give the APE and Gaussian smearing parameters used for each ensemble. 

  \begin{table}[t!]
  \begin{tabular}{c|c|c|c|c|c}
    \hline\hline
    Ensemble & $n_G$ & $\alpha_G$ & $n_{\rm APE}$ & $\alpha_{\rm APE}$ & $\sqrt{\langle r^2\rangle_\psi}$ [fm]\\
    \hline
    \texttt{cB211.072.64} & 125 & 0.2 & 50 & 0.5 & 0.461(2) \\
    \texttt{cC211.060.80} & 140 & 1.0 & 60 & 0.5 & 0.516(2) \\
    \texttt{cD211.054.96} & 200 & 1.0 & 60 & 0.5 & 0.502(3) \\    
    \hline
  \end{tabular}
\caption{The number of Gaussian smearing iterations $n_G$ and the Gaussian smearing coefficient $\alpha_G$ used for each ensemble. We also provide the number of APE-smearing iterations $n_{\rm APE}$ and parameter $\alpha_{\rm APE}$ applied to the links that enter the Gaussian smearing hopping matrix. The resulting source r.m.s.  obtained is given in the last column, where the error is due to the uncertainty in the lattice spacing.}\label{tab:smearing params}
\end{table}

For the construction of the three-point correlation function, the current is inserted at time slice $t_{\rm ins}$ between the time of the creation and annihilation of the states with the nucleon quantum numbers, $t_0$ and $t_s$, respectively. The expression for the three-point function is given by
\begin{eqnarray}
 && C_{\mu}(\Gamma_k,\vec{q},\vec{p}\,';t_s,t_{\rm ins},t_0) {=}
 \hspace{-0.1cm} {\sum_{\vec{x}_{\rm ins},\vec{x}_s}} \hspace{-0.1cm} e^{i (\vec{x}_{\rm ins} {-} \vec{x}_0)  \cdot\vec{q}}  e^{-i(\vec{x}_s {-} \vec{x}_0)\cdot \vec{p}\,'} {\times} \nonumber \\
  && \hspace{1.cm} \tr \left[ \Gamma_k \langle {\cal J}_N(t_s,\vec{x}_s) j_{\mu}(t_{\rm ins},\vec{x}_{\rm ins}) \bar{\cal J}_N(t_0,\vec{x}_0) \rangle \right],
  \label{Eq:3pf}
\end{eqnarray}
where $\Gamma_k = i \Gamma_0 \gamma_5 \gamma_k$ and  $j_\mu$ is either the axial-vector current $A_\mu$ needed for computing the matrix elements in Eq.~\eqref{Eq:DecompA} or $P$ for computing  the pseudoscalar form factor in Eq.~\eqref{Eq:Decomp5}. The Euclidean momentum transfer squared is given by $Q^2 = - q^2 = - (p' -p)^2$.
The connected three-point functions are computed using sequential propagators inverted through the sink, i.e. using the so-called~\emph{fixed-sink} method. This requires new sequential inversions for each sink momentum. Therefore, we restrict to $\vec{p}\,'=0$, meaning the source momentum $\vec{p}$ is determined via momentum conservation by the momentum transfer as $\vec{p} = -\vec{q}$ and in the following we drop the usage of $\vec{p}\,'$. Without loss of generality, we also take, in the following, $t_s$ and $t_{\rm ins}$ relative to the source time $t_0$, or equivalently $t_0$ is set to zero. 

\subsection{Excited states contamination and large time limit}\label{ssec:ExcStates}

The interpolating field in Eq.~(\ref{Eq:IntF}) creates a tower of states with the quantum numbers of the nucleon. The spectral decomposition of the  two- and three-point functions are given respectively by
\begin{align}
C(\Gamma_0,\vec{p},t_s) &= \sum_{i}^{N_{st}-1}c_i(\vec{p}) e^{-E_i(\vec{p}) t_s}
\label{Eq:Twp_tsf}\quad\text{and}\\
C_{\mu}(\Gamma_k,\vec{q},t_s,t_{\rm ins}) &= \sum_{i,j}^{N_{st}-1}{\cal A}^{i,j}_{\mu}(\Gamma_k,\vec{q}) e^{-E_i(\vec{0})(t_s-t_{\rm ins})-E_j(\vec{q})t_{\rm ins}}.
\label{Eq:Thrp_tsf}
\end{align}
The coefficients of the exponential terms in the two-point function of Eq.~(\ref{Eq:Twp_tsf}) are overlap terms given by 
\begin{equation}
c_i(\vec{p}) = \tr[ \Gamma_0 \langle \Omega \vert {\cal J}_N | N_i(\vec{p}) \rangle \langle N_i(\vec{p}) \vert \bar{{\cal J}}_N | \Omega \rangle],
\label{Eq:c2pf}
\end{equation}
where spin indices are suppressed. The $i$-index  denotes the $i^{\rm th}$  state with the quantum numbers of the nucleon that may also include multi-particle states.
The coefficients ${\cal A}^{i,j}$ appearing in the three-point function of Eq.~(\ref{Eq:Thrp_tsf}) are given by 
\begin{align}
    {\cal A}^{i,j}_\mu(\Gamma_k,\vec{q}) = \tr[& \Gamma_k \langle \Omega \vert {\cal J}_N | N_i(\vec{0})\rangle \langle N_i(\vec{0}) | A_\mu | N_j(\vec{p}) \rangle  \nonumber \\
    & \langle N_j(\vec{p}) | \bar{{\cal J}}_N | \Omega \rangle],
    \label{Eq:A3pf}
\end{align}
where  $\langle N_i(\vec{0}) | A_\mu | N_j(\vec{p}) \rangle$ is the matrix element between $i^{\rm th}$ and $j^{\rm th}$  states. In practice, one truncates the sums in Eqs.~\eqref{Eq:Twp_tsf} and~\eqref{Eq:Thrp_tsf} up to some state $N_{st}$. Finally, $E_i(\vec{p})$ is the energy of the $i^{\rm th}$ state carrying momentum $\vec{p}$. For the ground state, we use the dispersion relation to obtain $E_0(\vec{p})$, namely
\begin{equation}\label{Eq:E_N}
E_0(\vec{p}) = E_N(\vec{p}) = \sqrt{m_N^2 + \vec{p}\,^2},
\end{equation}
where the nucleon mass $m_N$ is determined from the zero momentum projected two-point function.

To extract the nucleon matrix element that we are interested in, any contribution from nucleon excited states and/or multi-particle states has to be sufficiently suppressed.
How fast ground state dominance is achieved, depends on the smearing procedure applied on the interpolating fields and the current type entering the three-point function. Since the noise increases exponentially with increasing $t_s$, establishing from the data convergence to the asymptotic ground state matrix element is very difficult. For this reason, we employ a multi-state analysis by fitting the explicit contribution of the first $N_{st}-1$ excited states. Our fitting strategy is described in Sec.~\ref{sec:fit_strategy} and aims at determining reliably the values of $c_0$ and  ${\cal A}^{0,0}_\mu$.

In order to cancel unknown overlaps of the interpolating field in Eq.~(\ref{Eq:IntF}) with the nucleon state, one commonly constructs an appropriate ratio
of three- to a combination of two-point functions~\cite{Alexandrou:2013joa,Alexandrou:2011db,Alexandrou:2006ru,Hagler:2003jd},
\begin{eqnarray}
&&  R_{\mu}(\Gamma_{k},\vec{q};t_s,t_{\rm ins}) = \frac{C_{\mu}(\Gamma_k,\vec{q};t_s,t_{\rm ins}\
)}{C(\Gamma_0,\vec{0};t_s)} \times \nonumber \\
&&  \sqrt{\frac{C(\Gamma_0,\vec{q};t_s-t_{\rm ins}) C(\Gamma_0,\vec{0};t_{\rm ins}) C(\Gamma_0,\vec{0};t_s)}{C\
(\Gamma_0,\vec{0};t_s-t_{\rm ins}) C(\Gamma_0,\vec{q};t_{\rm ins}) C(\Gamma_0,\vec{q};t_s)}}.
\label{Eq:ratio}
\end{eqnarray}
The ratio in Eq.~(\ref{Eq:ratio}) is constructed such that it converges to the nucleon ground state matrix element in the limit of large time separations $\Delta E (t_s-t_{\rm ins}) \gg 1$ and $\Delta E \,t_{\rm ins} \gg 1$, where $\Delta E$ is the energy difference between the first excited state and ground state, namely
\begin{equation}
  R_{\mu}(\Gamma_k;\vec{q};t_s,t_{\rm ins})\xrightarrow[\Delta E \,t_{\rm ins} \gg 1]{\Delta E (t_s-t_{\rm ins}) \gg 1}\Pi_{\mu}(\Gamma_k;\vec{q})\,.
\end{equation} 
By substituting Eqs.~\eqref{Eq:Twp_tsf} and~\eqref{Eq:Thrp_tsf} into Eq.~\eqref{Eq:ratio} we obtain
\begin{equation}
\label{Eq:gs}
  \Pi_{\mu}(\Gamma_k;\vec{q})=\frac{{\cal A}^{0,0}_\mu(\Gamma_k,\vec{q})}{\sqrt{c_0(\vec{0})c_0(\vec{q})}}\,.
\end{equation} 
The ground-state matrix elements are extracted from the ratio of the amplitude ${\cal A}^{0,0}_\mu(\Gamma_k,\vec{q})$ of the three-point function and the amplitudes $c_0(\vec{0})$ and $c_0(\vec{q})$ of the two-point functions. In this work, we determine these amplitudes from a simultaneous fit to two- and three-point functions as described in Sec.~\ref{sec:fit_strategy}, so that the energy spectrum is the same for two- and three-point functions. For visualization purposes only, we use the following variant of the ratio in Eq.~\eqref{Eq:ratio},
\begin{equation}
\begin{aligned}
R'_{\mu}(\Gamma_k;\vec{q};t_s,t_{\rm ins}) &= \frac{C_{\mu}(\Gamma_k,\vec{q};t_s,t_{\rm ins}\
)}{\sqrt{C(\Gamma_0,\vec{0};t_s)C(\Gamma_0,\vec{q};t_s)}},
\label{Eq:ratio_new}
\end{aligned}
\end{equation}
which has the same large time-separation limit as the ratio of Eq.~(\ref{Eq:ratio}) when $t_{\rm ins} = t_s/2$ while avoiding potential excited state contaminations in the two-point functions for small values of $t_{\rm ins}$.

\subsection{Analysis of nucleon correlators}
\label{sec:FF}
The ground state matrix elements, $\Pi_\mu$, are decomposed into form factors. In the following, we provide their decomposition in Euclidean space and for $\vec{p}\,'=0$. In the case of the  matrix element of the axial-vector current we have
	\begin{eqnarray}
	\Pi_i(\Gamma_k,\vec{q}) &=& \frac{i \mathcal{K}}{4 m_N} \left[ \frac{q_k q_i}{2 m_N} G_P(Q^2)- \delta_{i,k} (m_N+E_N) G_A(Q^2) \right] \nonumber\\~~
	\label{Eq:Aik_decomp}
	\end{eqnarray}
	for the case that  $\mu=i$. For the temporal direction, the corresponding expression is
	\begin{eqnarray}
	\Pi_0(\Gamma_k,\vec{q}) &=& -\frac{q_k\mathcal{K}}{2m_N} \left [G_A(Q^2) + \frac{(m_N-E_N)}{2 m_N}G_P(Q^2)\right]. \nonumber\\
	\label{Eq:A0k_decomp}
	\end{eqnarray}
	One can then form a  $2\times2$ matrix  of kinematical coefficients multiplying $G_A(Q^2)$ and $G_P(Q^2)$, given by
	\begin{equation}
	{\cal G}_{\mu}(\Gamma_k;\vec{q}) = 
	\begin{pmatrix}
	- q_k\frac{\mathcal{K}}{2 m_N} & - q_k\frac{ \mathcal{K} (m_N-E_N)}{4 m_N^2} \\
	-i \delta_{i,k}\frac{ \mathcal{K} (m_N+E_N)}{4 m_N}  &   i  q_k q_i\frac{\mathcal{K}}{8 m_N^2}
	\end{pmatrix},
	\label{Eq:coeffs}
	\end{equation}
	where the first row  of the matrix is for $\mu=0$ and  the second row for $\mu=i$, while the first column gives the kinematic coefficients multiplying  $G_A(Q^2)$ and the second column those multiplying $G_P(Q^2)$.	
	For the case of the  matrix element of the pseudoscalar current we have
	\begin{eqnarray}
	\Pi_5(\Gamma_k,\vec{q}) = -\frac{i q_k\mathcal{K} }{2 m_N} G_5(Q^2).
	\label{Eq:g5_decomp}
	\end{eqnarray}
In the above expressions,  $E_N$ is the energy of the nucleon and $\mathcal{K}$ is a kinematic factor given by
\begin{equation}
  \mathcal{K} = \sqrt{\frac{2 m_N^2}{E_N (E_N +m_N)}}.
  \label{Eq:K_coeff}
\end{equation}

 Given the above momentum-dependence of the decomposition, we can average over all momentum components for a given $Q^2$ value, namely
\begin{align}
\overline{\Pi}_{0}(Q^2) &= \displaystyle
-\mean\limits^{k}_{q_k \neq 0} \frac{1}{q_k}\Pi_0(\Gamma_k,\vec{q})\nonumber\\\label{eq:P0}
&=
\frac{\mathcal{K}}{2m_N}\left(G_A(Q^2) + \frac{m_N-E_N}{2m_N}G_P(Q^2)\right)\\
\overline{\Pi}_{AP}(Q^2,p^2) &=  \displaystyle
i\mean\limits^{k}_{q_k^2=p^2} \Pi_k(\Gamma_k,\vec{q})\nonumber\\\label{eq:PAP}
&=
\frac{\mathcal{K}}{4m_N}\left((E_N+m_N)G_A(Q^2) -\frac{p^2}{2m_N}G_P(Q^2)\right)\\
\label{eq:PP}
\overline{\Pi}_{P}(Q^2) &= \displaystyle
-i\mean\limits^{i,k}_{\substack{i\neq k\\q_kq_i\neq0}}\frac{1}{q_kq_i}\Pi_i(\Gamma_k,\vec{q})
=
\frac{\mathcal{K}}{8m^2_N}G_P(Q^2)\\
\label{eq:P5}
\overline{\Pi}_5(Q^2) & = \displaystyle
i\mean\limits^{k}_{q_k \neq 0} \frac{1}{q_k}\Pi_5(\Gamma_k,\vec{q})
=
\frac{\mathcal{K}}{2m_N}G_5(Q^2)\,,
\end{align}
where $p^2$ runs over the possible $q_k^2$ values, the symbol $\overline{\Sigma}$ stands for the average and we indicate above the symbol the indices of the sum, which are always spatial, and below the symbol the conditions to be satisfied such that values are included in the sum. We note that, while $G_P(Q^2)$ and $G_5(Q^2)$ can be extracted directly from $\overline{\Pi}_P$ and $\overline{\Pi}_5$, respectively, $G_A(Q^2)$ is always coupled to $G_P(Q^2)$ in  $\overline{\Pi}_0$ and $\overline{\Pi}_{AP}$ for $Q^2>0$. On the other hand, $G_A(Q^2)$ is the only form factor accessible at zero momentum transfer, while all others need to be extrapolated to $Q^2=0$. Our strategy for extracting the three form factors is to perform a combined fit of the $\overline{\Pi}$s at fixed $Q^2$ and express the ground state matrix elements in Eq.~\eqref{Eq:gs} in terms of the above linear combinations of form factors.

\subsection{Renormalization}
Matrix elements computed in lattice QCD need to be renormalized in order to relate to physical observables.   In the twisted mass fermion formulation, we need the renormalization functions $Z_S$  for the renormalization of the pseudoscalar form factor $G_5(Q^2)$, $Z_P$  for the renormalization of the bare quark mass and $Z_A$ for the renormalization of the axial-vector current. 
 We note that we do not use $Z_S$ since $G_5(Q^2)$ is evaluated in the scale-independent and ultra-violet finite combination $m_q G_5(Q^2)$. In Figs.~\ref{fig:G5_fits} and~\ref{fig:FFcomparison} where $G_5(Q^2)$ is shown without $m_q$, it is only done for visualization purposes. In those cases, we use  $Z_S$ computed  as $Z_P/(Z_P/Z_S)$ with $Z_P$ computed in the RI$^\prime$ scheme. This is because a direct evaluation of $Z_S$ in RI$^\prime$ is more difficult than for $Z_P$ due to increased hadronic contamination effects observed in the case of $Z_S$.
 
We use methods based on Ward identities or on the universality of renormalized hadronic matrix elements, which are often referred to as hadronic methods, in order to compute ultra-violet finite renormalization factors, such as $Z_A$ and $Z_P/Z_S$. Hadronic methods are fully non-perturbative and require no gauge fixing, unlike the RI$^\prime$ scheme. For more details, we refer to Appendix B of Ref.~\cite{Alexandrou:2022amy}, where this approach is used to extract the renormalization constants for the ensembles employed here. This approach is preferred to the usual RI$^\prime$ scheme because it provides much more accurate results on $Z_A$ and $Z_P/Z_S$. The RI$^\prime$ scheme is employed for the determination of $Z_P$, as discussed in Ref.~\cite{ExtendedTwistedMass:2021gbo}. For completeness, the values of the renormalization constants used in this work are collected in Table~\ref{tab:Zfacs}.

\begin{table}[t!]
  \begin{tabular}{c|c|c|c}
    \hline\hline
    Ensemble & $Z_A$ & $Z_P/Z_S$ & $Z_P$ [$\overline{\text{MS}}$ 2 GeV]\\
    \hline
    \texttt{cB211.072.64} & 0.74294(24) & 0.79018(35) & 0.4746(49) \\
    \texttt{cC211.060.80} & 0.75830(16) & 0.82308(23) & 0.4771(49)\\
    \texttt{cD211.054.96} & 0.77395(12) & 0.85095(18) & 0.4871(49)\\    
    \hline
  \end{tabular}
\caption{Values of the scheme-independent renormalization constants $Z_A$ and $Z_P/Z_S$ taken from  Ref.~\cite{Alexandrou:2022amy} and of the scheme-dependent $Z_P$ given in $\overline{\rm MS}$ at $\mu_{\rm ref}=2$~GeV computed in Ref.~\cite{ExtendedTwistedMass:2021gbo}.\label{tab:Zfacs}}
\end{table}

In what follows we will denote by $G_A(Q^2)$ and $G_P(Q^2)$ the renormalized form factors obtained by multiplying the lattice three-point functions of the axial-vector current by $Z_A$. For $G_5(Q^2)$ we consider the combination $m_qG_5(Q^2)$ that renormalizes with  $\mu Z_S/Z_P$, involving only the ratio $Z_S/Z_P$ that is determined accurately from hadronic matrix elements.    
The light bare quark mass $\mu$ takes values $a\mu = 0.00072$, $0.00060$, and $0.00054$ for the \Bens{}, \Cens{}, and \Dens{} ensembles, respectively.

\section{Extraction of form factors}\label{sec:fit_strategy}
As described in Sec.~\ref{sec:FF}, bare form factors at each value of  $Q^2$ are extracted from combined fits to the values of the two- and three-point functions, after we construct the averages given in Eqs.~\eqref{eq:P0}-\eqref{eq:P5}. Two-point functions are available for all source-sink separations, $t_s$, while three-point functions are measured at selected values of $t_s$, listed in Table~\ref{tab:statistics}, and available for all $t_{\rm ins}\in [0,t_s]$. Since the optimal fit range in $t_s$ and $t_{\rm ins}$ may vary for each case, as well as the number of states needed to describe the correlation functions, we explore a wide parameter space in the fitting ranges and number of excited states included. Results are then combined using model averaging as described below. Specifically, at each value of $Q^2$, we use the following  fitting approach:
\begin{description}
    \item[$N_{\rm st}$] We perform either two- or three-state fits of all quantities, cutting the sum in Eqs.~\eqref{Eq:Twp_tsf} and~\eqref{Eq:Thrp_tsf}, to a maximum of $i_{\rm max}=N_{st}-1$ with $N_{st} \in \{2,3\}$.
    \item[$t_{\rm 2pt,\,min}$] We vary $t_{\rm 2pt,\,min}$, the lower bound in the fit of the two-point functions.  The upper bound is taken to be the source-sink separation where the correlator becomes compatible with zero within $5\sigma$. This upper maximum value varies from approximately 2.5~fm at $Q^2=0$ to 1.5~fm at $Q^2=1$~GeV$^2$. 
    \item[$t_{\rm 3pt,\,min}$] We vary $t_{\rm 3pt,\,min}$, the smallest value  of $t_s$ used for fitting the three-point functions. We fit to all $t_s\ge t_{\rm 3pt,\,min}$ available.
    \item[$t_{\rm ins,\,0}~{\rm and}~t_{\rm ins,\,S}$] We vary the number of insertion time slices from the source and the sink kept in the fit,  using $t_{\rm ins}\in [t_{\rm ins, \,0}, t_s-t_{\rm ins,\,S}]$. We only allow for $t_{\rm ins, \,0}\ge t_{\rm ins, \,S}$ since the energy gap at the source, where we have momentum, is expected to be smaller than the energy gap at the sink, where there is no momentum. At $Q^2=0$ we fix $t_{\rm ins, \,0}=t_{\rm ins, \,S}$.
    \item[$N_{O}$] We vary the number of exponential terms when we perform three-state fits to the three-point functions, since certain overlaps may be sufficiently suppressed.  The suppression rate is ordered according to the energy gaps of the first and second excited state energies. Beyond the ground state  ${\cal A}^{0,0}_\mu$, the suppression increases for the terms containing the overlaps ${\cal A}^{1,0}_\mu$, ${\cal A}^{0,1}_\mu$, ${\cal A}^{1,1}_\mu$, ${\cal A}^{2,0}_\mu$, ${\cal A}^{0,2}_\mu$, ${\cal A}^{2,1}_\mu$, ${\cal A}^{1,2}_\mu$, ${\cal A}^{2,2}_\mu$. We use either the first four, six, or all parameters, namely, we take  $N_O\in\{4,6,9\}$. $N_O=4$ corresponds to a full two-state fit.
\end{description}
In summary, $N_{st}$ and $N_O$ affect the number of parameters in the fit, while $t_{\rm 2pt,\,min}$, $t_{\rm 3pt,\,min}$ $t_{\rm ins,\,0}$, and $t_{\rm ins,\,S}$ the number of data used in the fit. We fit together the data for $Q^2=0$ and for the lowest non-zero value of $Q^2$ obtained when the momentum transfer in one spatial direction is $2\pi/L$. After performing the model averaging for the zero and for the lowest non-zero value of $Q^2$, we extract $m_N$ and use it as a prior to fit independently each larger $Q^2$ value. We note that the summation method is not used since taking into account only the ground state fails to describe the results in the case of $G_P(Q^2)$ and $G_5(Q^2)$, where excited state effects are large. We find that for the summation method to converge one would require source-sink time separations  larger than 2 fm which are not available.

\subsection{Model average}

Results obtained using the different fit approaches are averaged using the Akaike Information Criterion (AIC) and we refer to Ref.~\cite{Jay:2020jkz,Neil:2022joj} for a detailed introduction to the method. In the following, we summarize the practical aspects of our implementation. To each fit $i$, we assign a weight $w_i$, defined as
\begin{equation}\label{eq:AICw}
\log(w_i) = -\frac{\chi^2_i}{2} + N_{{\rm dof},i},
\end{equation}
where $N_{\rm dof} = N_{\rm data} - N_{\rm params}$ is the number of degrees of freedom, given as the difference between the number of data, $N_{\rm data}$, and the number of parameters, $N_{\rm params}$, used in the corresponding fit. We use correlated fits and, therefore, the $\chi^2$ is defined as
\begin{equation}
\chi^2_i = {\vec{r}_i}^{\,T} C_i^{-1} \vec{r}_i\quad\text{with}\quad \vec{r}_i=\vec{y}_i-f_i(\vec{x}_i),
\end{equation}
where, for each fit $i$, $C_i$ is the covariance matrix between the selected data $\vec{y}_i$ and $\vec{r}_i$ is the  residual computed using the selected fit approach $f_i$ evaluated on the selected data range $x_i$.
From the weights in Eq.~\eqref{eq:AICw}, we define the probability
\begin{equation}
p_i = \frac{w_i}{Z}\quad\text{with}\quad Z=\sum_i{w_i}.
\end{equation}
The model-averaged value of an observable $\mathcal{O}$ is given as
\begin{equation}\label{eq:model_average}
\begin{aligned}
\langle\mathcal{O}\rangle = \text{mean}\,(\text{error})\quad\text{with}\quad\text{mean}= \sum_i {\bar{\mathcal{O}}}_i p_i\\
\text{and}\quad\text{error}^2 = \sum_i (\sigma_i^2+\bar{\mathcal{O}}_i^2) p_i- \text{mean}^2
\end{aligned}
\end{equation}
where  ${\bar{\mathcal{O}}}_i$ and $\sigma_i$ are, respectively, the central value and the error of the observable $\mathcal{O}$ measured using the parameters of the $i^{\rm th}$ fit.

\subsection{Selection of data and fits}

\begin{figure}[t!]
    \centering
    \includegraphics[width=0.95\linewidth]{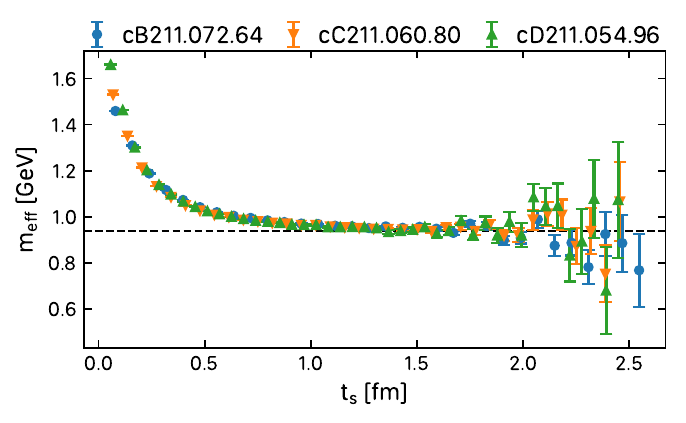}
    \caption{Nucleon effective mass using the three physical point ensembles. The dashed line is the value of the nucleon mass $m_N=0.938$~GeV.}
    \label{fig:m_eff_comparison}
    \includegraphics[width=1.0\linewidth]{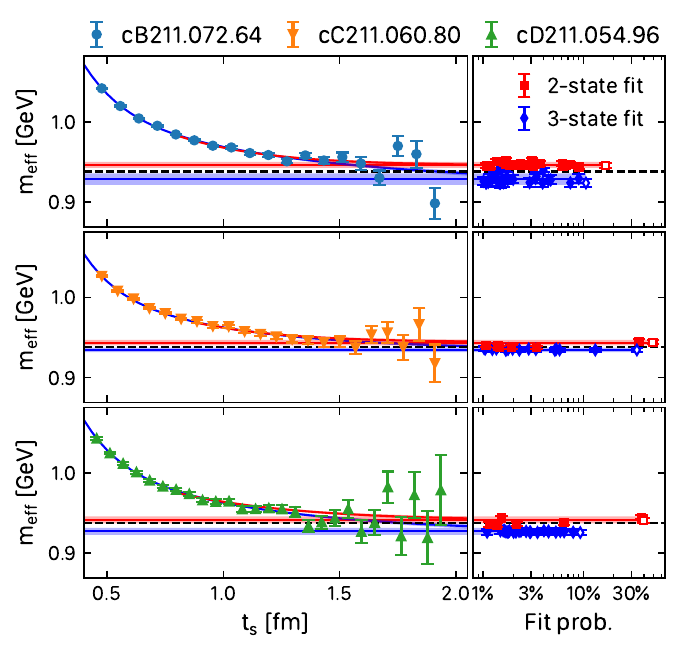}
    \caption{Nucleon effective mass versus the time separation (left column) and results for the nucleon mass obtained via fits that have a model probability larger than 1\% (right column). The horizontal bands spanning both the left and right panels are the results of the model average among all fits using two states (red points and red band) and three states (blue points and blue band). The most probable fit is depicted with open symbols. In the left panel, we show for all ensembles, the result of the most probable fit using two states (red curve) and three states (blue curve) over the range used in the fit. Panels from top to bottom are for the \Bens{}, \Cens{}, and \Dens{} ensembles, respectively.
    \label{fig:m_eff_fits}}
\end{figure}

We first illustrate our fitting procedure by considering the zero-momentum nucleon two-point function.
In Fig.~\ref{fig:m_eff_comparison}, we show the nucleon effective mass for each ensemble. We observe an impressive agreement among the data using the three ensembles showing very mild cut-off effects and compatible excited state contamination. This confirms that maintaining the radius constant of the Gaussian smearing, as shown in Table~\ref{tab:smearing params}, is a good strategy.

In Fig.~\ref{fig:m_eff_fits}, we show the nucleon effective mass separately for each of the three ensembles, as well as the values of the nucleon mass obtained via fits to two- and three-point functions keeping only those fits with model probability $\ge 1\%$. Note that since we perform a combined fit of the nucleon two- and three-point functions, for $Q^2=0$ and the lowest non-zero value of $Q^2$ as described at the beginning of this section, the values depicted in the figure are not obtained by fitting only the nucleon effective mass data shown in these figures. We observe a good distribution of the probabilities of the fits with the most probable fit having a probability between 10\% to 50\%.

\begin{figure}[t!]
    \centering
    \includegraphics[width=1.0\linewidth]{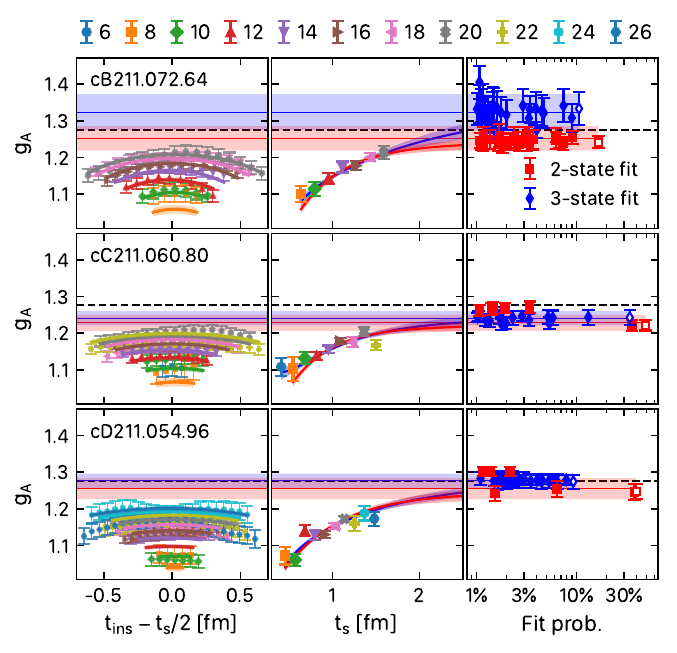}
    
    \caption{The ratio $R'$ of Eq.~\eqref{Eq:ratio_new} that yields $g_A$ versus $t_{\rm ins}-t_s/2$ (left column) and versus $t_s$ for $t_{\rm ins}=t_s/2$ (middle column). In the header of the figure, we give the symbols used to denote the various values of $t_s/a$. In the right column, we show the value of the nucleon isovector axial charge obtained via fits, as in Eq.~\eqref{Eq:gs}, versus the fit probability, using the notation of Fig.~\ref{fig:m_eff_fits}. In the left and middle panels, the curves correspond to the fit results, which have the largest probability among all two- (blue) or three- (red) state fits.
      \label{fig:gA_fits}}

    \includegraphics[width=1.0\linewidth]{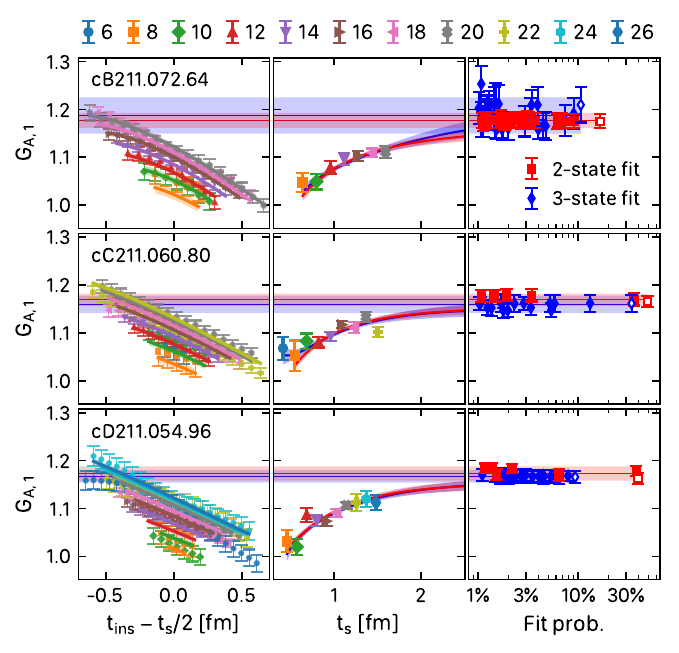}

      \caption{The same as in Fig.~\ref{fig:gA_fits}, but for
        $G_A(Q^2)$ obtained via $\overline{\Pi}_{AP}(Q^2,0)$ for the
        lowest non-zero value of $Q^2$.\label{fig:GA1_fits} }
\end{figure}

\begin{figure}[t!]
    \centering
    \includegraphics[width=1.0
\linewidth]{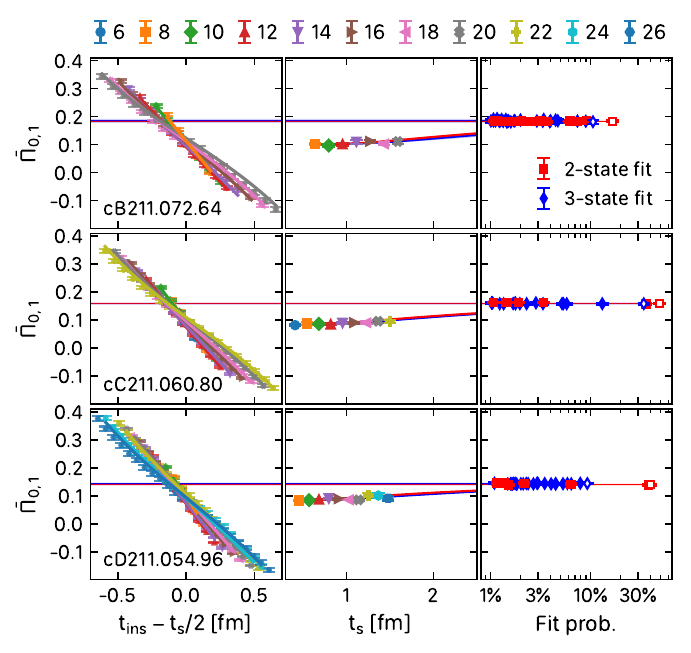}
    \caption{$\overline{\Pi}_{0}(Q^2)$ for the lowest non-zero value
      of $Q^2$ using the notation of
      Fig.~\ref{fig:gA_fits}.\label{fig:P0_fits}}
    \includegraphics[width=1.0 \linewidth]{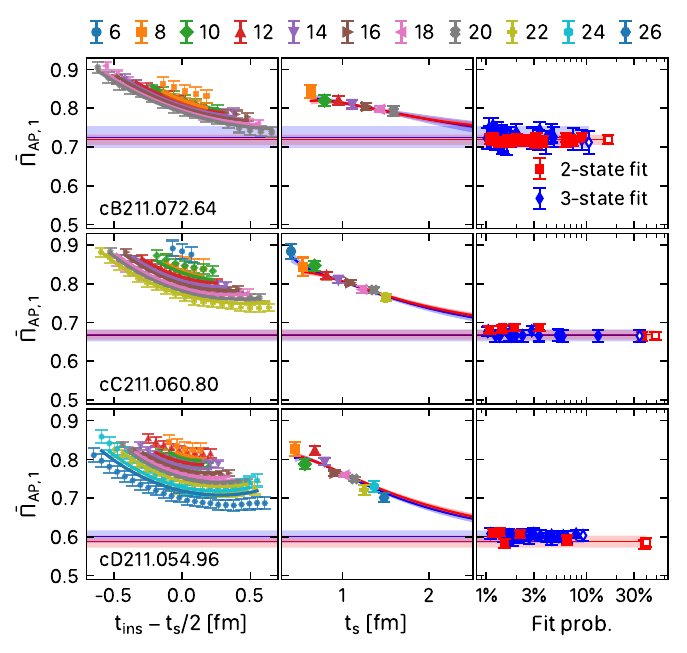}
    \caption{$\overline{\Pi}_{AP}(Q^2)$ for the lowest non-zero value
      of $Q^2$ using the notation of
      Fig.~\ref{fig:gA_fits}.\label{fig:PAP_fits}}
\end{figure}

\begin{figure}[t!]
    \centering
    \includegraphics[width=1.0\linewidth]{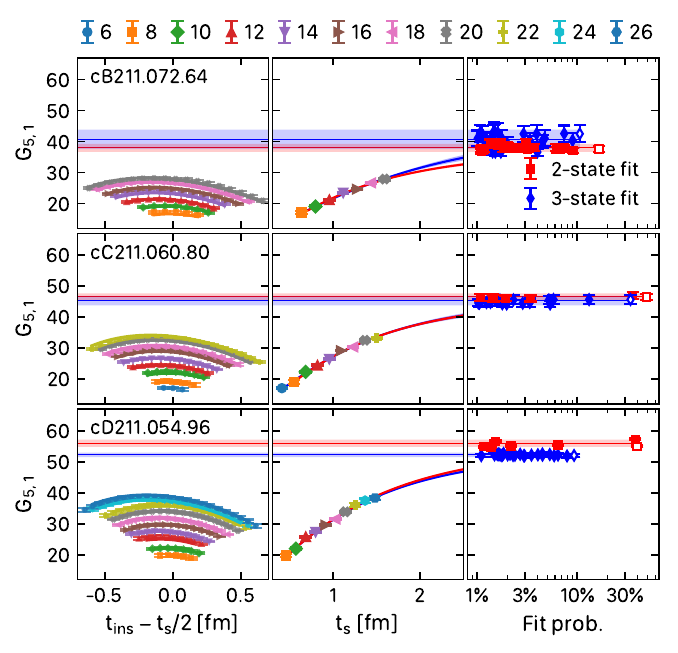}
    \caption{$G_5(Q^2)$ for the lowest non-zero value
      of $Q^2$ using the notation of
      Fig.~\ref{fig:gA_fits}.\label{fig:G5_fits}}
\end{figure}

Similarly, in Figs.~\ref{fig:gA_fits} and~\ref{fig:GA1_fits}, we show, respectively, results for $Q^2=0$ and the lowest momentum transfer for the axial form factor. The data shown in these figures are those obtained from the ratio $\overline{\Pi}_{AP}(Q^2,0)$ defined in Eq.~\eqref{eq:PAP}. We note that the results for the lowest non-zero value of the momentum transfer also have information from $\overline{\Pi}_{0}(Q^2)$ and $\overline{\Pi}_{AP}(Q^2,(2\pi/L)^2)$ given in Eqs.~\eqref{eq:P0} and~\eqref{eq:PAP}. Results on the latter two are shown in Figs.~\ref{fig:P0_fits} and~\ref{fig:PAP_fits}, respectively, from which $G_P(Q^2)$ is extracted for the lowest non-zero value of $Q^2$. In Fig.~\ref{fig:G5_fits} we show the corresponding results for $G_5(Q^2)$ for the lowest non-zero $Q^2$ value.

\begin{figure}[t!]
    \centering
    \includegraphics[width=1.0\linewidth]{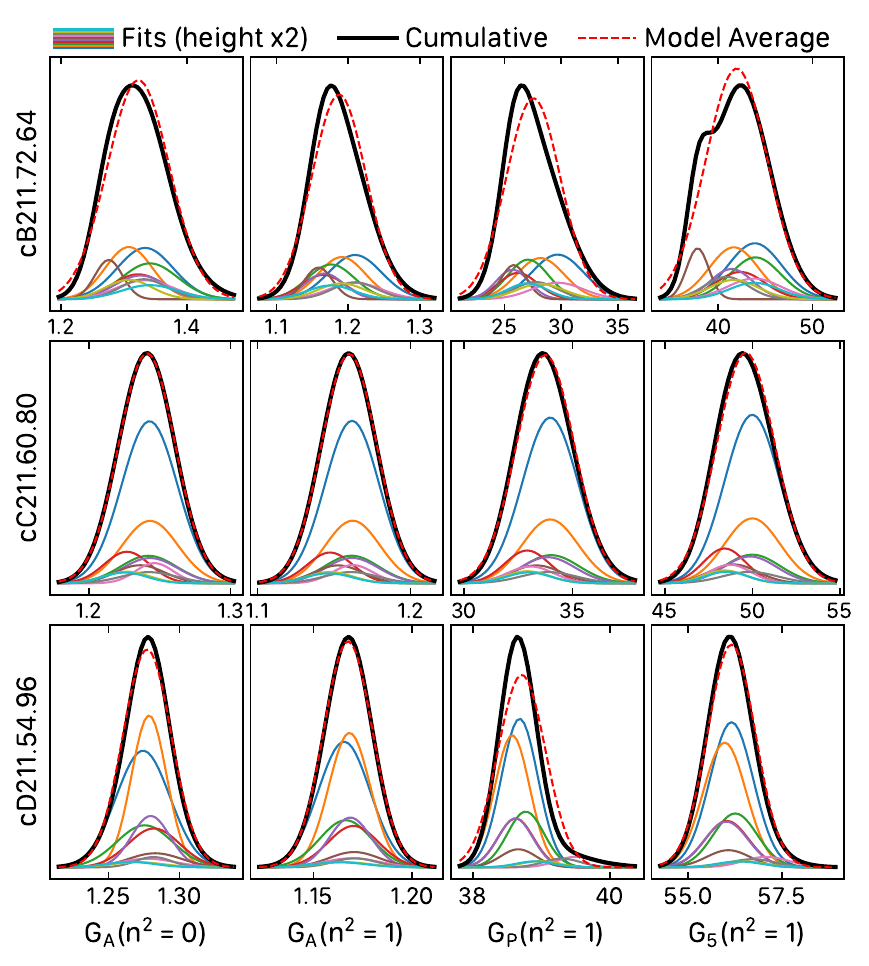}
    \caption{Cumulative distributions of the results from three-state fits weighted by the fit probability. The black curve represents the cumulative distribution; the colored curves are the Gaussian distributions associated with each fit; and the dashed red curve depicts the Gaussian distribution associated with the outcome of the model average given in Eq.~\eqref{eq:model_average}. For each panel, all distributions are normalized with the maximum value of the cumulative distribution and we have doubled the height of the colored curves for visualization purposes.\label{fig:distributions}}
\end{figure}

We note that in most cases the model average gives a central value and error that is compatible with the corresponding ones of the most probable fit. In the few cases where there is a discrepancy, as e.g. in the three-state fit results for $G_{A,1}$ in Fig.~\ref{fig:GA1_fits}, the outcome of the model average has a Gaussian distribution and thus the model average procedure, given in Eq.~\eqref{eq:model_average}, properly accounts for the systematics arising from the model choices. We demonstrate this by depicting in Fig.~\ref{fig:distributions} the cumulative distribution of the Gaussians associated with each fit and the resulting Gaussian distribution outcome of the model average. We observe an excellent agreement.

\subsection{Determination of axial charge and radius}
\label{sec:direct}
\begin{figure}[t!]
    \centering
    \includegraphics[width=0.95\linewidth]{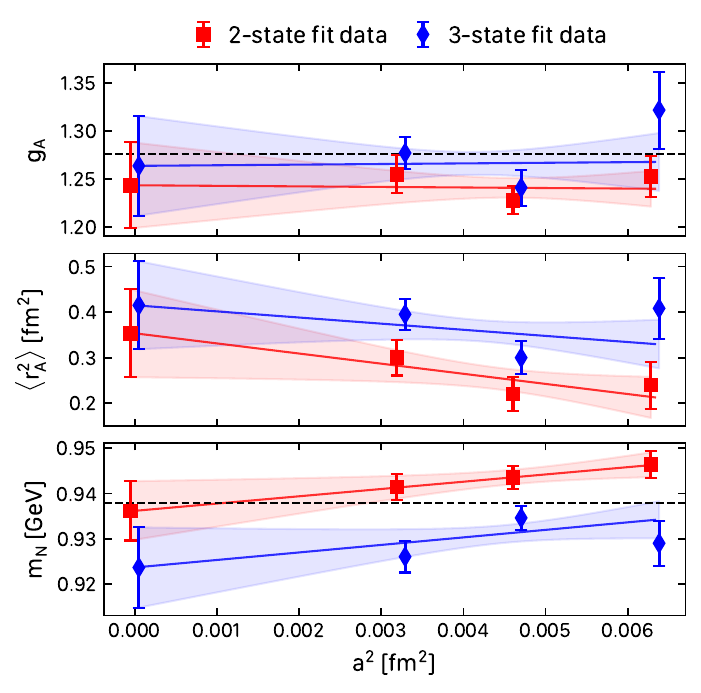}
    \caption{Continuum limit of the nucleon isovector axial charge $g_A$ (top), radius $r_A$ (middle), and nucleon mass $m_N$ (bottom) using a linear extrapolation in $a^2$. The dashed line in the top panel is the experimental value $g_A=1.27641(56)$~\cite{Markisch:2018ndu} and in the bottom panel the Nucleon mass $m_N=938$~MeV.}
    \label{fig:gA_zero_mom}
\end{figure}

Before presenting the analysis of the $Q^2$-dependence of form factors, we perform fits to the zero and the lowest non-zero momentum transfers. For $Q^2=0$ only  $G_A(Q^2)$ can be extracted, yielding the isovector axial charge $g_A$. Computing the slope using the values of $G_A(Q^2)$ at these two values  yields the radius $\langle r_A^2\rangle$, namely
\begin{equation}
    \langle r_A^2\rangle = -\frac{6}{Q^2_1}\left(\frac{G_{A}(Q^2_1)}{G_A(0)}-1\right)\,,
\end{equation}
where $Q^2_1$ is the lowest non-zero momentum transfer squared.
Results obtained using two- or three-state fits are analyzed separately. The results on $g_A$ and  $r_A^2$ extracted after model averaging for each ensemble are collected in Table~\ref{tab:zero_mom} and depicted in Fig.~\ref{fig:gA_zero_mom}. We also include the results obtained for $m_N$. We perform a linear extrapolation in $a^2$ to the continuum limit.  We observe a very good agreement at the continuum limit between the results from fits using two- and three-states for all three quantities. On the other hand, there are slight deviations at finite lattice spacings between two- and three-state fits. For this reason, we analyze separately all quantities using two- and three-state fits.
We take as our mean value the one extracted using the two-state fit and we give as a systematic error the difference between the central values using two- and three-state fits. We find
\begin{equation}
\begin{aligned}
    g_A&=1.244(45)(20)\\
    r_A^2&=0.354(96)(61)~{\rm fm}^2.
\end{aligned}\quad
\text{(direct approach)}
\label{Eq:gA_rA_direct}
\end{equation}

\begin{table}[t!]
    \centering
    \begin{tabular}{c|c|c|c|c}
    \cmidrule{2-5}\morecmidrules\cline{2-5}
    & Ensemble & $g_A$ & $\langle r_A^2\rangle$  [rm$^2$] & $m_N$ [GeV]  \\
\hline\parbox[t]{2mm}{\multirow{4}{*}{\rotatebox[origin=c]{90}{2-state fit}}} & \texttt{cB211.72.64} & 1.253(21) & 0.240(52) & 0.9464(30) \\
 & \texttt{cC211.60.80} & 1.228(14) & 0.220(37) & 0.9436(25) \\
 & \texttt{cD211.54.96} & 1.255(20) & 0.300(39) & 0.9414(29) \\
\cline{2-5} & $a=0$ & 1.244(45) & 0.354(96) & 0.9362(65) \\
\hline\hline\parbox[t]{2mm}{\multirow{4}{*}{\rotatebox[origin=c]{90}{3-state fit}}} & \texttt{cB211.72.64} & 1.322(40) & 0.408(67) & 0.9290(50) \\
 & \texttt{cC211.60.80} & 1.241(19) & 0.300(37) & 0.9346(27) \\
 & \texttt{cD211.54.96} & 1.277(17) & 0.395(34) & 0.9261(35) \\
\cline{2-5} & $a=0$ & 1.264(52) & 0.415(97) & 0.9237(89) \\
\hline\hline
    \end{tabular}
    \caption{Values for the nucleon isovector axial charge $g_A$,
      radius $\langle r_A^2\rangle$, and nucleon mass $m_N$ for each
      ensemble and extrapolated to the continuum limit using a linear
      function in $a^2$. These results are referred to as obtained via
      the ``direct approach'' in the text. Results are given
      separately for values extracted from a two- and a three-state
      fit analysis.}
    \label{tab:zero_mom}
\end{table}

\subsection{Energy spectrum and dispersion relation}

\begin{figure}[t!]
    \centering
    \includegraphics[width=0.95\linewidth]{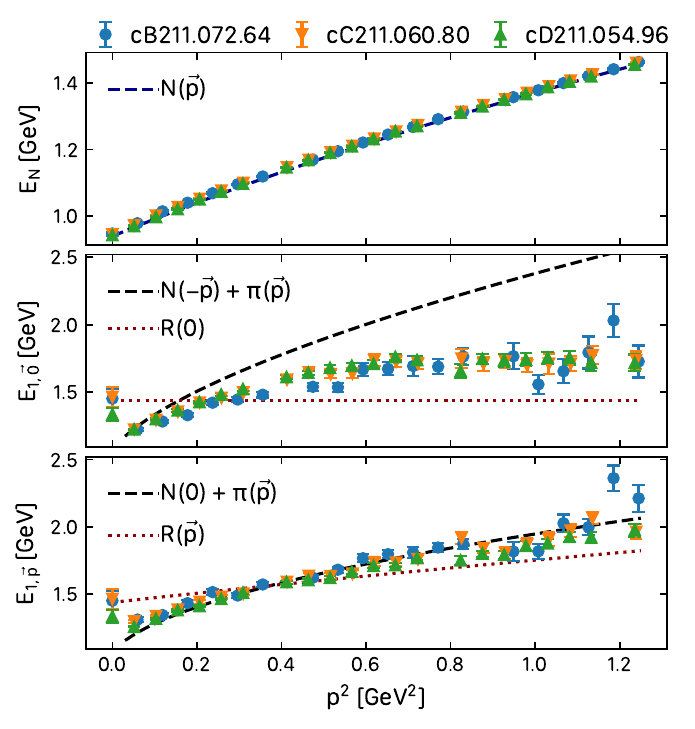}
    \caption{Nucleon  $E_N(\vec{p})=\sqrt{m_N^2+\vec{p}^2}$ (top) and first excited state $E_{1,\vec{0}}$ (middle) at sink and  $E_{1,\vec{p}}$ (bottom) at source as a function of  $\vec{p}^2$. These are determined by performing a two-state fit to the two- and three-point functions. To guide the eye, we depict the expected dispersion relations: the energy of a boosted nucleon $E_N$ (top), a non-interacting $\pi N$ state with either total momentum zero, $N(-\vec{p})+\pi(\vec{p})$ (middle), or with momentum $\vec{p}$, $N(\vec{0})+\pi(\vec{p})$ (bottom); and the energy or the first excited state of the nucleon, the Roper with $m_R=1.44$~GeV, either at rest ($R(0)$, middle) or moving with momentum $\vec{p}$ ($R(\vec{p})$, bottom).}
    \label{fig:E_p2}
\end{figure}

As customarily done in similar studies~\cite{Jang:2019vkm, RQCD:2019jai, Alexandrou:2020okk, Jang:2023zts}, we analyze the first excited state at the source and sink, $E_{1,\vec{p}}$ and $E_{1,\vec{0}}$, respectively. These are obtained using two-state fits to the three-point functions. Our results are depicted in Fig.~\ref{fig:E_p2} for the three ensembles, where we also depict the dispersion relation for the nucleon energy $E_N$. We observe the following:
\begin{itemize}
    \item Since the dispersion relation is included using priors when fitting each $Q^2$ value larger than the lowest non-zero it is not surprising that we see excellent agreement between the extracted energy and the dispersion relation;
    \item The first excited state at zero momentum transfer is compatible with the Roper; 
      \item $E_{1,\vec{0}}$ for low values of the momentum is compatible with  the lowest energy  of the $\pi N$ system in the rest frame, namely $\pi$ and $N$   moving with momentum $p$ back to back; The dependence on the momentum is due to the strong enhancement of the $\pi N$ excited state due to the pion pole~\cite{RQCD:2019jai}; As the momentum grows the lowest energy of $\pi N$ becomes larger than the mass of the Roper and then $E_{1,\vec{0}}$ becomes approximately constant somewhat above the mass of the Roper which is expected since in a two-state fit the first excited energy is contaminated by higher states;
    \item $E_{1,\vec{p}}$ is compatible with   $N(0)+\pi(\vec{p})$ for all non-zero values of $\vec{p}^2\le 0.6$~GeV$^2$. After that, the energy of the Roper denoted by   $R(\vec{p})$ becomes smaller and the results tend to be in between the energy of the Roper and the energy of the $N(0)+\pi(\vec{p})$ system. These are, indeed, the two lowest one-particle and two-particle excited state energies;
    \item Excited state contamination is similar for all three ensembles and this is in line with the observation of mild cut-off effects for the nucleon mass.
\end{itemize}

\subsection{Comparison of results extracted with two- and three-state fits}

\begin{figure*}[t!]
    \centering
    \includegraphics[width=0.95\linewidth]{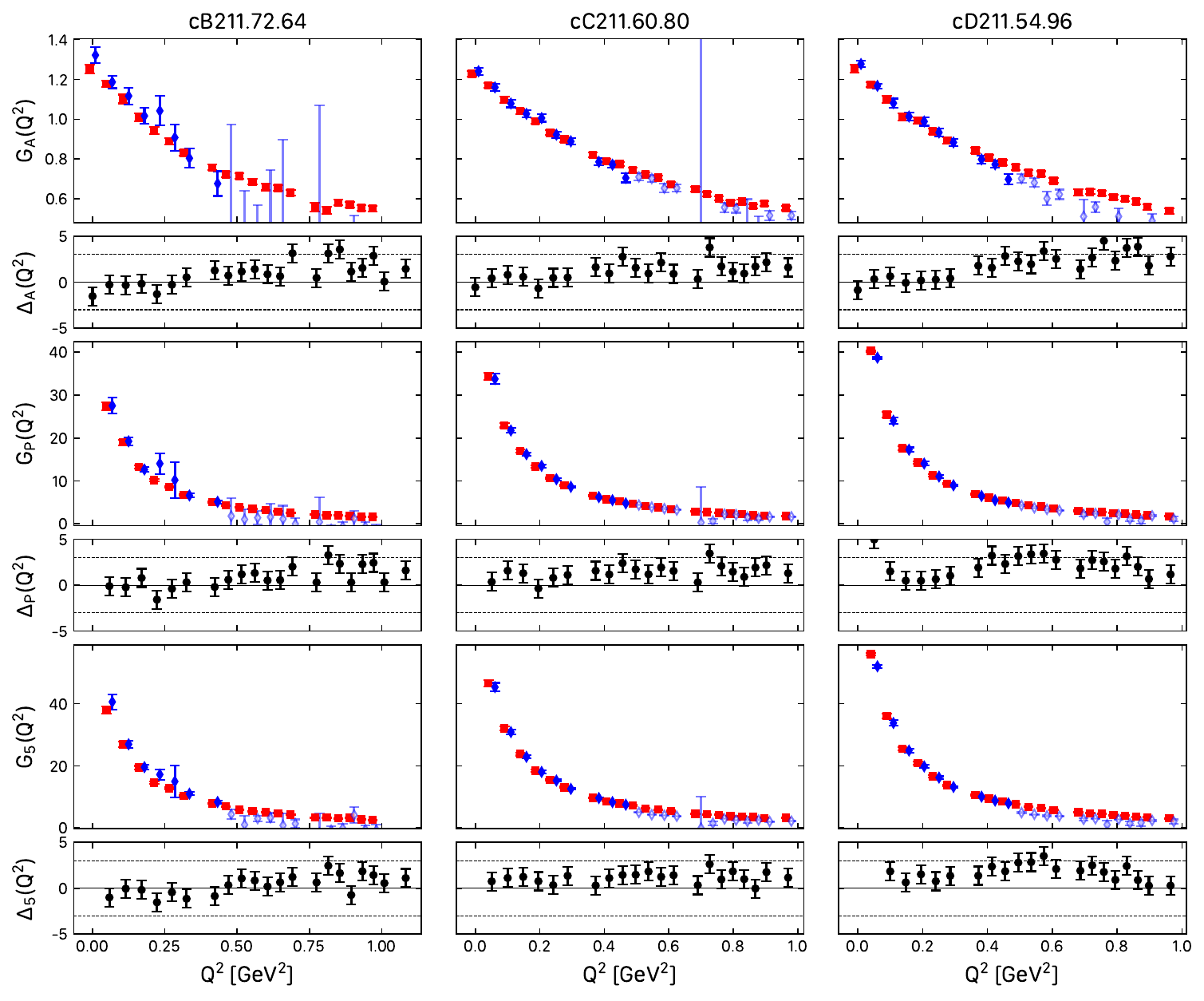}
    \caption{Results for the three form factors obtained using a two- (red squares) or a three-state (blue crosses) fit analysis. From left to right we show results for the \Bens{}, \Cens{}, and \Dens{} ensembles. From top to bottom, we show results for the axial, induced pseudoscalar, and pseudoscalar form factors. In the lower panel of each plot, we include the difference $\Delta$ defined in Eq.~\eqref{Eq:delta} between the results extracted using two- and three-state fits. The difference $\Delta$ is normalized such that errors are unity and the dashed line represents a three standard deviation difference. Three-state fits are stable for  $Q^2<0.465$~GeV$^2$ for all three ensembles. For larger values the fits become unstable. We display these points by using lighter blue color and we thus do not use them in the extraction of form factors. 
    }
    \label{fig:FFcomparison}
\end{figure*}

The renormalized form factors obtained using two- and three-state fits are compared in Fig.~\ref{fig:FFcomparison}, where we also depict the difference $\Delta$ between the results extracted using two- and three-state fit normalized such that errors are unity, namely
\begin{equation}\label{Eq:delta}
    \delta \equiv \langle G_{\rm 2st}-G_{\rm 3st}\rangle\quad\text{and}\quad \Delta(Q^2) \equiv\frac{\delta(Q^2)}{\sigma_{\delta}(Q^2)},
\end{equation}
with $\sigma_{\delta}$ the jackknife error on the difference $\delta$.
We observe a very good agreement between the results extracted using two- and three-state fits with all differences $\Delta$ lying within three standard deviations. We note that we only perform three-state fits to extract the form factors up to  $Q^2\simeq 0.5~{\rm GeV}^2$ despite the fact that the stability of the three-state fits improves because time separations become more dense as the lattice spacing decreases. Since we observe consistency between the results from two- and three-state fits and the two-state fits do not suffer from instabilities at large values of $Q^2$, we opt to take the results from two-state fits up to $Q^2\simeq 1~{\rm GeV}^2$. The three-state fits are only considered up to $Q^2\simeq 0.5~{\rm GeV}^2$, which is the largest $Q^2$ where the three-state fits for the \Bens{} are stable.

\section{$Q^2$-dependence and continuum limit} \label{sec:Q2Fit}

We will first discuss the parameterization of the $Q^2$ dependence of form factors that are free from the pion pole such as the axial form factor. Typically two functional forms are employed, the  dipole Ansatz and the model
independent $z$-expansion~\cite{Hill:2010yb,Bhattacharya:2011ah}.
The dipole Ansatz, given by
\begin{equation}
G(Q^2) = \frac{g}{(1+\frac{Q^2}{m^2})^2},
\label{Eq:dipole}
\end{equation}
 has two parameters, the charge $g$ and the dipole mass $m$. 
In this case, the radius defined in Eq.~\eqref{Eq:radius} is given by
\begin{equation}
    r^2 = \frac{12}{m^2}.
\end{equation}
We parameterize cut-off effects of the charge and of the radius using a linear function in $a^2$, namely
\begin{equation}\label{{Eq:ch_r_a}}
    g(a^2)=g_0+a^2g_2\quad\text{and}\quad
    r^2(a^2)=r_0^2+a^2r^2_2    
\end{equation}
and obtain for the dipole Ansatz the following combined $(Q^2,a^2)$-dependence
\begin{equation}
G(Q^2,a^2) = \frac{g(a^2)}{(1+\frac{Q^2}{12}r^2(a^2))^2},
\label{Eq:dipole_a}
\end{equation}
which we will use to fit all form factors after factoring in any pion pole dependence for a given lattice spacing. 

In the case of the $z$-expansion,  the form factor is parameterized as,
\begin{equation}
G(Q^2) = \sum_{k=0}^{k_{\rm max}} a_k\; z^k(Q^2), 
\label{Eq:zExp}
\end{equation}
where 
\begin{equation}
z(Q^2) = \frac{\sqrt{t_{\rm cut} + Q^2} - \sqrt{t_{\rm cut}+t_0} }{ \sqrt{t_{\rm cut} + Q^2} + \sqrt{t_{\rm cut}+t_0} }
\label{Eq:zform}
\end{equation}
with $-t_{\rm cut} < t_0 < \infty$ and $t_0$ an arbitrary number and  $t_{\rm cut}$ the particle production threshold.  For  $t_{\rm cut}$, we use the three-pion production threshold, namely   $t_{\rm cut}=\left(3 m_\pi\right)^2$~\cite{Bhattacharya:2011ah} with $m_\pi=0.135$~GeV. For $t_0$ we use a vanishing value such that the charge is given by $a_0$ and the radius is proportional to the ratio $a_1/a_0$, namely
\begin{equation}
    g = a_0\quad\text{and}\quad r^2 = -\frac{3a_1}{2a_0t_{\rm cut}}\quad\text{with}\quad t_0=0.
\end{equation}
We introduce the dependence on the lattice spacing by writing
\begin{equation}
G(Q^2,a^2) = g(a^2)\sum_{k=0}^{k_{\rm max}} c_k(a^2)\; z^k(Q^2), 
\label{Eq:zExp_a}
\end{equation}
where $c_k=a_k/a_0$ and
\begin{equation}\label{Eq:zexp_coeffs}
\begin{aligned}    
    c_0(a^2)&=1\,,\quad c_1(a^2)=-\frac{2t_{\rm cut}}{3}r^2(a^2)\quad\text{and}\\
    c_k(a^2)&=c_{k,0}+a^2c_{k,2}\quad\text{for}\quad k\ge2.
\end{aligned}
\end{equation}
The coefficients $c_k$ can be further constrained by requiring that the $z$-expansion converges smoothly to zero at infinite momentum, namely~\cite{Lee:2015jqa}
\begin{equation}
    \sum_{k=0}^{k_{\rm max}} c_k\; \frac{d^nz^k}{dz^n}\bigg|_{z=1} \!\!\!\!= 0\quad\text{with}\quad n=0,1,\dots
    \label{Eq:z_conv}
\end{equation}
This suggests that priors centered around zero should be used to help enforce this condition at various orders with a width that falls like $1/k$~\cite{Meyer:2016oeg}.
Additionally, an examination of the explicit spectral functions and scattering data~\cite{Bhattacharya:2011ah} motivates the bound of $|c_k|\le 5$. We, therefore, use the following Gaussian priors 
\begin{equation}\label{Eq:zexp_priors}
    c_{k,0}\sim 0(w/k),\quad c_{k,2}\sim 0(20w/k)\quad\text{for}\quad k\ge2,
\end{equation}
where $w\le5$ is a fitting parameter that we vary together with the order of the expansion $k_{\max}\in[1,4]$.

In both the dipole and $z$-expansion fits that follow, we will refer to one- and two-step fits. In two-step fits we first fit the $Q^2$ dependence for each lattice spacing separately and then take the continuum limit of the parameters, while in the one-step fits the three ensembles are fitted together. The one-step approach provides for a global $\chi^2$.

\section{The axial form factor $G_A(Q^2)$}\label{sec:GA}

We first present the analysis of the axial form factor, which at $Q^2=0$ yields the axial charge already discussed in  Sec.~\ref{sec:AP_ME}.

\subsection{Dipole Ansatz}
\label{sec:dipole}
\begin{figure}[t!]
    \centering
    \includegraphics[width=0.95\linewidth]{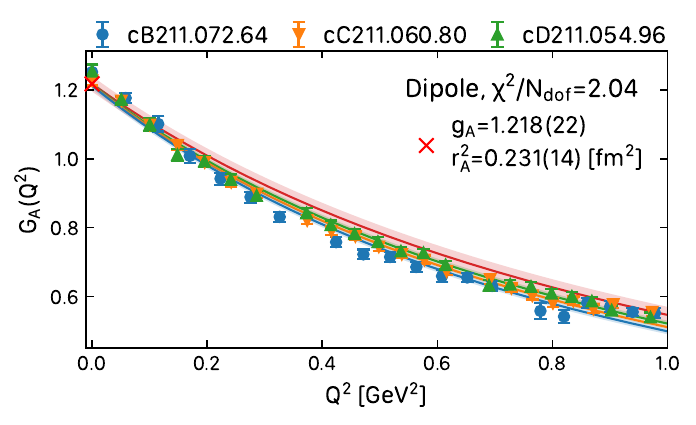}
    \caption{The axial form factor obtained on each of the three ensembles using two-state fits (blue circles, orange downwards pointing triangles, and green triangles). The continuum limit form factor (red curve and band) and value for the axial charge (red cross) are obtained via dipole fits within the one-step approach. Also shown are the form factor curves obtained at the three values of the lattice spacing (blue, orange, and green curves, respectively).}
    \label{fig:GA_dipole}
\end{figure}

An example fit using the dipole Ansatz is shown in Fig.~\ref{fig:GA_dipole}, where we depict the results of using Eq.~\eqref{Eq:dipole_a} to perform a combined fit of the form factors for the three ensembles, i.e. following a one-step approach. The values for $G_A(Q^2)$ shown in this figure are obtained using two-state fits in the range $0\le Q^2<1$~GeV$^2$.

\begin{figure}[t!]
    \centering
    \includegraphics[width=0.9\linewidth]{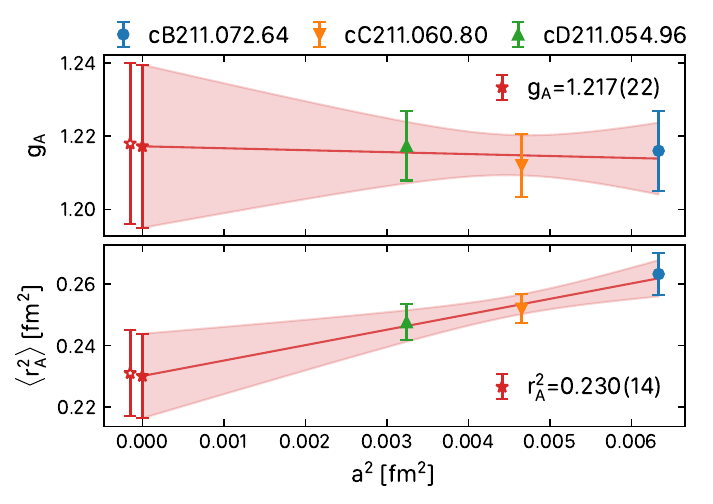}
    \caption{The axial charge (top) and radius (bottom) obtained via a dipole fit to each ensemble (blue circles, orange downwards pointing triangles, and green triangles). The filled red asterisks and corresponding bands show the continuum limit using the two-step approach, i.e. via fits linear in $a^2$ to the axial charge and radius obtained at each value of $a^2$. At $a^2=0$ we compare with results obtained from a one-step approach (open asterisks), as described in the text.}
    \label{fig:gA_rA_dipole}
\end{figure}

As already mentioned, an alternative to the two-step approach is to perform a simultaneous fit to the $Q^2$ dependence for all three ensembles. To demonstrate that taking a global fit in a one-step approach is equivalent to performing a two-step approach, we show in Fig.~\ref{fig:gA_rA_dipole} the continuum limit of the axial charge and the radius extracted from the one- and two-step approaches. In Table~\ref{tab:gA_rA_dipole}, we give the corresponding values extracted when using the one- and two-step approaches, including their reduced $\chi^2$. As can be seen, the continuum values are in perfect agreement both in terms of the central value and in terms of the error. The reduced $\chi^2$ for the two-step procedure only refers to the linear extrapolation to the continuum limit. The one-step approach provides for a single value of $\chi^2$ that reflects the quality of the fit to the combined $Q^2$- and $a^2$-dependence and it is thus more practical to compute the relative weights between the various fits when carrying out our model averaging. Therefore, from now on we will proceed with the one-step approach.
\begin{table}[t!]
    \centering
    \begin{tabular}{c|c|c|c}
    \hline\hline Ensemble & $g_A$ & $\langle r^2_A \rangle$ [fm$^2$] & $\chi^2/N_{\rm dof}$ \\
\hline\texttt{cB211.72.64} & 1.216(11) & 0.2632(67) & 3.72\\
\texttt{cC211.60.80} & 1.2120(85) & 0.2519(47) & 1.61\\
\texttt{cD211.54.96} & 1.2174(95) & 0.2476(59) & 1.16\\
\hline $a=0$, 1-step & 1.218(22) & 0.231(14) & 2.04\\
$a=0$, 2-step & 1.217(22) & 0.230(14) & 0.19\\
\hline\hline
    \end{tabular}
    \caption{Values for the nucleon isovector axial charge $g_A$, radius $r_A$, and reduced $\chi^2$ obtained using a dipole Ansatz to fit each ensemble (first three rows). We also give the continuum limit values using the one-step (second to last row) and two-step (last row) approaches.}
    \label{tab:gA_rA_dipole}
\end{table}

\begin{figure}[t!]
    \centering
    \includegraphics[width=0.95\linewidth]{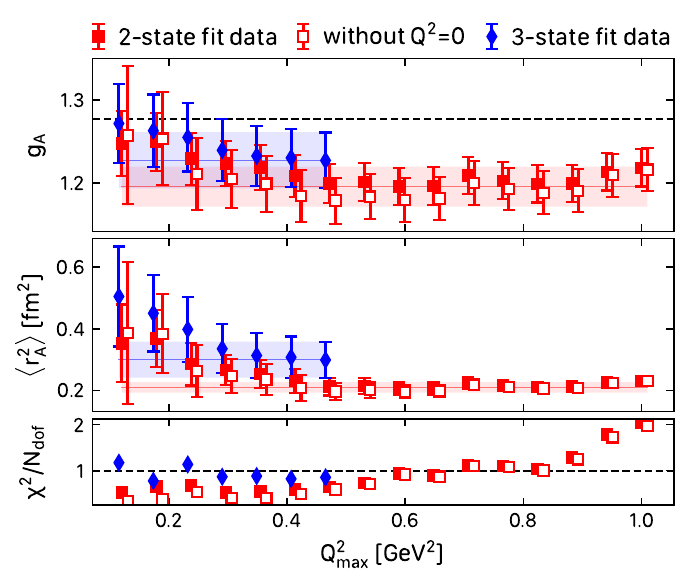}
    \caption{Results for the axial charge and radius extracted using the dipole Ansatz, versus $Q^2_{\rm max}$, i.e. the maximum value of $Q^2$ included in the fit. We show separately dipole fits to results obtained from two- (red squares) and three- (blue diamonds) state fits to the three- and two-point correlation functions. For the two-state fit case, we include a variation in which the values of the form factors at $Q^2=0$ are not included in the fit (open squares).}
    \label{fig:gA_rA_dipole_qmax}
\end{figure}

Using the one-step approach, we perform dipole fits to results obtained using two- and three-state fits to the correlators. We vary the largest $Q^2$ included in the fits, and for the two-state fit results, which are more precise, we also repeat the fits omitting the result at $Q^2=0$. The reasoning is that at $Q^2=0$ only $G_A(0)$ survives, which can affect the determination of the energy extracted for the first excited state, as already shown in Fig.~\ref{fig:E_p2}. A comparison of the results obtained using these variations is shown in Fig.~\ref{fig:gA_rA_dipole_qmax}. We perform a model average of the results separately for the case of using two- and three-state fits on the correlators. We find
\begin{equation}
\begin{aligned}
    g_A&=1.196(24) & \text{(2-state)}\\
    &=1.228(34) &\text{(3-state)}\\
    \langle r^2_A \rangle&=0.210(17)~{\rm fm}^2 &\text{(2-state)}\\
    &=0.300(59)~{\rm fm}^2\,. &\text{(3-state)}
\end{aligned} \quad\text{(dipole Ansatz)}
\end{equation}
Since the values are compatible, we opt to quote the model-averaged values obtained from data that were extracted using two-state fits to the correlators.  We then give as a systematic error the difference between the central values of the model-averaged results obtained from data extracted using two- and three-state fits to the correlators. We find
\begin{equation}
\begin{aligned}
    g_A&=1.196(24)(32)\\
    \langle r^2_A \rangle&=0.210(17)(90)~{\rm fm}^2\,.
\end{aligned}
\quad\text{(dipole Ansatz)}
\label{Eq:gA_rA_dipole}
\end{equation}

\subsection{ $z$-expansion}
\subsubsection{First order $z$-expansion}
\label{sec:z1_exp}
\begin{figure}[t!]
    \centering
    \includegraphics[width=\linewidth]{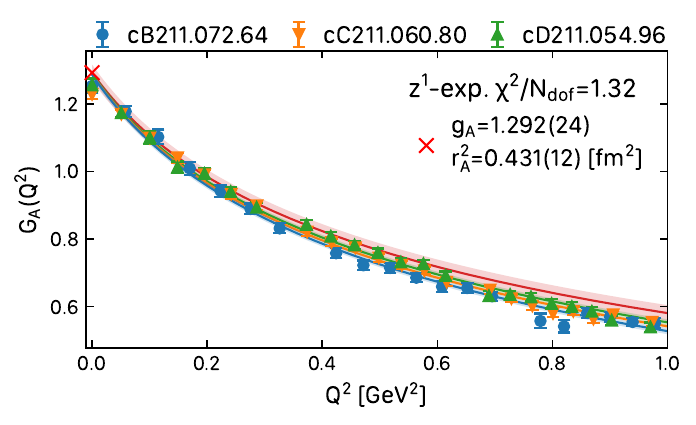}
    \caption{The same as in Fig.~\ref{fig:gA_rA_dipole}, but using the
      first-order $z$-expansion to fit the $Q^2$ dependence.}
    \label{fig:GA_zexp1}
\end{figure}

\begin{figure}[t!]
    \centering
    \includegraphics[width=0.9\linewidth]{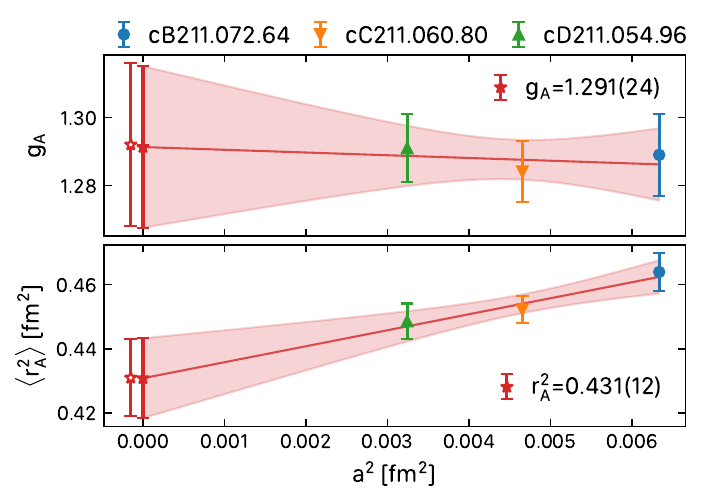}
    \caption{The axial charge (top) and radius (bottom) obtained via
      first-order $z$-expansion fits to each ensemble. The notation is
      the same as in Fig.~\ref{fig:gA_rA_dipole}.}
    \label{fig:gA_rA_zexp1}
\end{figure}

\begin{figure}[t!]
    \centering
    \includegraphics[width=0.95\linewidth]{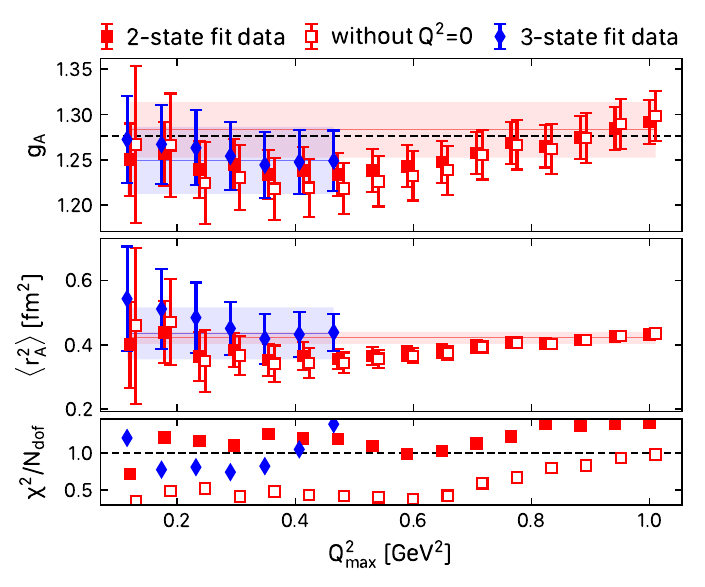}
    \caption{Results for the axial charge and radius extracted using
      the first-order $z$-expansion, versus $Q^2_{\rm max}$. The
      notation is the same as in Fig.~\ref{fig:gA_rA_dipole_qmax}.}
    \label{fig:gA_rA_zexp1_qmax}
\end{figure}

We repeat the same procedure using a first-order $z$-expansion that has the same number of parameters as the dipole Ansatz, and where no priors are employed. In Fig.~\ref{fig:GA_zexp1} we demonstrate the one-step approach as an example, with the same notation as Fig.~\ref{fig:gA_rA_dipole}; in Fig.~\ref{fig:gA_rA_zexp1} and Table~\ref{tab:gA_rA_zexp1} we similarly demonstrate that our one-step approach is equivalent to the two-step approach; and in Fig.~\ref{fig:gA_rA_zexp1_qmax} we depict the results as a function of $Q^2_{\rm max}$ for the case of data extracted using two- and three-state fits to the correlators.  After model averaging we find
\begin{equation}
\begin{aligned}
    g_A&=1.283(31) & \text{(2-state)}\\
    &= 1.249(37)& \text{(3-state)}\\
    \langle r^2_A \rangle&=0.421(18)~{\rm fm}^2 & \text{(2-state)}\\
    &=0.435(81)~{\rm fm}^2\,. &\text{(3-state)}
\end{aligned}\quad (z^1\text{-expansion)}
\end{equation}
Again, we observe that the values are compatible whether we use results from the two- or three-state fits to the correlators and thus we quote the model-averaged value for the case of using two-state fits and take as systematic error the difference between the central values obtained when using two- and three-state fits to the correlators. We find
\begin{equation}\label{Eq:gA_rA_zexp1}
\begin{aligned}
    g_A&=1.283(31)(34)\\
    \langle r^2_A \rangle&=0.421(18)(14)~{\rm fm}^2\,. 
\end{aligned}\quad (z^1\text{-expansion)}
\end{equation}

\begin{table}[t!]
    \centering
    \begin{tabular}{c|c|c|c}
    \hline\hline Ensemble & $g_A$ & $\langle r^2_A \rangle$ [fm$^2$] & $\chi^2/N_{\rm dof}$ \\
\hline\texttt{cB211.72.64} & 1.277(12) & 0.4349(54) & 1.43\\
\texttt{cC211.60.80} & 1.2739(90) & 0.4248(41) & 1.34\\
\texttt{cD211.54.96} & 1.281(10) & 0.4214(52) & 0.79\\
\hline $a=0$, 1-step & 1.292(24) & 0.431(12) & 1.32\\
$a=0$, 2-step & 1.291(24) & 0.431(12) & 0.29\\
\hline\hline
    \end{tabular}
    \caption{Values for the nucleon isovector axial charge $g_A$, radius $r_A$, and reduced $\chi^2$ obtained using a first-order $z$-expansion fit on each ensemble and extrapolated to the continuum limit using our one- or two-step approach.}
    \label{tab:gA_rA_zexp1}
\end{table}

\subsubsection{Convergence of the $z$-expansion}
\label{sec:zk_exp}
\begin{figure}[t!]
    \centering
    \includegraphics[width=0.95\linewidth]{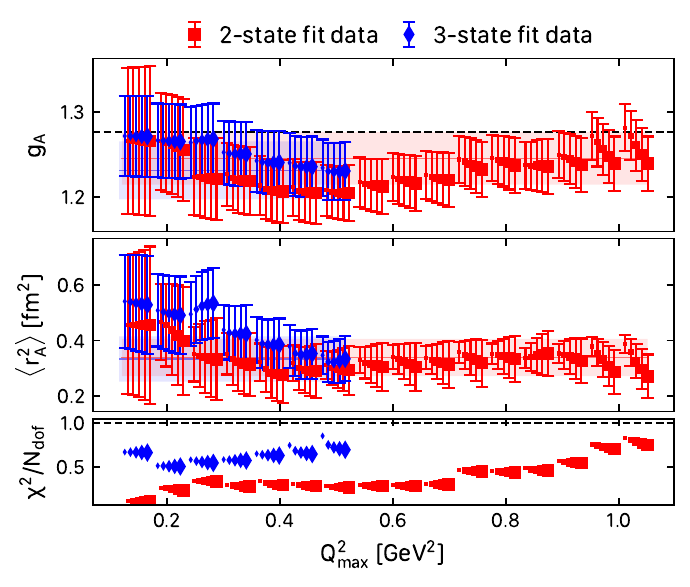}
    \caption{Results for the axial charge and radius as a function of $Q^2_{\rm max}$ obtained from using the  $z^3$-expansion to parameterize the $Q^2$-dependence. For each $Q^2_{\max}$ we depict five points having prior width $w=1,2,3,4,5$. The points are shifted to the right as $w$ increases with an increasing symbol size.}
    \label{fig:gA_rA_zexpN_qmax}
\end{figure}

In order to check the stability of the $z$-expansion fits, we study the convergence of the $z$-expansion as a function of the order and the amplitude of the priors used in Eq.~\eqref{Eq:zexp_priors}. Priors are used to ensure convergence of the $z$-expansion~\cite{Lee:2015jqa} being centered around zero with a width that decreases with $1/k$.
We observe convergence for $k_{\max}\ge 3$ for all cases. We vary the amplitude of the prior using $w\in[1,5]$. In Fig.~\ref{fig:gA_rA_zexpN_qmax} we depict the extracted axial charge and radius as a function of the prior width and $Q^2_{\rm max}$, the largest $Q^2$ used in the fit. We observe that the results obtained by changing the width of the priors are all consistent. The result of the model average is
\begin{equation}\label{eq:gA_rA_zexpk}
\begin{aligned}
    g_A&=1.245(28) & \text{(2-state)}\\
    &= 1.231(34)& \text{(3-state)}\\
    \langle r^2_A \rangle&=0.339(48)~{\rm fm}^2 & \text{(2-state)}\\
    &=0.333(72)~{\rm fm}^2\,. &\text{(3-state)}
\end{aligned}\quad (z^3\text{-expansion)}
\end{equation}
quoting again the value from the data extracted from two-state fits with a systematic error in the difference between the central values of the  model-averaged results when using data from  two- and three-state fit to the correlators
\begin{equation}\label{Eq:gA_rA_zexpk}
\begin{aligned}
    g_A&=1.245(28)(14)\\
    \langle r^2_A \rangle&=0.339(67)(06)~{\rm fm}^2\,. 
\end{aligned}\quad (z^3\text{-expansion)}
\end{equation}

\begin{figure}[t!]
    \centering
    \includegraphics[width=\linewidth]{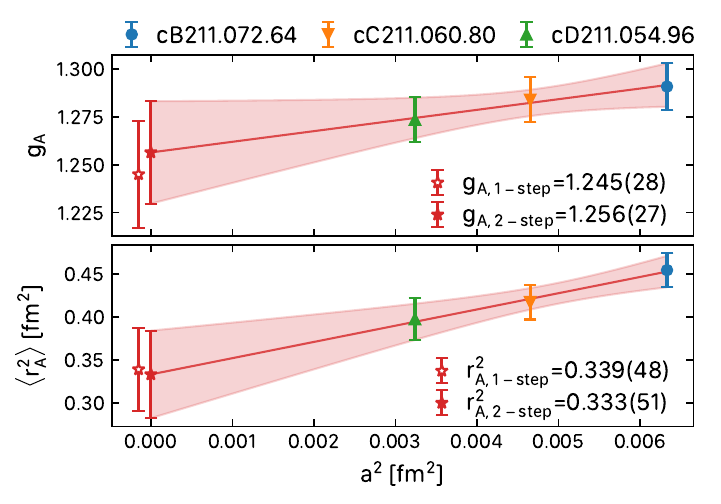}
    \caption{The axial charge (top) and radius (bottom) obtained via
    third-order $z$-expansion fits and model average for each ensemble. The notation is the same as that in Fig.~\ref{fig:gA_rA_dipole}.}
    \label{fig:gA_rA_zexpN}
\end{figure}

Following a similar procedure to the one-step approach we perform the fits in two steps, namely we first perform the fits and model average for each ensemble and then take the continuum limit.  This two-step approach yields results that are compatible with the one-step approach, as depicted in Fig.~\ref{fig:gA_rA_zexpN} with results provided in Table~\ref{tab:gA_rA_zexpN}.

\begin{table}[t!]
    \centering
    \begin{tabular}{c|c|c|c}
    \hline\hline Ensemble & $g_A$ & $\langle r^2_A \rangle$ [fm$^2$] & $\chi^2/N_{\rm dof}$ \\
\hline\texttt{cB211.72.64} & 1.291(12) & 0.455(20) & 0.80\\
\texttt{cC211.60.80} & 1.284(12) & 0.417(20) & 0.64\\
\texttt{cD211.54.96} & 1.274(12) & 0.398(25) & 0.58\\
\hline $a=0$, 1-step & 1.245(28) & 0.339(48) & 0.69\\
$a=0$, 2-step & 1.256(27) & 0.333(51) & 0.05\\
\hline\hline
    \end{tabular}
    \caption{Values for the nucleon isovector axial charge $g_A$ and radius $r_A$, and reduced $\tilde{\chi}^2$ of the best fit obtained using a model average over third-order $z$-expansion fits on each ensemble and extrapolated to the continuum limit using the one- or two-step approaches.}
    \label{tab:gA_rA_zexpN}
\end{table}

\subsection{Final results}\label{sec:GA_final}

\begin{figure}[t!]
    \centering
    \includegraphics[width=0.95\linewidth]{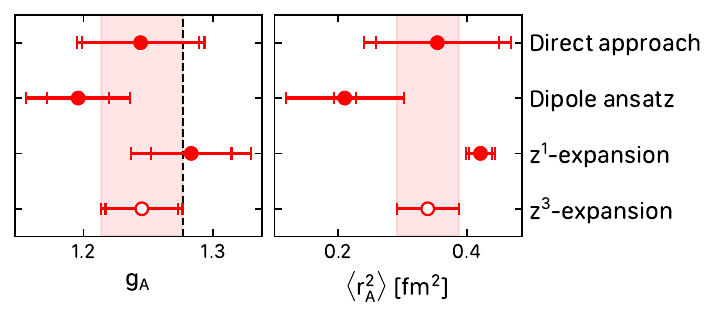}
    \caption{Results for the isovector axial charge and radius extracted using four different approaches as described in the text. Numerical values are given in Table~\ref{tab:gA_rA_final}.}
    \label{fig:gA_rA_final}
\end{figure}

Having presented the variations used to extract the axial charge and radius at the continuum limit, we continue here to discuss the consistency among them and how we choose our final values. To summarize the variations, we have used: i) in Sec.~\ref{sec:direct} a direct determination using the matrix element at $Q^2=0$ and, for the radius the matrix element at the lowest non-zero $Q^2$ value yielding the results of Eq.~\eqref{Eq:gA_rA_direct}; ii) in Sec.~\ref{sec:dipole} the dipole Ansatz to describe the $Q^2$-dependence resulting in the values given in Eq.~\eqref{Eq:gA_rA_dipole}; iii) in Sec.~\ref{sec:z1_exp} the first-order $z$-expansion to describe the $Q^2$-dependence resulting in the values given in Eq.~\eqref{Eq:gA_rA_zexp1}; and iv) in Sec.~\ref{sec:zk_exp} the higher-order $z$-expansion with the resulting values given in Eq.~\eqref{Eq:gA_rA_zexpk}. In Table~\ref{tab:gA_rA_final}, we collect these values and depict them in Fig.~\ref{fig:gA_rA_final}. In all cases, the central value and first error are obtained via model averaging of the two-state fit data, while the second error is a systematic obtained as the difference between the central values when model averaging the two- or three-state fit data.  We note the following
\begin{itemize}
\item We opt to take as our final
values the results extracted from using the two-state fits because they are in general more stable, having fewer fit
parameters and smaller errors and, thus, allowing a better study of the convergence to the ground state. We reiterate that we
observe a very good agreement between results extracted using data determined by fitting correlators to two-  or three-states,  as depicted in Fig.~\ref{fig:FFcomparison}.
\item  All results agree with the value extracted using the $z^k$-expansion with $k=3$ within error bars. Results using the dipole Ansantz and the $z^1$-expansion yield respectively smaller or larger values as compared to those using the $z^3$-expansion.  This observation is compatible with what has been found in another  study~\cite{Jang:2023zts}.
\item Furthermore,  the results using the $z^3$-expansion are completely consistent with those from the direct approach. Since the direct approach uses the matrix element at $Q^2=0$ for $g_A$ and for $\langle r_A^2\rangle$ the slope using also the lowest non-zero value of $Q^2$, it does not depend on any Ansatz used to fit the $Q^2$-dependence of the form factor. The fact that the $z^3$-expansion yields the same results shows that indeed it provides a model-independent approach to extract the same information on these two quantities. The error when using the $z^3$-expansion is smaller compared to the errors when using the direct approach since the $z$-expansion makes use of more information.
\end{itemize}
 Given the above observations, we quote as our final values the
 results from the $z^k$-expansion that has shown convergence for $k=3$
 and is model-independent.  Thus, we take as our final values for the
 axial charge and radius
\begin{equation}\label{Eq:gA_rA_final}
\begin{aligned}
    g_A&=1.245(28)(14)[31]\\
    \langle r_A^2\rangle &=0.339(48)(06)[48]~{\rm fm}^2\,, 
\end{aligned}\quad \text{(final value)}
\end{equation}
where in the square brackets we have combined quadratically the two errors. We also collect all our final values in the conclusions in Eq.~\eqref{Eq:final_values_conclusions}.

\begin{table}[t!]
    \centering
    \begin{tabular}{l|c|c}
    \hline\hline
       Method  &  $g_A$ & $\langle r^2_A\rangle$ [fm$^2$] \\
       \hline
       Direct approach  & 1.244(45)(20) & 0.354(96)(61) \\
       Dipole Ansatz & 1.196(24)(32) &  0.210(17)(90) \\
       $z^1$-expansion & 1.283(31)(34) & 0.421(18)(14) \\
       $z^3$-expansion & 1.245(28)(14) & 0.339(48)(06) \\
       \hline\hline
    \end{tabular}
    \caption{Results for the isovector axial charge and radius extracted using four different approaches as described in the text. The values are depicted in Fig.~\ref{fig:gA_rA_final}.}
    \label{tab:gA_rA_final}
\end{table}

\begin{figure}[t!]
    \centering
    \includegraphics[width=0.95\linewidth]{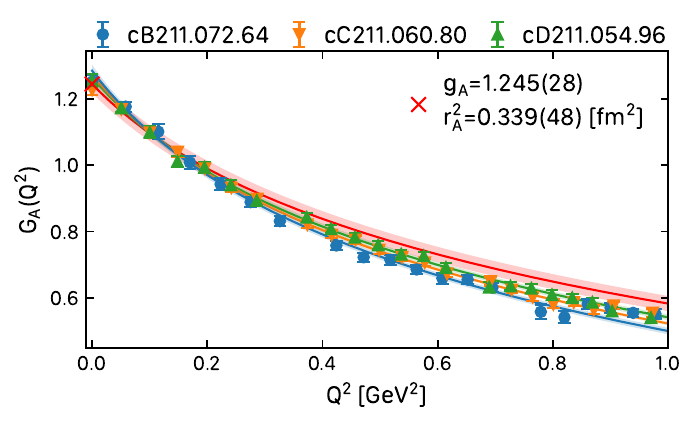}
    \caption{Continuum limit of $G_A(Q^2)$ using the $z^3$-expansion and data from the two-state fit analysis of the correlators up to $Q^2=1$~GeV$^2$ for the three ensembles with the symbols as indicated in the header of the figure.\label{fig:GA_order2}}
    \includegraphics[width=0.95\linewidth]{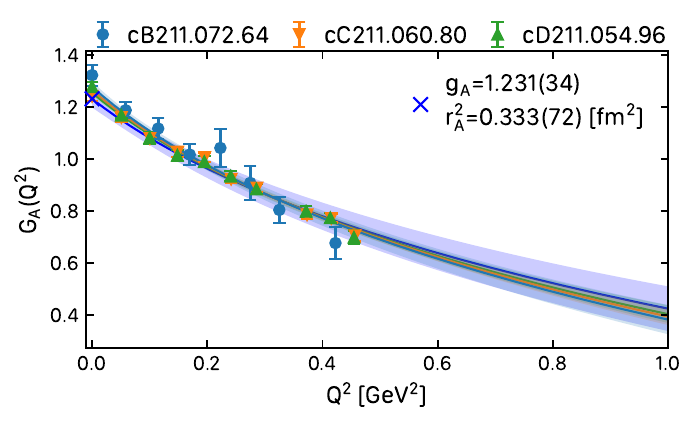}
    \caption{Continuum limit of $G_A(Q^2)$  using the $z^3$-expansion and data from the three-state fit analysis of the correlators up to $Q^2=0.47$~GeV$^2$ for the three ensembles with the symbols as indicated in the header of the figure.\label{fig:GA_order3}}
\end{figure}

\begin{figure}[t!]
    \centering
    \includegraphics[width=0.95\linewidth]{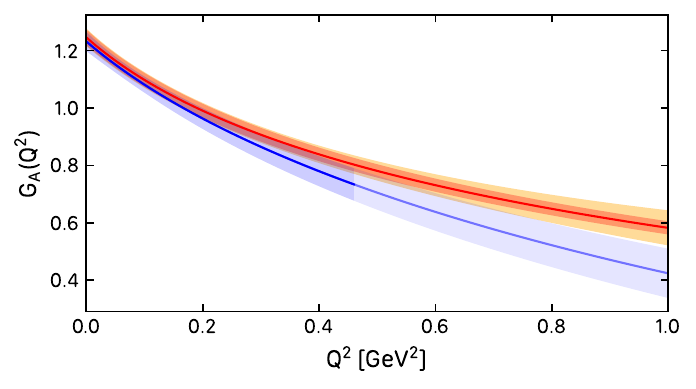}
    \caption{Results on $G_A(Q^2)$ at the continuum limit when fitting data extracted from the two- (red band) and three- (blue band) state fit analysis of the correlators. The darker blue curve indicates up to which $Q^2$ we had data for the three-state fit analysis. The yellow band is when we added systematic errors to the parameters that define the red curve as discussed in the text.   The  parameters of the fit are given in Eq.~\eqref{Eq:GA_params}.\label{fig:GA_order23_comparison}\label{fig:GA_final}}
\end{figure}

In Figs.~\ref{fig:GA_order2} and~\ref{fig:GA_order3} we show the model-averaged results as a function of $Q^2$ when using the $z^3$-expansion for either two- or three-state fits to the correlators, respectively.
As can be seen, in Fig.~\ref{fig:FFcomparison} the data extracted using two- and three-state fits are compatible. However, small statistical fluctuations and the lack of data in the case of the three-state fit analysis for $Q^2>0.5$~GeV$^2$ can affect the fits of the $Q^2$-dependence and thus the continuum limit, given that we only have three lattice spacings. We remind the reader that we perform simultaneously fits to the $Q^2$-dependence for each ensemble and at the same time take the continuum limit.
We compare the resulting $G_A(Q^2)$ in the continuum limit for these two cases in  Fig.~\ref{fig:GA_order23_comparison}, where we give the continuum fits only. The fit parameters of the curves corresponding to the standard form of the $z$-expansion in Eq.~\eqref{Eq:zExp} with $k_{\max}=3$, $t_{\rm cut} = (3 m_\pi)^2$, $m_\pi=0.135$~GeV and $t_0=0$~GeV$^2$ are
\begin{equation}\label{Eq:GA_params_st}
 \begin{aligned}
 \vec{a}_\text{2-state} &= [1.245(28),-1.19(18),-0.54(55),-0.13(59)]\\
\vec{a}_{\rm 3-state} &= [1.231(34),-1.16(27),-0.80(47),-1.23(58)]\,.
\end{aligned}
\end{equation}
As can be seen, the resulting curve for the three-state fit case is in agreement with that for the two-state fit for $Q^2\leq 0.5$~GeV$^2$. Also, the parameters of the two fits are in good agreement. Since however, the three-state fits become unstable for $Q^2>0.5$~GeV$^2$, we take $a_\text{2-state}$ as our central values and the difference between the central values of  $a_\text{2-state}$ and $a_{\rm 3-state}$ as the systematic error to account for systematics due to excited states. 
Our final parameterization for the form factor is then
\begin{equation}\label{Eq:GA_params}
\begin{aligned}
    \vec{a}_A = &\big[1.245(28)(14)[31],-1.19(18)(03)[18],\\&-0.54(55)(26)[61],-0.13(59)(1.1)[1.3]\big]\\
    \text{corr}_{\vec{a},A} =& \begin{pmatrix}
1.0 & -0.421 & 0.247 & -0.246 \\
-0.421 & 1.0 & -0.918 & 0.799 \\
0.247 & -0.918 & 1.0 & -0.952 \\
-0.246 & 0.799 & -0.952 & 1.0 \\
    \end{pmatrix}\,,
\end{aligned}
\end{equation}
where we have used the correlation matrix of the parameters from two-state fit data. More information on the form factors at the continuum limit is provided in Appendix~\ref{sec:appendix}.

We include the resulting $G_A(Q^2)$ when we assign this systematic error to the parameters of the continuum fit in Fig.~\ref{fig:GA_final}. We consider the values of $G_A(Q^2)$ including the systematic uncertainty as our final results. Our final results for the form factor $G_A(Q^2)$ are given in Table~\ref{tab:GA}. We will adopt the same strategy for the analysis of the other two form factors and for checking the PCAC and PPD relations.

\section{Induced pseudoscalar $G_P(Q^2)$ and pseudoscalar $G_5(Q^2)$ form factors}\label{sec:GPandG5}

 We perform a similar analysis to determine $G_P(Q^2)$ and $G_5(Q^2)$ to the one discussed in detail above for $G_A(Q^2)$. The additional complication in the case of $G_P(Q^2)$ and $G_5(Q^2)$ is that both form factors have a pole, at $Q^2=-m_\pi^2$, which needs to be removed before proceeding to apply similar fit functions to the ones applied for $G_A(Q^2)$. Therefore, before proceeding with the $Q^2$-dependence analysis of these form factors, we present a detailed study of pion pole dominance.

\subsection{Pion pole dominance (PPD)}
\label{sec:PPD}

\begin{figure}[t!]
    \centering
    \includegraphics[width=0.95\linewidth]{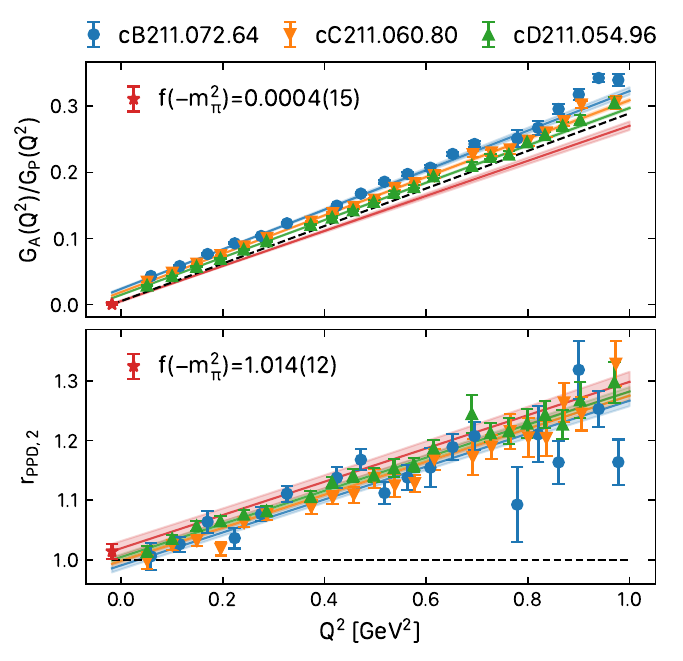}
    \caption{$G_A(Q^2)/G_P(Q^2)$ (top) and $r_{\rm PPD, 2}$ (bottom) for the three ensembles. The blue, orange, and green curves show the results of combined linear fit in $Q^2$ for the \Bens{}, the \Cens{} and \Dens{}, respectively, using the form of Eq.~(\ref{a-ratios}) to take into account cut-off effects. The red band is the continuum extrapolation after performing the model average as described in the text. The dashed line shows the expected value of the ratios if PPD close to the pion pole is satisfied, namely a slope of $1/4m_N^2$ from Eq.~\eqref{Eq:GA_GP} for the first ratio and unity for the second.}
    \label{fig:FF_ratios}
\end{figure}

The pion pole dominance (PPD) hypothesis  introduced in Sec.~\ref{sec:AP_ME} can be tested by forming two ratios of form factors, one of which is 
\begin{align}\label{Eq:GA_GP}
    \frac{G_A(Q^2)}{G_P(Q^2)} = \frac{Q^2+m_\pi^2}{4m_N^2} &\bigg|_{Q^2\rightarrow -m^2_\pi},
\end{align}
  arising from Eq.~\eqref{Eq:PPD} and  the second  $r_{\rm PPD, 2}$ derived in Eq.~\eqref{Eq:rPPD2} assuming a non-zero Goldberger-Treiman discrepancy. Using the results for the form factors from the two-state fits to the correlators we find the ratios depicted in Fig.~\ref{fig:FF_ratios}. We indeed observe for both ratios a linear dependence in $Q^2$, as expected from Eq.~\eqref{Eq:GA_GP} and Eq.~\eqref{Eq:rPPD2_AP}, respectively. We also observe clear cut-off effects for the first ratio whereas for the second the results from the three ensembles are consistent among them. To capture the $a$-dependence  we fit the ratios using the functional form
\begin{equation}
    f(Q^2,a^2)=b_{0}+b_2a^2+(c_0+c_2a^2)Q^2,
    \label{a-ratios}
\end{equation}
where we include the leading order $a$ dependence to both the intercept and the $Q^2$-slope. We also perform fits where we set $b_2$ and $c_2$ to zero to account for the fact that the second ratio $r_{\rm PPD, 2}$ shows no detectable cut-off effects to the accuracy of our data.  We then perform a model average over all fits where we both include and exclude cut-off effects as well as change the largest $Q^2$ value, $Q^2_{\rm max}$, used in the fit. The resulting continuum fits are shown in Fig.~\ref{fig:FF_ratios}.

The conclusions drawn from Fig.~\ref{fig:FF_ratios} are
\begin{itemize}
    \item The pion pole dominance hypothesis  is satisfied for both ratios at the  pole since we obtain the expected value at $Q^2=-m_\pi^2$, namely
    \begin{equation}
    \begin{aligned}
        \frac{G_A(-m_\pi^2)}{G_P(-m_\pi^2)} &= 0.0004(15) \approx 0\\    
        r_{\rm PPD, 2}(-m_\pi^2) &= 1.015(12) \approx 1\,.\\    
    \end{aligned}
    \end{equation}
    \item For the first ratio, $G_A(Q^2)/G_P(Q^2)$, we find a slope and intercept in the continuum limit consistent with the PPD hypothesis. Namely, from the intercept at $Q^2=-m_\pi^2$, we find a value of the pion pole mass at the continuum limit  of
    \begin{equation}
        m_\pi^{\rm pole} = 0.141(20)~{\rm GeV} \approx 0.135~{\rm GeV} 
    \end{equation}
    compatible with the physical pion mass; and from the $Q^2$-slope, whose value is expected to be $1/4m_N^2$, we find a  nucleon mass of
    \begin{equation}
        m_N =  0.9401(39)~{\rm GeV}\approx 0.938~{\rm GeV}\,,
    \end{equation} when the fit is done  with $Q^2_{\rm max}=0.3$~GeV$^{2}$.
   Thus, we conclude that the ratio $G_A(Q^2)/G_P(Q^2)$ satisfies the PPD relation close to the pole.
   
   \begin{table}[t!]
    \centering
    \begin{tabular}{c|c|c|c}
    \hline
    Ensemble & $m^{\rm pole}_{\pi}$ [MeV] & $m^{\rm TM}_{\pi}$ [MeV] & $m_{\pi}^{\rm OS}$ [MeV]   \\
  \hline
  \Bens{} & 299.3(4.5) & 140.2(2) & 297.5(7)   \\
  
  \Cens{} & 266.7(3.2) & 136.6(2) & 248.9(5)  \\
  
  \Dens{} & 235.8(4.8) & 140.8(3) & 210.0(4)   \\
  \hline
    \end{tabular}
    \caption{\label{tab:pole_masses} Pion pole mass extracted from the ratio $G_A(Q^2)/G_P(Q^2)$ for each ensemble, compared to the simulated unitary pion mass and the Osterwalder-Seiler (OS) pion mass.}
\end{table}
   \item Examining the pion pole mass determined from the fits to a given ensemble at finite lattice spacing, we find a significantly different pion pole mass. It is well-known that the twisted mass fermion formulation has significant cut-off effects in the pion mass~\cite{Dimopoulos:2009qv} but much milder ones in other quantities such as other hadron masses~\cite{Alexandrou:2017xwd} or hadronic operator matrix elements~\cite{Dimopoulos:2009qv}. In Table~\ref{tab:pole_masses}, we give the pion masses that we find per ensemble as well as the values of the unitary pion mass, denoted by $m_\pi^{\rm TM}$ and the Osterwalder-Seiler (OS) pion mass~\cite{Osterwalder:1977pc,Frezzotti:2004wz}, denoted by $m_\pi^{\rm OS}$. 
   We would like to clarify that the difference between the charged pion mass $m_\pi^{\rm TM ,+/-}$ and its neutral counterpart $m_\pi^{\rm TM, 0}$ is an ${\cal{O}}(a^2)$ artifact due to the breaking of isospin symmetry in the clover twisted-mass fermion lattice action formulation. We find that the mass difference between charged and neutral unitary pions $m_\pi^{{\rm TM},-/+}-m_\pi^{\rm TM,0}$ is of order $20$--$40$~MeV at the lattice spacings employed here. The uncertainty in the determination of this mass difference arises due to the statistical error of the disconnected quark contribution entering in the computation of $m_\pi^{\rm TM,0}$.  While, the larger difference between $m_\pi^{\rm TM, +/-,0}$ and $m_\pi^{\rm OS}$ is also an ${\cal{O}}(a^2)$  cutoff effect of the OS mixed action, the coefficient is different.
    Technically, the difference between the neutral $m_\pi^{\rm TM,0}$  and $m_\pi^{\rm OS}$ can be traced back to the presence in the two-point correlator of the neutral TM pion of quark-disconnected contributions of ${\cal{O}}(a^2)$ that are absent in the two-point correlator of the OS pion.
    
   We observe that the pion pole mass that we find is very close to $m_\pi^{\rm OS}$. This is because, in our evaluation of $G_P(Q^2)$, we use the flavor diagonal isovector current and neglect the noisy ${\cal{O}} (a^2)$ quark disconnected contributions, which in turn corresponds to computing the three-point correlators in the OS mixed action formulation. We, thus, obtain a neutral pion pole with mass given by the OS pion mass, $m_\pi^{\rm OS}$. In this way, indeed, we can understand the large cut-off effects observed for form factors that have a pion pole behavior within the twisted mass formulation with OS-type valence quarks in contrast to other fermion discretization schemes that observe similar cut-off effects in both $G_A(Q^2)$ and $G_P(Q^2)$, as when e.g. using Clover Wilson fermions~\cite{Jang:2023zts}, although in their formulation the form factors have ${\cal O} (a)$ discretization errors. 

    \item We demonstrate that cut-off effects are due to the pion pole in  Fig.~\ref{fig:PPD_ratios}, where we show results for the ratio $r_{\rm PPD, 1}$ defined in Eq.~\eqref{Eq:rPPD1}  removing the pole using either the unitary or the OS pion mass. Data obtained using the OS pion mass shows, indeed, very mild cut-off effects and yield a value of $r_{\rm PPD, 1}$  close to unity as expected by PPD and as observed by other groups using clover fermions.
    
\begin{figure}[t!]
    \centering
    \includegraphics[width=0.95\linewidth]{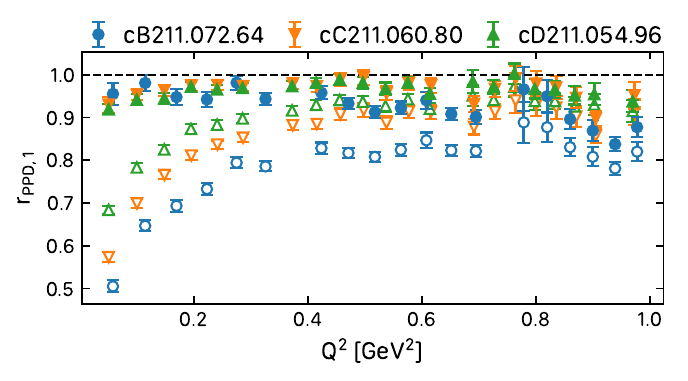}
    \caption{The ratio $r_{\rm PPD, 1}$ for the \Bens{} (blue circles), \Cens{} (orange down-pointing triangles) and \Dens{} (green upwards-pointing triangles) ensembles when using the unitary pion mass $m_\pi^{\rm TM}$  (open symbols) and the OS pion mass $m_\pi^{\rm OS}$ (filled symbols) to remove the pion pole. The pion mass values are given in Table~\ref{tab:pole_masses}. The dashed line shows the  expected value $r_{\rm PPD, 1}=1$ based on PPD.}
    \label{fig:PPD_ratios}
\end{figure}

    \item  The deviation from unity observed for  the  ratio, $r_{\rm PPD, 2}$, is connected to the Goldberger-Treiman discrepancy as given in Eq.~\eqref{Eq:rPPD2_GT} where we followed  a similar analysis to that of  Ref.~\cite{Park:2021ypf}. Using the slope of our fit at the continuum limit to extract $\Delta_{\rm GT}$ from Eq.~\eqref{Eq:rPPD2_GT} and $\bar{d}_{18}$ from Eq.~\eqref{Eq:d18}, we find
    \begin{equation}
    \begin{aligned}
    \Delta_{\rm GT}&=-0.0213(38) \, {\rm or}\,\approx 2\% \\
    \bar{d}_{18} &= -0.73(13)~{\rm GeV}^{-2}.
    \end{aligned}
    \end{equation}
    We note that in extracting these values we use our final values for $g_A$ and $\langle r_A^2 \rangle$ given in Eq.~\eqref{Eq:gA_rA_final}. For both  $\Delta_{\rm GT}$ and the low energy constant $\bar{d}_{18}$ we find values that are compatible with chiral perturbation theory, which predicts for $\Delta_{\rm GT} \sim 2\%$ and for $-1.40(24)\,{\rm GeV}^{-2}<\bar{d}_{18}<-0.78(27)\,{\rm GeV}^{-2}$ depending on the type of fit used in the determination~\cite{Fettes:1998ud}. Our determination gives a more precise value and provides valuable input for chiral perturbation theory.

    \item The mild cut-off effects observed for the ratio  $r_{\rm PPD, 2}$ in Fig.~\ref{fig:FF_ratios} is understood by the fact that this ratio involves $G_P(Q^2)$ and $G_5(Q^2) $ both of which have the same pion pole mass dependence, thus canceling the cut-off effects. 
\end{itemize}

\subsection{Parametrizations for the fits at the continuum of $G_P(Q^2)$ and $G_5(Q^2)$ }

In the previous section, we made three important observations, namely: i) $G_P(Q^2)$ and $G_5(Q^2)$ have the same pion pole mass dependence at each lattice spacing; 
ii) the pion pole mass obtained using valence OS quarks in the mixed action twisted fermion mass formulation shows significant cut-off effects, much larger than the mass splitting between the unitary charged and neutral pion;
and iii) in the continuum limit we obtained a pion pole consistent with the physical pion mass of $m_\pi=0.135$~{\rm GeV}. Based on these observations, we use the following  functional form
\begin{equation}
    G_{\rm w\,pole}(Q^2,a^2) = \frac{1}{Q^2+m_\pi^2+ba^2} G_{\rm res}(Q^2,a^2)
\end{equation}
to fit $G_P(Q^2)$ and $G_5(Q^2)$, where for $G_{\rm res}$ we used the $z^k$-expansion and repeat the same analysis presented  for $G_A(Q^2)$ in Sec.~\ref{sec:Q2Fit}. Instead of fitting $G_5(Q^2)$ we  fit the scaled form factor $\tilde{G}_5(Q^2)$ given by
\begin{equation}\label{Eq:G5_fit}
    \tilde{G}_{5}(Q^2) = \frac{4m_N}{m_\pi^2} m_q G_5(Q^2).
\end{equation}
The combination $m_q G_5(Q^2)$ is scale-independent and renormalizes with $Z_S/Z_P$, which is accurately determined. Furthermore, scaling by $1/m_\pi^2$ takes into account the slight difference in the unitary pion mass for each ensemble (see Table~\ref{tab:ens}) and by $m_N$  makes the whole combination dimensionless. 
Since the pion pole mass is the same for both $G_P(Q^2)$ and $G_5(Q^2)$, we in addition, perform a combined fit of both form factors taking the parameter ``$b$'' of the pole to be a common fit parameter. Since, as demonstrated in the previous section, the PPD relation is satisfied at the continuum limit, we perform fits where we enforce the value of $g_{\pi NN}$ extracted from both form factors to take the same value at the continuum limit. This is implemented  by using a $z^3$-expansion with $t_0=-m_\pi^2$ and, therefore,
\begin{equation}
    G_{\rm res}(-m_\pi^2,0) = a_0\,,
\end{equation}
since,  according to  Eq.~\eqref{Eq:zform}, $z(-m_\pi^2)=0$ and  $a_0$ is a fit parameter as given in Eq~\eqref{Eq:zExp}.
The coupling constant $g_{\pi NN}$ is then the same for both form factors if
\begin{equation}
    a_0^{(P)} = \tilde{a}_0^{(5)} = a_0\,,
\end{equation}
namely by making $a_0$ a  common fit parameter for  both $G_P(Q^2)$ and $\tilde{G}_{5}(Q^2)$.

\subsection{Convergence of the $z$-expansion}

\begin{figure}[t!]
    \centering
    \includegraphics[width=0.95\linewidth]{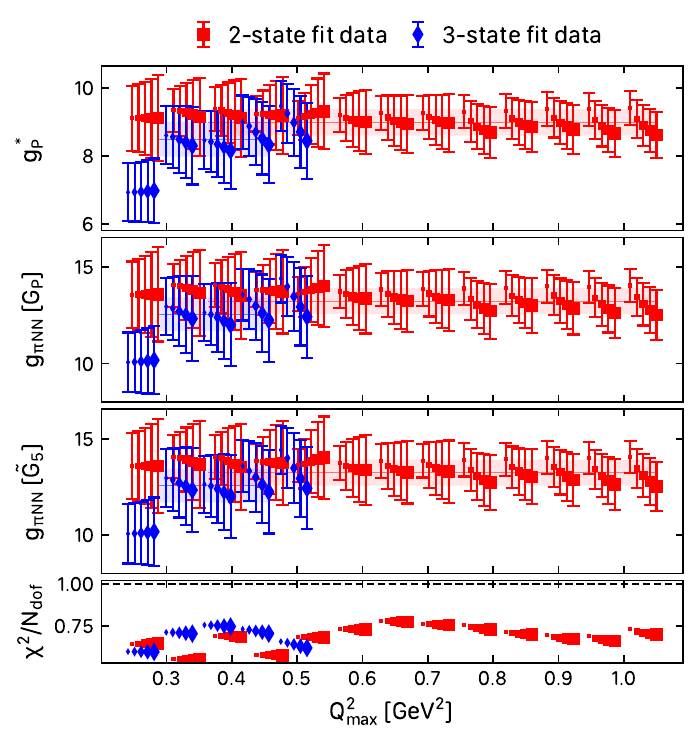}
    \caption{Induced pseudoscalar coupling, $g_P^*$,
    and pion-nucleon coupling, $g_{\pi N N}$ from a $z^3$-expansion fit as a function of the largest $Q^2$ used in the fit, $Q^2_{\max}$, and the width of the priors. For each $Q^2_{\max}$ we depict five points having prior width of $w=1,2,3,4,5$. The points are shifted to the right as $w$ increases with an increasing symbol size.}
    \label{fig:gp_zexpN_Q2}
\end{figure}

As for the analysis performed for $G_A(Q^2)$, we study the convergence of the $z^k$-expansion as a function of the order $k$, the width of the priors used, and the largest $Q^2$ employed in the fit. We first discuss results when we fit separately $G_P(Q^2)$ and $\tilde{G}_5(Q^2)$ without enforcing to have the same pole or the same value of $g_{\pi NN}$ and we monitor convergence by looking at the values it takes fitting  $G_P(^2)$, and $\tilde{G}_5(Q^2)$. We also monitor the value of $g_P^*$, which is extracted from $G_P(Q^2)$ using  Eq.~\eqref{Eq:gP*}. In Fig.~\ref{fig:gp_zexpN_Q2}, we show results on these quantities when we use data determined from the two- and three-state fit analysis to the correlators. We observe convergence for $k_{\max}\ge 3$, stability in the values we extract as a function of $Q^2_{\rm max}$ and the width of the priors for both data from the two- and three-case analysis fits. After model averaging we find
\begin{equation}
\begin{aligned}
    g^*_P&=8.87(66) \quad(\text{2-state})\\
    &=8.9(1.1) \quad(\text{3-state})\\
    g_{\pi NN} &= 13.0(1.2)\quad(\text{2-state, from }G_P)\\
    &= 13.5(1.3)\quad(\text{2-state, from }\tilde{G}_5)\\
    &= 13.3(2.0)\quad(\text{3-state, from }G_P)\\
    &= 11.9(1.7)\quad(\text{3-state, from }\tilde{G}_5)\,.\\
\end{aligned}
\end{equation}

All values determined using data from the two- and three-state fit analyses are in good agreement with each other. The values of $g_{\pi NN}$ extracted from $G_P(Q^2)$ and $\tilde{G}_5(Q^2)$ are also in agreement within error bars.

\begin{figure}[t!]
    \centering
    \includegraphics[width=0.95\linewidth]{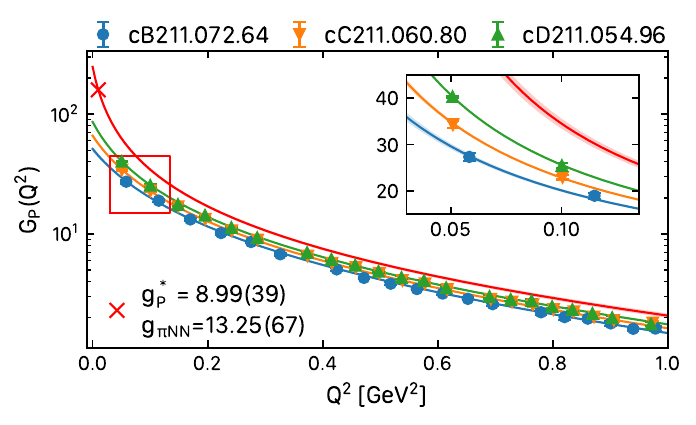}
    \caption{\label{fig:GP_final} Results on $G_P(Q^2)$ for each ensemble (blue band for the \Bens{}, orange band for the \Cens{} and green band for the \Dens{} ensemble) and at the  continuum limit (red band) using the $z^3$-expansion to fit the $Q^2$-dependence of the data determined  from the two-state fit analysis up to $Q^2=1$~GeV$^2$. The inner panel shows a zoom-in of the region marked by the red square.}
    \includegraphics[width=0.95\linewidth]{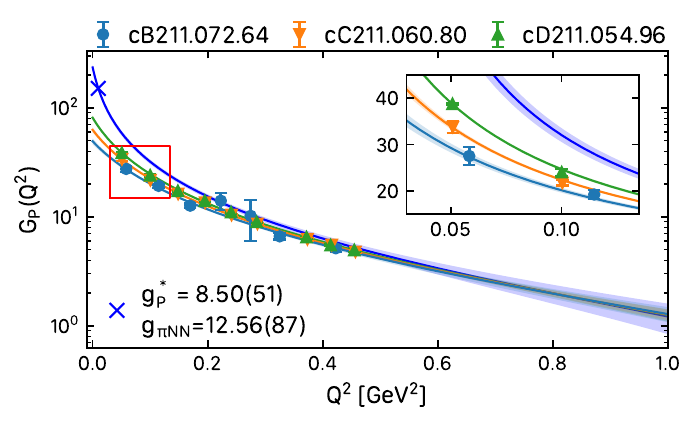}
    \caption{Results on $G_P(Q^2)$  using the $z^3$-expansion to fit the $Q^2$-dependence of the data determined from the three-state fit analysis of the correlators up to $Q^2=0.47$~GeV$^2$. The notation is the same as that for Fig.~\ref{fig:GP_final}\label{fig:GP_order3}.}
\end{figure}

\begin{figure}[t!]
    \centering
    \includegraphics[width=0.95\linewidth]{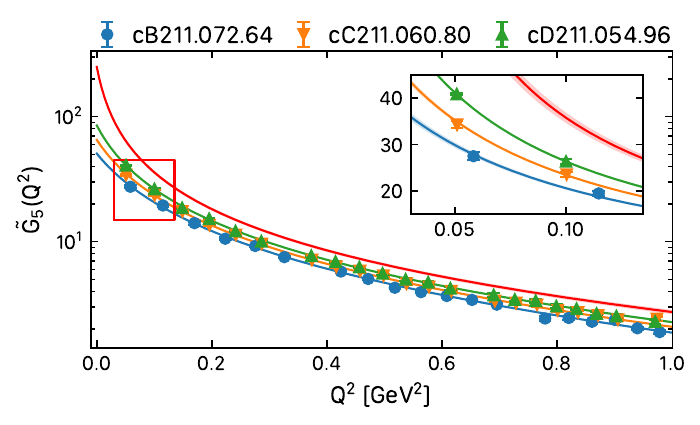}
    \caption{\label{fig:G5_final} Results on $\tilde{G}_5(Q^2)$ as defined in Eq.~\eqref{Eq:G5_fit} for each ensemble (blue band for the \Bens{}, orange band for the \Cens{} and green band for the \Dens{} ensemble) and at the  continuum limit (red band) using the $z^3$-expansion to fit the $Q^2$-dependence of the data determined  from the two-state fit analysis up to $Q^2=1$~GeV$^2$. The inner panel shows a zoom-in of the region marked by the red square.}
    \includegraphics[width=0.95\linewidth]{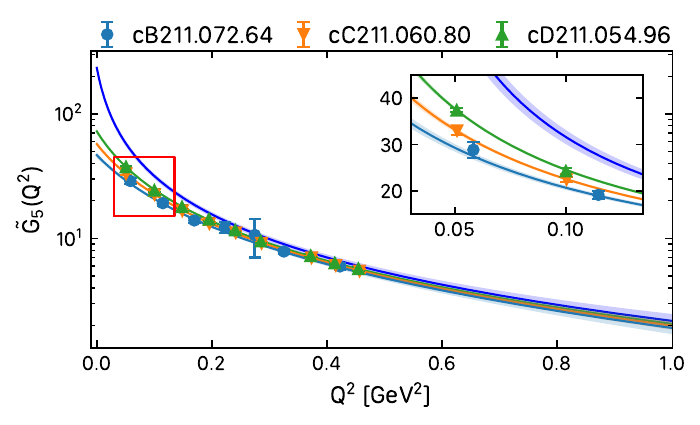}
    \caption{Results on $\tilde{G}_5(Q^2)$ as defined in Eq.~\eqref{Eq:G5_fit} using the $z^3$-expansion to fit the $Q^2$-dependence of the data determined from the three-state fit analysis of the correlators up to $Q^2=0.47$~GeV$^2$. The notation is the same as that of Fig.~\ref{fig:G5_final}\label{fig:G5_order3}.}
\end{figure}

If we enforce the pion pole and $g_{\pi NN}$ to have the same value as determined from  $G_P(Q^2)$ and $G_5(Q^2)$ and perform the same analysis, we obtain
\begin{equation}
\begin{aligned}
    g^*_P&=8.99(39) \quad(\text{2-state})\\
    &=8.50(51) \quad(\text{3-state})\\
    g_{\pi NN} &= 13.25(67)\quad(\text{2-state})\\
    &= 12.56(87)\quad(\text{3-state})\,.
\end{aligned}
\end{equation}
These results are in agreement with those where we did not enforce the value of $g_{\pi NN}$ and have smaller uncertainties thanks to the combined fit approach. Since we have demonstrated that the PPD relation is satisfied at the continuum limit, we opt to quote these as our final results for these quantities.
We follow the same strategy as for $G_A(Q^2)$ and quote the model-averaged value determined from using the data from the two-state fit analysis of the correlators and take as a systematic error the difference between the model-averaged central values of the data from the two- and three-state fits. We find the following values
\begin{equation}\label{Eq:gp_combined}
\begin{aligned}
    g^*_P&=8.99(39)(49)[63]\\
    g_{\pi NN} &= 13.25(67)(69)[96]
\end{aligned}\quad\text{(final value)\,,}
\end{equation}
where in the square brackets we have summed in quadrature the two errors in parentheses.
In Figs.~\ref{fig:GP_final} and~\ref{fig:GP_order3} we depict results on $G_P(Q^2)$ obtained after taking the model average using the data from the two- or three-state fit analysis, respectively. In Figs.~\ref{fig:G5_final} and~\ref{fig:G5_order3} we show the corresponding results for $G_5(Q^2)$.

\begin{table}[t!]
    \centering
    \begin{tabular}{c|c|c|c|c}
    \cmidrule{2-5}\morecmidrules\cline{2-5}
    & Ensemble & $m^{G_P}_{\pi}$ [MeV] & $m^{G_5}_{\pi}$ [MeV] & $m^{{\rm G_P,G_5}}_{\pi}$ [MeV]  \\
\hline\parbox[t]{2mm}{\multirow{3}{*}{\rotatebox[origin=c]{90}{2-state}}} & \texttt{cB211.72.64} & 279(27) & 295(27) & 284(21) \\
 & \texttt{cC211.60.80} & 249(22) & 262(22) & 254(17) \\
 & \texttt{cD211.54.96} & 221(17) & 231(17) & 224(13) \\
\hline\hline\parbox[t]{2mm}{\multirow{3}{*}{\rotatebox[origin=c]{90}{3-state}}} & \texttt{cB211.72.64} & 306(45) & 317(44) & 292(35) \\
 & \texttt{cC211.60.80} & 271(37) & 281(37) & 260(29) \\
 & \texttt{cD211.54.96} & 238(29) & 246(29) & 229(23) \\
\hline\hline
    \end{tabular}
    \caption{Pion pole masses per ensemble extracted from the individual or combined fit of $G_P$ and $\tilde{G}_5$  from two- or three-state fit data.}
    \label{tab:pion_pole}
\end{table}

\begin{figure}[t!]
    \centering
    \includegraphics[width=0.9\linewidth]{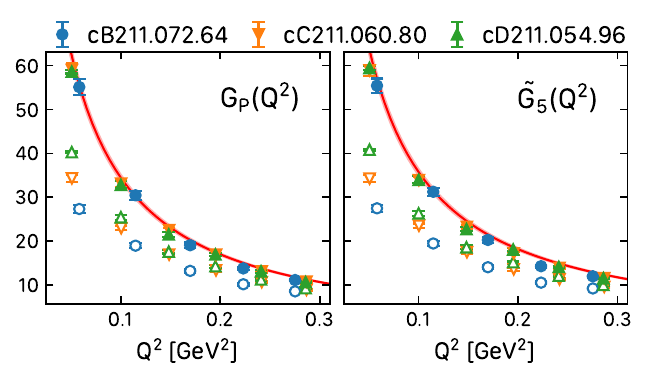}
    \caption{\label{fig:G5P_corrected} 
    Results per gauge ensemble for $G_P (Q^2)$ (left) and $\tilde{G}_5(Q^2)$ (right) when using the data for $G_{\rm w\,pole}(Q^2)$ (open symbols) compared to those when using $G_{\rm improved}(Q^2)$ (filled symbols) of Eq.~\eqref{Eq:FF_improved} by correcting for the pole OS pion mass. The continuum limit form factors (red band) are those determined in Fig.~\ref{fig:GP_final} and Fig.~\ref{fig:G5_final} using the data for $G_{\rm w\,pole}$.
    }
\end{figure}

In Table~\ref{tab:pion_pole}, we quote the values of the pion pole masses per ensemble as extracted from the individual or combined fit of $G_P$ and $\tilde{G}_5$ from two- or three-state fit data. We observe an overall good agreement within errors, and the values confirm the agreement already discussed in Sec.~\ref{sec:PPD} with the OS pion mass reported in Table~\ref{tab:pole_masses}.

One can drastically reduce cut-off effects by considering for $G_P(Q^2)$ and $G_5(Q^2)$ the following modified expression
\begin{equation}\label{Eq:FF_improved}
    G_{\rm improved}(Q^2,a^2) = \frac{Q^2+m_{\pi, \rm OS}^2}{Q^2+m_{\pi,\rm TM}^2} G_{\rm w\,pole}(Q^2,a^2)\,.
\end{equation}
In Fig.~\ref{fig:G5P_corrected}, we show the improved expressions for form factors per ensemble.
We observe that upon using the improved expression defined in Eq.~\eqref{Eq:FF_improved}, the results per ensemble are compatible with each other and with those obtained in the continuum limit by extrapolating $G_{\rm w pole}(Q^2)$. These findings further confirm the interpretation that the sizable cut-off artifacts in $G_P(Q^2)$ and $G_5(Q^2)$ stem from the cutoff effects in using the OS pion mass for the pole. Since at finite $a$ we neglect disconnected $O(a^2)$ terms in our form factor computation this is indeed the expected behavior and fully justifies our fit Ansatz in Eq. (85) for the continuum extrapolation of the data when using $G_{\rm w pole}$.

\subsection{Continuum  results for $G_P(Q^2)$ and $G_5(Q^2)$ }

\begin{figure}[t!]
    \centering
    \includegraphics[width=0.95\linewidth]{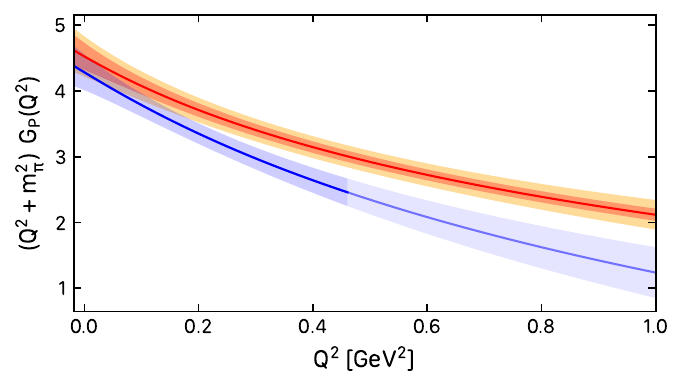}
    \caption{Results on $(Q^2+m_\pi^2)\,G_P(Q^2)$ at the continuum limit when fitting data extracted from the two- (red band) and three- (blue band) state fit analysis of the correlators. The darker blue curve indicates up to which $Q^2$ we had data for the three-state fit analysis. The yellow band is when we added systematic errors to the parameters that define the red curve as discussed in the text. The parameters of the fit are given in Eq.~\eqref{Eq:GP_params}.
    \label{fig:GP_order23_comparison}}
\end{figure}

We follow the same procedure as the one for $G_A$ in Sec.~\ref{sec:GA_final} to arrive at the $Q^2$ parameterization of $G_P(Q^2)$ and $G_5(Q^2)$. In particular, in Fig.~\ref{fig:GP_order23_comparison}, we show the corresponding results for $G_P(Q^2)$ as those shown for $G_A(Q^2)$ in Fig.~\ref{fig:GA_order23_comparison} showing the comparison of results obtained when using data from the two- and three-state fit analysis after removing the pole, namely we show results for $(Q^2+m_\pi^2) G_P(Q^2)$. As in the case of $G_A(Q^2)$, the data from the three-state analysis of the correlators are in agreement with those from the two-state fit analysis. However, after the continuum extrapolated results using the data from the three-state fit analysis yield systematically smaller values for $G_P(Q^2)$ for higher $Q^2$ values. As we already pointed out, the three-state fit analysis becomes unstable for $Q^2>0.5$~GeV affecting the fits to the $Q$-dependence. Since cut-off effects are larger for $G_P(Q^2)$ the slope in linear $a^2$ extrapolation is larger and thus more affected by small fluctuations in the data given that we also only have three lattice spacings. This explains why the continuum results in the two cases differ by up to a standard deviation at large $Q^2$  while the lattice data for the three ensembles are compatible.  Having higher statistics will enable us to extract more reliable results using the three-state fit procedure and having more lattice spacings will better control the continuum extrapolation, something that we plan to do in the future when more computational resources are available. 

In the following, we provide parameters for the standard form of the $z$-expansion in Eq.~\eqref{Eq:zExp} with $k_{\max}=3$, $t_{\rm cut} = (3 m_\pi)^2$, $m_\pi=0.135$~GeV and $t_0=-m_\pi^2$. 
The fit parameters of the two- and three-state fit data curves are given by
\begin{equation}\label{Eq:GP_params_st}
 \begin{aligned}
 \vec{a}_\text{2-state} &= [4.62(23),-3.0(1.2),-4.7(2.5),-0.1(2.4)]\\
\vec{a}_\text{3-state} &= [4.38(30),-3.1(1.5),-5.9(2.6),-2.9(2.0)]\,.
\end{aligned}
\end{equation}
As can be seen, the parameters are in agreement albeit some carry large statistical errors and thus, we follow the same strategy as for $G_A(Q^2)$ for determining the best parametrization of the continuum results and for estimating the errors. Our final parameterization that takes into account systematic errors is
\begin{equation}\label{Eq:GP_params}
\begin{aligned}
    \vec{a}_{P} = &\big[4.62(23)(24)[33],-3.0(1.2)(0.1)[1.2],\\
    &-4.7(2.5)(1.2)[2.8],-0.1(2.4)(2.8)[3.7]\big]\\
    \text{corr}_{\vec{a},P} = &\begin{pmatrix}
1.0 & -0.812 & 0.414 & 0.151 \\
-0.812 & 1.0 & -0.819 & 0.23 \\
0.414 & -0.819 & 1.0 & -0.713 \\
0.151 & 0.23 & -0.713 & 1.0 \\
    \end{pmatrix}\,.
\end{aligned}
\end{equation}
The values of $G_P(Q^2)$ that result from this parametrization are given in Table~\ref{tab:GP} of the Appendix.

\begin{figure}[t!]
    \centering
    \includegraphics[width=0.95\linewidth]{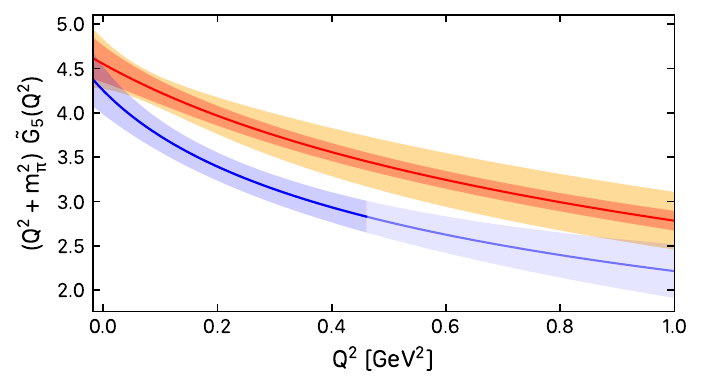}
    \caption{Results on $(Q^2+m_\pi^2)\,\tilde{G}_5(Q^2)$ at the continuum limit when fitting data extracted from the two- (red band) and three- (blue band) state fit analysis of the correlators. The darker blue curve indicates up to which $Q^2$ we had data for the three-state fit analysis. The yellow band is when we added systematic errors to the parameters that define the red curve as discussed in the text. The parameters of the fit are given in Eq.~\eqref{Eq:G5_params}.\label{fig:G5_order23_comparison}}
\end{figure}
 
Repeating the same analysis for $\tilde{G}_5(Q^2)$, we find the results shown in Fig.~\ref{fig:G5_order23_comparison}, after removing the pole, namely we show results for $(Q^2+m_\pi^2) \tilde{G}_5(Q^2)$. The behavior of the continuum limit results is the same as that observed for $G_P(Q^2)$ since both have similar cut-off effects due to the pion pole dominance.
The fit parameters of the two- and three-state fit data curves are given by
\begin{equation}\label{Eq:G5_params_st}
 \begin{aligned}
 \vec{a}_\text{2-state} &= [4.62(23),-2.2(1.2),-2.9(2.4),-1.2(2.4)]\\
\vec{a}_\text{3-state} &= [4.38(30),-4.3(1.6),-0.1(2.7),-0.7(2.0)]\,.
\end{aligned}
\end{equation}
Our final parameterization that takes into account systematic errors is
\begin{equation}\label{Eq:G5_params}
\begin{aligned}
    \vec{a}_{5} = &\big[4.62(23)(24)[33],-2.2(1.2)(2.1)[2.5],\\&-2.9(2.4)(2.8)[3.7],-1.2(2.4)(0.5)[2.4]\big]\\
    \text{corr}_{\vec{a},5} = &\begin{pmatrix}
1.0 & -0.804 & 0.435 & 0.14 \\
-0.804 & 1.0 & -0.825 & 0.217 \\
0.435 & -0.825 & 1.0 & -0.694 \\
0.14 & 0.217 & -0.694 & 1.0 \\
    \end{pmatrix}\,.
\end{aligned}
\end{equation}
The values of $\tilde{G}_5(Q^2)$ that result from this parametrization are given in Table~\ref{tab:G5}
of the Appendix~\ref{sec:appendix}, where we also provide more information on the form factors at the continuum limit.

\subsection{Continuum limit of the PCAC and PPD relations}
\begin{figure}[t!]
    \centering
    \includegraphics[width=0.95\linewidth]{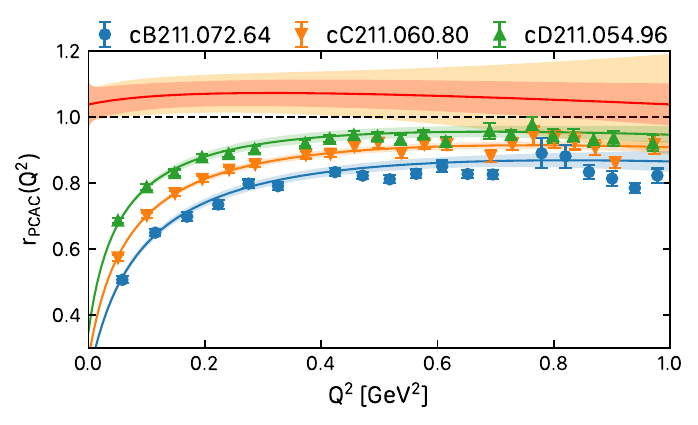}
    \includegraphics[width=0.95\linewidth]{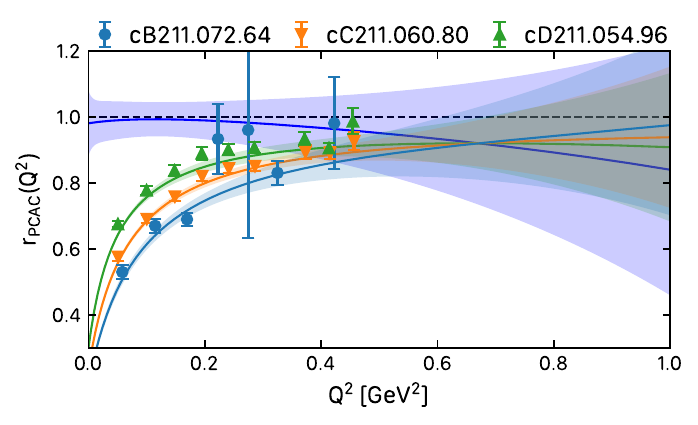}
    \caption{Top: $r_{\rm PCAC}$ as  defined in Eq.~\eqref{Eq:rPCAC}. The blue, orange, and green curves are the result of the combined fits to $G_A(Q^2)$, $(m_\pi^2+Q^2)G_P(Q^2)$, and $(m_\pi^2+Q^2)\tilde{G}_5(Q^2)$ for the \Bens{}, \Cens{} and \Dens{} ensembles, respectively.  The fits are done using the $z^3$-expansion to fit the $Q^2$-dependence of the data determined from the two-state fit analysis up to $Q^2=1$~GeV$^2$. The red curve and band show the results at the continuum limit. The yellow band shows the errors on the curve when systematic effects are taken into account, i.e. using the final parameterization of the form factors given in Eqs.~\eqref{Eq:GA_params},~\eqref{Eq:GP_params}, and~\eqref{Eq:G5_params} for $G_A(Q^2)$, $(m_\pi^2+Q^2)G_P(Q^2)$, and $(m_\pi^2+Q^2)\tilde{G}_5(Q^2)$, respectively.
    Bottom: Same as top panel, but using three-state fit data and the respective $z^3$-expansion to fit the $Q^2$-dependence up to $Q^2 \sim 0.5$~GeV$^2$.}
    \label{fig:PCAC_final}
    \includegraphics[width=0.95\linewidth]{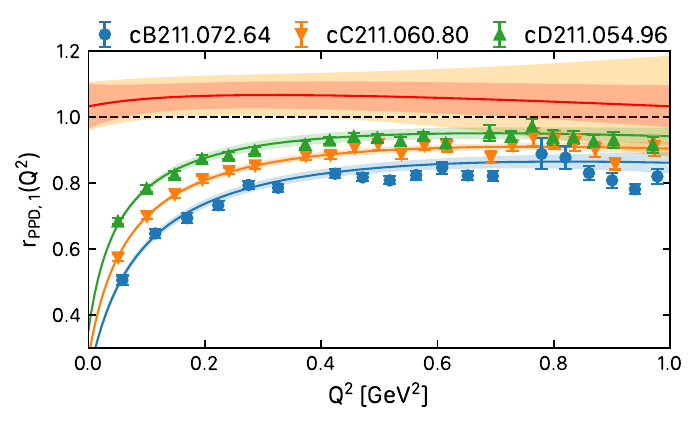}
    \caption{$r_{\rm PPD,1}$ as defined in Eq.~\eqref{Eq:rPPD1}. The notation is the same as that in the top panel of Fig.~\ref{fig:PCAC_final}.}
    \label{fig:PPD_final}
\end{figure}

Having determined the three form factors $G_A(Q^2)$, $G_P(Q^2)$, and $G_5(Q^2)$, we can check the PCAC relation at the continuum limit. We use the values of the fit parameters of the  $z^3$-expansion to the $Q^2$-dependence after taking the model average for each ensemble.  We use the form factors extracted from the two-state fits to correlators. We also repeat using the three-state fits correlators. In both cases, we also take the continuum limit of the parameters determined at each lattice spacing, as previously discussed. In Fig.~\ref{fig:PCAC_final}, we depict the resulting $r_{\rm PCAC}$ as a function of $Q^2$ using data from the two- and three-state fit analysis, upper and lower panels, respectively.   As can be seen, in both cases the PCAC relation is recovered in the continuum limit. In addition, we obtain the PCAC ratio in the continuum limit by using the final parameterizations of the form factors that take into account the systematic uncertainty due to how we treat excited states, i.e. difference of central values when we use two- or three-state fits, namely the results shown by the yellow band of Figs.~\ref{fig:GA_final}, \ref{fig:GP_final} and \ref{fig:G5_final} for $G_A(Q^2)$, $(m_\pi^2+Q^2)G_P(Q^2)$, and $(m_\pi^2+Q^2)\tilde{G}_5(Q^2)$, respectively. As expected, the PCAC relation is recovered but the systematic error due to the treatment of excited states increases the error band.  For comparison, we plot in Fig.~\ref{fig:PPD_final} in the same format, the results for the ratio $r_{\rm PPD,1}$. It is no surprise that it also fulfills the PPD dominance in the continuum limit, as already discussed in relation to Fig.~\ref{fig:FF_ratios}. As in the case of $r_{\rm PCAC}$  we show both the continuum limit curve extracted using the data from the two-state fit analysis and the one when we include the systematic uncertainty difference between the central values of the fit parameters determined by using data from to the two- and three-state fit analysis.

\section{Comparison with other results} \label{sec:comparison}

\begin{figure}[t!]
    \includegraphics[width=0.95\linewidth]{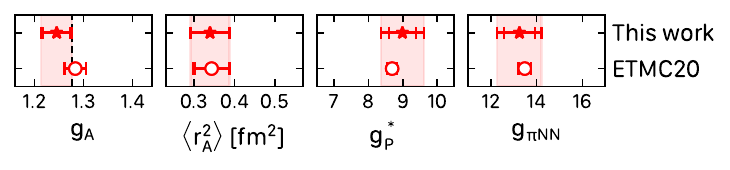}
    \vspace{-0.4cm}
    \caption{From left to right we show our lattice QCD results from this work on $g_A$, $\langle r_A^2\rangle$, $g_P^*$ and $g_{\pi NN}$ (red stars). The open circles show results extracted using only the \Bens{}~\cite{Alexandrou:2020okk} and the PCAC and PPD relations to extract $g^*_P$ and $g_{\pi NN}$ from $G_A(Q^2)$. \label{fig:Bens-compare}}\medskip
    \includegraphics[width=\linewidth]{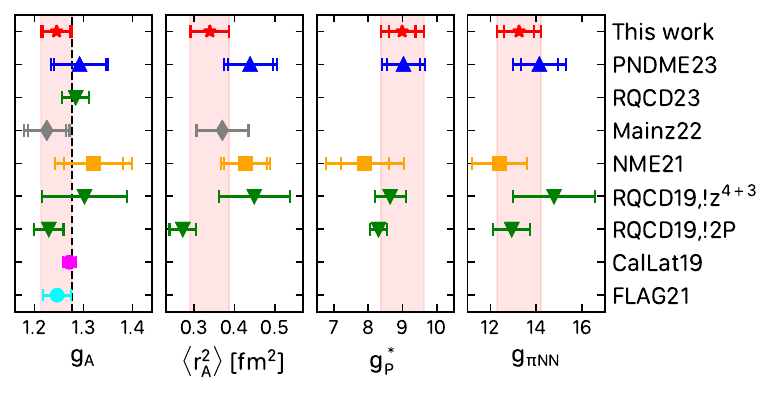}
    \vspace{-0.4cm}
    \caption{From left to right we show recent lattice QCD results on $g_A$, $\langle r_A^2\rangle$, $g_P^*$ and $g_{\pi NN}$. Our results are shown with the red star and red error band.  The blue triangles show the recent results by PNDME~\cite{Jang:2023zts}, the green triangles by RQCD~\cite{RQCD:2019jai,Bali:2023sdi}, the yellow squares by NME~\cite{Park:2021ypf}, the gray diamonds by the Mainz group~\cite{Djukanovic:2022wru} and the magenta square by CalLat~\cite{Chang:2018uxx}. The cyan circle shows the FLAG21 average of lattice results published at the time of the report~\cite{FlavourLatticeAveragingGroupFLAG:2021npn}.  }
    \label{fig:comparison}
\end{figure}
Before comparing with other lattice QCD studies, we compare in Fig.~\ref{fig:Bens-compare} our older results~\cite{Alexandrou:2020okk} where only the \Bens{} ensemble was used to the ones obtained in this work. As can be seen, the central values are in agreement showing that cut-off effects are mild for these quantities. The error on $g_A$ increases after taking the continuum limit, while the error on the axial radius is approximately the same. The fact that the errors on $g_P^*$ and $g_{\pi NN}$ are much smaller is a combination of two things: i) taking the continuum limit and ii) in our previous work we used the PCAC and PPD relation and lattice QCD data on $G_A(Q^2)$ which is more precisely determined. The reason was that with one lattice spacing, we could not account for the large cut-off effects on $G_P(Q^2)$ and $G_5(Q^2)$ leading to a violation of the PCAC relation. In this work,  $g_P^*$ and $g_{\pi NN}$ are determined directly from our data on $G_P(Q^2)$  and $G_5(Q^2)$, although, as shown in this work, in the continuum limit the PCAC relation holds and could be used to determine them. We note that the trend that we observe of errors becoming larger in a number of quantities highlights the importance of having results using ensembles with smaller lattice spacings. However, as it is well known, simulations for lattice spacing $a<0.05$~fm become difficult due to the increase in the autocorrelation time. There are ongoing efforts to address the critical slowing down of Hybrid Monte Carlo simulations~\cite{Kanwar:2020xzo, Finkenrath:2022ogg, Bacchio:2022vje}.

The nucleon axial charge and radius as well as the coupling constants $g_P^*$ and $g_{\pi NN}$ are compared to other recent lattice QCD results in Fig.~\ref{fig:comparison}. We selected studies that provide results at the continuum limit and at the physical pion mass, either computed directly like ours or via a combined chiral and continuum extrapolation.  There is a nice agreement among all lattice QCD results on these quantities which are defined either at $Q^2=0$ or at the limit $Q^2\rightarrow 0$.  

\begin{figure}[t!]
    \centering
    \includegraphics[width=\linewidth]{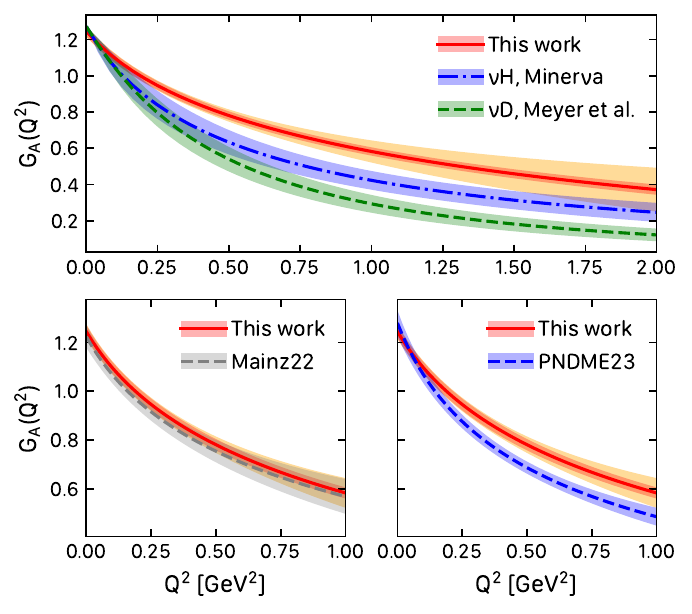}
    \vspace{-0.6cm}
    \caption{Top: $G_A(Q^2)$ determined within this work using the parameters $a_\text{2-state}$ (red solid line and band) and when including the systematic uncertainty as the difference in the central values of $a_\text{2-state}$ and $a_\text{3-state}$ (yellow band). We compare to the fit to the deuterium bubble-chamber data~\cite{Meyer:2016oeg} shown by the green dashed line with error band and with the fit to the recent MINER$\nu$A antineutrino-hydrogen data~\cite{MINERvA:2023avz} shown by the blue dot-dashed line with error band. 
    Bottom: $G_A(Q^2)$ determined within this work compared to two recent lattice QCD calculations: i) by the Mainz group~\cite{Djukanovic:2022wru} shown with the gray dashed line with its error band,  and ii)  by PNDME~\cite{Jang:2023zts} shown with the blue dashed line with its error band.} 
    \label{fig:GA_comparison}
\end{figure}

\begin{figure}[t!]
    \centering
    \includegraphics[width=\linewidth]{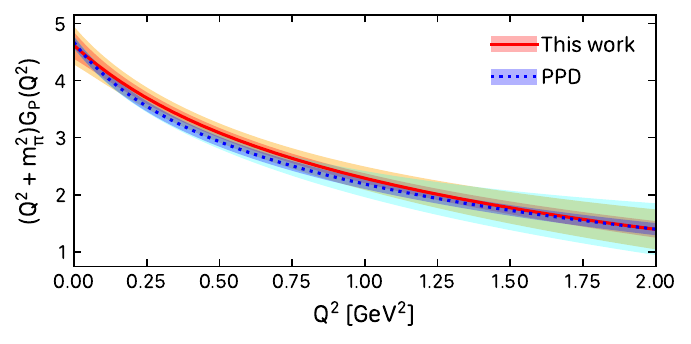}
    \vspace{-0.6cm}
    \caption{Results on $G_P(Q^2)$ determined directly from our lattice data using the parameters $a_\text{2-state}$ (red solid line and band) and when including the systematic uncertainty as the difference in the central values of $a_\text{2-state}$ and $a_\text{3-state}$ (yellow band) are compared to those obtained by using PPD given in Eq.~\eqref{Eq:PPD} and to our data on $G_A(Q^2)$ (dark blue and light blue when including the systematic error).}
    \label{fig:GP_comparison}
\end{figure}

In Fig.~\ref{fig:GA_comparison} we compare our final parameterization of $G_A(Q^2)$ given in Eq.~\eqref{Eq:GA_params} with fits to experimental data on $G_A(Q^2)$ and with fits to data computed by other lattice QCD groups. When compared to experimental data, our results fall off slower than the fits to experimental data. While our results are within two standard deviations as compared to the recent results from the Miner$\nu$a experiment~\cite{MINERvA:2023avz}, they show more tension with the fit to the deuterium bubble-chamber data~\cite{Meyer:2016oeg}. In addition, our results are in good agreement with the results by the Mainz group~\cite{Djukanovic:2022wru} and close to the results by PNDME~\cite{Jang:2023zts} and NME~\cite{Park:2021ypf, Tomalak:2023pdi}.

We comment below on some aspects of the lattice QCD calculations:
\begin{itemize}
\item The results of this work are the only ones that are extrapolated to the continuum limit using only ensembles simulated directly with physical pion mass.
\item The rest of the collaborations combined results extracted using ensembles simulated with larger than physical pion masses to extrapolate to the physical pion mass and to the continuum limit. Specifically, NME~\cite{Park:2021ypf} uses no physical point ensembles, the Mainz group~\cite{Djukanovic:2022wru} uses one with their physical point results having large errors and  RQCD~\cite{RQCD:2019jai} and PNDME~\cite{Jang:2023zts} use two physical pion mass ensembles.
\item In the case of the RQCD~\cite{RQCD:2019jai} and PNDME~\cite{Jang:2023zts}, results using the physical pion mass ensembles have larger statistical errors as compared to those of ensembles with heavier than physical mass.  This means that results at the physical point weigh less in the extrapolation. Additionally, in both studies,  the form factors are computed using the physical pion mass ensembles only for $Q^2\lesssim 0.3$~GeV$^2$ and information at high $Q^2$ is provided by a subset of the ensembles. We instead compute the form factors up to $Q^2=1$~GeV$^2$.
\item
The PNDME collaboration~\cite{Jang:2023zts} uses a $N_f=2+1+1$ mixed action of clover fermions on a staggered sea. They have employed thirteen  ensembles simulated at four values of the lattice spacing, three values of the pion mass (135, 220, and 310
MeV), and volumes with $3.7 \le m_\pi L \le 5.5$. Their axial-vector current is unimproved which means they have ${\cal O}(a)$ cut-off effects. Nevertheless, they observe mild dependence on the lattice spacing. They also do not observe any significant lattice volume dependence, while they do see a stronger dependence on the pion mass. 
In this work, we use ensembles with approximately the same volume, namely with  $3.6\le m_\pi L \le 3.9$. Given the volume study by  PNDME~\cite{Jang:2023zts}, we expect finite size effects on our results to be small.

PNDME also carried out an elaborate study of excited states highlighting the effects of  $\pi N$ states and concluded that an approach compatible with the one employed in this work is the most suitable. Namely, they propose performing a combined fit of all matrix elements at the same $Q^2$ using common fit parameters for the excited states. However, they do not include a systematic error due to excited states and this explains why their results have a smaller error band.
\item The Mainz group~\cite{Djukanovic:2022wru} has used fourteen CLS $N_f=2+1$ ensembles simulated with clover-improved Wilson fermions at four values of the lattice spacing, pion masses in the range $130\,{\rm MeV} \le m_\pi \le 350\,{\rm MeV}$, and volumes with $3.9 \le m_\pi L \le 5.9$.  Their current is ${\cal O}(a)$-improved and they observe a mild dependence on the lattice spacing and the lattice volume, while a stronger dependence on the pion mass, which requires the inclusion of higher order corrections that are not considered by PNDME. The Mainz group also includes a systematic error in a similar way to what we do.
\item The RQCD collaboration~\cite{RQCD:2019jai} uses the same CLS ensembles as the Mainz group, but thirty-seven of them, having five values of the lattice spacing, pion masses in the range $130\,{\rm MeV} \le m_\pi \le 420\,{\rm MeV}$ and volumes with $3.5 \le m_\pi L \le 6.4$. Their physical point limit also involves a limit to the physical strange quark mass that is not required in the set of ensembles used by the Mainz group. They perform a thorough study of excited states including the effect of $\pi N$ states. They also observe a strong dependence on the pion mass.  They have provided results using dipole fits (labeled $!2P$) or using $z$-expansion (labeled $!z^{4+3}$), without selecting one of the two as the final value. For this reason, we report two sets of points in Fig.~\ref{fig:comparison} for RQCD19. In their recent work~\cite{Bali:2023sdi}, they have provided a new value for the axial charge extracted from the analysis of matrix elements at zero momentum transfer and from an analysis of ten additional ensembles having four at the physical pion mass. 
\item NME~\cite{Park:2021ypf} has used seven $N_f=2+1$ Wilson-clover fermions ensembles simulated at five values of the lattice spacing, with pion masses in the range $166\,{\rm MeV} \le m_\pi \le 285\,{\rm MeV}$ and volumes with $3.9 \le m_\pi L \le 6.2$. Their axial-vector current is not $\mathcal{O}(a)$-improved and they observe strong lattice spacing effects and pion mass dependence. The analysis of the excited states is compatible with the one carried out by PNDME and they include an elaborated study of excited states using priors around the $\pi N$ excited states. For this case, we only show the values of the coupling constants and axial radius in Fig.~\ref{fig:comparison}.
\item The  PACS collaboration computed $G_A(Q^2)$, $G_P(Q^2)$ and $G_5(Q^2)$ using $N_f=2+1$ clover-improved fermions and a large spatial volume of length $L=8.1$~fm~\cite{Ishikawa:2018rew} and pion mass of 146~MeV and lattice spacing of $a=0.085$~fm. However, they only have time separations up to $t_s\sim 1.3$ fm and only perform plateau fits to individual form factors. They also have results for these form factors for lower $Q^2$ values up to $\sim 0.25$~GeV$^2$. Since their results are given only at one lattice spacing they are not included in the comparisons. 
\end{itemize}

While $G_A(Q^2)$ is determined by a number of lattice QCD collaborations, there are scarce results on $G_P(Q^2)$ and, to our knowledge, this work is the first to compute  $G_5(Q^2)$ at the continuum limit. Experimental studies also probe $G_A(Q^2)$ and one could use PCAC and PPD to estimate $G_P(Q62)$. In Fig.~\ref{fig:GP_comparison}, we show our results from a {\it direct} evaluation of $G_P(Q^2)$ in comparison with the ones obtain using our data on $G_A(Q^2)$ and the Eq.~\eqref{Eq:PPD} to extract $G_P(Q^2)$. As can be seen, the results are in perfect agreement with the uncertainties. Therefore, one would be justified to use Eq.~\eqref{Eq:PPD} and the experimental data on $G_A(Q^2)$ to estimate  $G_P(Q^2)$.  

\section{Conclusions}\label{sec:conclusions}

In this work, we present results on the axial, induced pseudoscalar, and pseudoscalar form factors in the continuum limit. This study is performed using three $N_f=2+1+1$ twisted-mass ensembles with all quark masses tuned to their physical value and simulated at three values of the lattice spacing. Our analysis is also done up to $Q^2=1$~GeV$^2$ as compared to some other lattice QCD studies where, for physical point ensembles, only smaller values of $Q^2$ were accessible. 
Our final values for the nucleon axial charge and radius as well as the coupling constants $g_P^*$ and $g_{\pi NN}$ are
\begin{equation}\label{Eq:final_values_conclusions}
    \begin{aligned}
    g_A&=1.245(28)(14)[31]\\
    \langle r_A^2\rangle &=0.339(48)(06)[48]~{\rm fm}^2\\
    g^*_P&=8.99(39)(49)[63]\\
    g_{\pi NN} &= 13.25(67)(69)[96]\,,
    \end{aligned}
\end{equation}
where the central values and the first error in the parenthesis are obtained from an analysis of data extracted from the two-state fits to the correlators, the second error is the systematic error due to the excited states computed as the difference between the central values from using the data extracted from the two- and three-state fit analysis of the correlators and the third error in the square brackets is the total error obtained by summing in quadrature the first two.
In Appendix~\ref{sec:appendix}, we provide the final parameterization and values of the form factors at the continuum limit with and without systematic uncertainties due to the excited states. 

From our analysis of the ratio $r_{\rm PPD,2}$ defined in  Eq.~\eqref{Eq:rPPD2_AP} we also determine the values of the  Goldberger-Treiman
discrepancy and the low-energy constant $\bar{d}_{18}$ 
    \begin{equation}
    \begin{aligned}
    \Delta_{\rm GT}&=2.13(38)\% \\
    \bar{d}_{18} &= -0.73(13)~{\rm GeV}^{-2}\,.
    \end{aligned}
    \end{equation}

Our results on $G_A(Q^2)$ are in good agreement with other recent lattice QCD studies.  Having taken the continuum limit using only ensembles at the physical point mass we avoid a chiral extrapolation that in the nucleon sector can lead to an uncontrolled systematic error.  An advantage of this setup is that it allows us to directly access cut-off effects. We find that for $G_A(Q^2)$, cut-off effects for the range of lattice spacings used are mild, ranging from not detectable within our errors at low $Q^2$ to slightly positive at high $Q^2$. 
On the other hand, the induced pseudoscalar and the pseudoscalar form factors exhibit similar large cut-off effects that can be traced back to the known $O(a^2)$ artifacts on the pion mass pole.  Such cut-off effects are expected in the twisted mass fermion formulation used in this work for the computation of these form factors and as such they can be conveniently parameterized in our continuum extrapolation fits. Alternatively, they can be also substantially reduced by considering modified expression for the pole of the form factors. As shown in this work, the important conclusion is that in the continuum limit, all cut-off effects are safely eliminated as expected. In particular, both the pion pole dominance close to $Q^2 = -m_\pi^2$, with $m_\pi=135$~MeV, and the fundamental PCAC relation that follows from QCD chiral Ward identities, are fully recovered. Regarding finite volume effects, we would like to point out that the PNDME collaboration~\cite{Jang:2023zts} and the CLS-Mainz group~\cite{Djukanovic:2022wru} investigated finite volume effects for these quantities and found that for the range of $m_\pi L> 3.5$ of our three ensembles, no detectable finite volume effects were observed within the accuracy of their lattice data, that is similar to ours.

\begin{acknowledgments}
We thank all members of ETMC for the most enjoyable collaboration.  C.A. acknowledges partial support by the project 3D-nucleon, id number EXCELLENCE/0421/0043, co-financed by the European Regional Development Fund and the Republic of Cyprus through the Research and Innovation Foundation and by the European Joint Doctorate AQTIVATE that received funding from the European Union’s research and innovation program under the Marie Skłodowska-Curie Doctoral Networks action and Grant Agreement No 101072344. S.B. is funded by the project QC4LGT, id number EXCELLENCE/0421/0019, co-financed by the European Regional Development Fund and the Republic of Cyprus through the Research and Innovation Foundation. J.F. acknowledges support by the German Research Foundation (DFG) research unit FOR5269 ``Future methods for studying confined gluons in QCD''. S.B. and J.F. also acknowledge funding from the EuroCC project (grant agreement No. 951740).   G.K. acknowledges partial support by the project NiceQuarks, id number EXCELLENCE/0421/0195, co-financed by the European Regional Development Fund and the Republic of Cyprus through the Research and Innovation Foundation. M.C. acknowledges financial support from the U.S. Department of Energy, Office of Nuclear Physics, Early Career Award under Grant No. DE-SC0020405. G.S. acknowledges financial support from the European Regional Development Fund and the Republic of Cyprus through the Research and Innovation Foundation under contract number EXCELLENCE/0421/0025.  This research was supported in part by grant NSF PHY-1748958 to the Kavli Institute for Theoretical Physics (KITP). This work was supported by grants from the Swiss National Supercomputing Centre (CSCS) under projects with ids s702 and s1174.  We also acknowledge PRACE for awarding us access to Piz Daint, hosted at CSCS, Switzerland, and Marconi100, hosted at CINECA, Italy.  The authors gratefully acknowledge the Gauss Centre for Supercomputing e.V. (www.gauss-centre.eu) for funding this project by providing computing time through the John von Neumann Institute for Computing (NIC) on the GCS Supercomputer JUWELS-Booster~\cite{JUWELS} at J\"ulich Supercomputing Centre (JSC). Part of the results were created within the EA program of JUWELS Booster also with the help of the JUWELS Booster Project Team (JSC, Atos, ParTec, NVIDIA). The authors also acknowledge the Texas Advanced Computing Center (TACC) at The University of Texas at Austin for providing HPC resources that have contributed to the research results. We thank the developers of the QUDA~\cite{Clark:2009wm, Babich:2011np, Clark:2016rdz} library for their continued support, without which the calculations for this project would not have been possible.
Ensemble production for this analysis made use of tmLQCD~\cite{Jansen:2009xp,Kostrzewa:2022hsv}, DD-$\alpha$AMG~\cite{Alexandrou:2016izb, Bacchio:2017pcp, Alexandrou:2018wiv}.
\end{acknowledgments}

\appendix

\section{Results on the axial and pseudoscalar form factors}\label{sec:appendix}

In this appendix, we collect our results on the two axial form factors $G_A(Q^2)$ and $G_P(Q^2)$ and the pseudoscalar form factor $G_5(Q^2)$ computed at the continuum limit. The $Q^2$-dependence of  the form factors is parameterized using a $z^3$-expansion of the form 
\begin{equation}
G(Q^2) = \sum_{k=0}^{3} a_k\; z^k(Q^2), 
\label{Eq:z3Exp}
\end{equation}
where 
\begin{equation}
z(Q^2) = \frac{\sqrt{t_{\rm cut} + Q^2} - \sqrt{t_{\rm cut}+t_0} }{ \sqrt{t_{\rm cut} + Q^2} + \sqrt{t_{\rm cut}+t_0} }
\end{equation}
with $t_{\rm cut}=\left(3 m_\pi\right)^2$, $m_\pi=0.135$~GeV and $t_0$ chosen at convenience as discussed below. As discussed, taking $k_{\rm max}=3$ we obtained results that are stable as compared to taking higher orders in the $z$-expansion.

\subsection{Axial form factor $G_A(Q^2)$}

\begin{table}[t!]
\begin{tabular}{c|c}
  \hline\hline
  $Q^2$ [GeV$^2$] & $G_A$ \\
  \hline
0.00 & 1.245(28)(13)[31] \\
0.05 & 1.164(25)(12)[28] \\
0.10 & 1.097(24)(12)[27] \\
0.15 & 1.040(24)(12)[27] \\
0.20 & 0.990(24)(12)[27] \\
0.25 & 0.946(24)(14)[28] \\
0.30 & 0.907(24)(14)[28] \\
0.35 & 0.871(24)(16)[29] \\
0.40 & 0.838(24)(18)[30] \\
0.45 & 0.808(23)(22)[32] \\
0.50 & 0.781(23)(25)[34] \\
0.55 & 0.755(23)(27)[36] \\
0.60 & 0.731(23)(30)[38] \\
0.65 & 0.709(23)(33)[41] \\
0.70 & 0.687(23)(36)[43] \\
0.75 & 0.668(23)(39)[46] \\
0.80 & 0.649(23)(43)[49] \\
0.85 & 0.631(23)(46)[52] \\
0.90 & 0.614(23)(49)[55] \\
0.95 & 0.598(23)(53)[58] \\
1.00 & 0.582(23)(56)[61] \\
  \hline\hline
\end{tabular}
\caption{Values for $G_A(Q^2)$ in the continuum limit as a function of $Q^2$. We provide values for 21 points uniformly distributed in the range $Q^2\in[0,1]$~GeV$^2$. The central values and first errors are obtained from the $z^3$-expansion fitted to the two-state fit data. The second error is the systematic error due to excited states, computed as explained in the text, namely by the difference between the central values of the  $z^3$-expansion parameters when fitting the two- or three-state fit data. The third error is the total error obtained by summing in quadrature the first two. }\label{tab:GA}
\end{table}

\begin{table}[t!]
\begin{tabular}{c|c}
  \hline\hline
  $Q^2$ [GeV$^2$] & $(Q^2+m_\pi^2)G_P(Q^2)$ \\
  \hline
-0.018 & 4.62(23)(23)[33] \\
0.033 & 4.37(17)(20)[27] \\
0.084 & 4.14(14)(18)[23] \\
0.135 & 3.94(12)(17)[21] \\
0.185 & 3.76(11)(15)[19] \\
0.236 & 3.59(10)(14)[18] \\
0.287 & 3.44(9)(14)[17] \\
0.338 & 3.30(9)(14)[17] \\
0.389 & 3.17(9)(14)[17] \\
0.440 & 3.05(9)(14)[17] \\
0.491 & 2.94(8)(15)[17] \\
0.542 & 2.84(8)(15)[17] \\
0.593 & 2.74(8)(15)[17] \\
0.644 & 2.64(8)(16)[18] \\
0.695 & 2.56(8)(16)[18] \\
0.745 & 2.47(8)(17)[19] \\
0.796 & 2.40(9)(17)[20] \\
0.847 & 2.32(9)(17)[20] \\
0.898 & 2.25(9)(18)[21] \\
0.949 & 2.18(9)(20)[22] \\
1.000 & 2.12(9)(21)[23] \\
  \hline\hline
\end{tabular}
\caption{Values for $(Q^2+m_\pi^2)\,G_P(Q^2)$. We provide values for 21 points uniformly distributed in the range $Q^2\in[-m_\pi,1]$~GeV$^2$. The notation is the same as that of Table~\ref{tab:GA}.}\label{tab:GP}
\end{table}

\begin{table}[t!]
\begin{tabular}{c|c}
  \hline\hline
  $Q^2$ [GeV$^2$] & $(Q^2+m_\pi^2)\tilde{G}_5(Q^2)$ \\
  \hline
-0.018 & 4.62(23)(23)[33] \\
0.033 & 4.44(17)(13)[22] \\
0.084 & 4.28(14)(11)[18] \\
0.135 & 4.13(12)(13)[18] \\
0.185 & 4.00(11)(16)[20] \\
0.236 & 3.88(11)(19)[22] \\
0.287 & 3.77(10)(21)[24] \\
0.338 & 3.67(10)(24)[26] \\
0.389 & 3.58(10)(25)[27] \\
0.440 & 3.49(10)(26)[28] \\
0.491 & 3.41(10)(27)[29] \\
0.542 & 3.33(10)(28)[30] \\
0.593 & 3.25(10)(28)[30] \\
0.644 & 3.18(10)(29)[31] \\
0.695 & 3.12(10)(29)[31] \\
0.745 & 3.06(10)(30)[32] \\
0.796 & 3.00(10)(30)[32] \\
0.847 & 2.94(10)(30)[32] \\
0.898 & 2.89(11)(30)[32] \\
0.949 & 2.83(11)(30)[32] \\
1.000 & 2.78(11)(31)[33] \\
  \hline\hline
\end{tabular}
\caption{Values for $(Q^2+m_\pi^2)\,\tilde{G}_5(Q^2)$. We provide values for 21 points uniformly distributed in the range $Q^2\in[-m_\pi,1]$~GeV$^2$. The notation is the same as that of Table~\ref{tab:GA}. }\label{tab:G5}
\end{table}

In Table~\ref{tab:GA}, we provide values for $G_A(Q^2)$ up to 1~GeV$^2$. For this form factor, we use $t_0=0$~GeV$^2$. The values of the fit parameters of two-state fit data are given by
\begin{equation}
\begin{aligned}
    \vec{a}_\text{2-state} = [1.245(28),-1.19(18),-0.54(55),-0.13(59)]\\
    \text{corr}_\text{2-state} = \begin{pmatrix}
1.0 & -0.421 & 0.247 & -0.246 \\
-0.421 & 1.0 & -0.918 & 0.799 \\
0.247 & -0.918 & 1.0 & -0.952 \\
-0.246 & 0.799 & -0.952 & 1.0 \\
    \end{pmatrix}\,.
\end{aligned}
\end{equation}
The fit parameters of three-state fit data are given by
\begin{equation}
\begin{aligned}
    \vec{a}_\text{3-state} = [1.231(34),-1.16(27),-0.80(47),-1.23(58)]\\
    \text{corr}_\text{3-state} = \begin{pmatrix}
1.0 & -0.575 & 0.116 & -0.051 \\
-0.575 & 1.0 & -0.5 & 0.046 \\
0.116 & -0.5 & 1.0 & -0.52 \\
-0.051 & 0.046 & -0.52 & 1.0 \\
    \end{pmatrix}\,.
\end{aligned}
\end{equation}
The fit parameters of the final curve, when we include the systematic error taken as the difference $|a_\text{2-state}-a_\text{3-state}|$ that quantifies systematic uncertainties in the analysis of the excited states, are given by
\begin{equation}
\begin{aligned}
    \vec{a}_\text{ final} = [1.245(31),-1.19(18),-0.54(61),-0.1(1.3)]\\
    \text{corr}_\text{ final} = \begin{pmatrix}
1.0 & -0.421 & 0.247 & -0.246 \\
-0.421 & 1.0 & -0.918 & 0.799 \\
0.247 & -0.918 & 1.0 & -0.952 \\
-0.246 & 0.799 & -0.952 & 1.0 \\
    \end{pmatrix}\,.
\end{aligned}
\end{equation}
The final form factor reproduces the quoted values of $g_A$ and $\langle r_A \rangle$, namely
\begin{equation}
\begin{aligned}
    g_A&=1.245(31)\\
    \langle r_A^2\rangle &=0.339(49)~{\rm fm}^2\,.
\end{aligned}
\end{equation}
$G_A(Q^2)$ and its derivative $G^\prime_A(Q^2)$ also have the correct limit as $Q^2\rightarrow\infty$ having the values
\begin{equation}
\begin{aligned}
    G_A(\infty) &= \sum_k a_k = -0.62(82)\\
    G'_A(\infty) &= \sum_k ka_k = -2.7(2.8),
\end{aligned}
\end{equation}
both compatible with zero.

\subsection{Induced pseudoscalar axial form factor $G_P(Q^2)$}

In Table~\ref{tab:GP}, we provide values for $(Q^2+m_\pi^2)\,G_P(Q^2)$ up to 1~GeV$^2$. For this form factor, we use $t_0=-m_\pi^2$. The values of the fit parameters of two-state fit data are given by
\begin{equation}
\begin{aligned}
    \vec{a}_\text{2-state} = [4.62(23),-3.0(1.2),-4.7(2.5),-0.1(2.4)]\\
    \text{corr}_\text{2-state} = \begin{pmatrix}
1.0 & -0.812 & 0.414 & 0.151 \\
-0.812 & 1.0 & -0.819 & 0.23 \\
0.414 & -0.819 & 1.0 & -0.713 \\
0.151 & 0.23 & -0.713 & 1.0 \\
    \end{pmatrix}\,.
\end{aligned}
\end{equation}
The fit parameters of three-state fit data are given by
\begin{equation}
\begin{aligned}
    \vec{a}_\text{3-state} = [4.38(30),-3.1(1.5),-5.9(2.6),-2.9(2.0)]\\
    \text{corr}_\text{3-state} = \begin{pmatrix}
1.0 & -0.795 & 0.342 & 0.214 \\
-0.795 & 1.0 & -0.712 & -0.292 \\
0.342 & -0.712 & 1.0 & 0.129 \\
0.214 & -0.292 & 0.129 & 1.0 \\
    \end{pmatrix}\,.
\end{aligned}
\end{equation}
The fit parameters of the final curve, when we include the systematic error taken as the difference $|a_\text{2-state}-a_\text{3-state}|$ that quantifies systematic uncertainties in the analysis of the excited states, are given by
\begin{equation}
\begin{aligned}
    \vec{a}_\text{ final} = [4.62(33),-3.0(1.2),-4.7(2.8),-0.1(3.7)]\\
    \text{corr}_\text{ final} = \begin{pmatrix}
1.0 & -0.812 & 0.414 & 0.151 \\
-0.812 & 1.0 & -0.819 & 0.23 \\
0.414 & -0.819 & 1.0 & -0.713 \\
0.151 & 0.23 & -0.713 & 1.0 \\
    \end{pmatrix}\,.
\end{aligned}
\end{equation}
The final form factor reproduces the quoted values of $g_{\pi NN}$ and $g_P^*$, namely
\begin{equation}
    g_{\pi NN} = 13.25(96)\quad\text{and}\quad
    g^*_P=8.99(63)\,.
\end{equation}

\subsection{Pseudoscalar form factor $G_5(Q^2)$}

In Table~\ref{tab:G5}, we provide values for $(Q^2+m_\pi^2)\,\tilde{G}_5(Q^2)$ up to 1~GeV$^2$. The $\tilde{G}_5(Q^2)$ is defined as
\begin{equation}
    \tilde{G}_{5}(Q^2) = \frac{4m_N}{m_\pi^2} m_q G_5(Q^2),
\end{equation}
$m_N=0.938$~GeV and $m_q= 3.636(89)$~MeV~\cite{ExtendedTwistedMass:2021gbo} in the $\overline{\rm MS}$(2 GeV) scheme at the continuum limit. For this form factor, we use $t_0=-m_\pi^2$. The values of the fit parameters of two-state fit data are given by
\begin{equation}
\begin{aligned}
    \vec{a}_\text{2-state} = [4.62(23),-2.2(1.2),-2.9(2.4),-1.2(2.4)]\\
    \text{corr}_\text{2-state} = \begin{pmatrix}
1.0 & -0.804 & 0.435 & 0.14 \\
-0.804 & 1.0 & -0.825 & 0.217 \\
0.435 & -0.825 & 1.0 & -0.694 \\
0.14 & 0.217 & -0.694 & 1.0 \\
    \end{pmatrix}\,.
\end{aligned}
\end{equation}
The fit parameters of three-state fit data are given by
\begin{equation}
\begin{aligned}
    \vec{a}_\text{3-state} = [4.38(30),-4.3(1.6),-0.1(2.7),-0.7(2.0)]\\
    \text{corr}_\text{3-state} = \begin{pmatrix}
1.0 & -0.782 & 0.422 & 0.265 \\
-0.782 & 1.0 & -0.802 & -0.338 \\
0.422 & -0.802 & 1.0 & 0.117 \\
0.265 & -0.338 & 0.117 & 1.0 \\
    \end{pmatrix}\,.
\end{aligned}
\end{equation}
The fit parameters of the final curve, when we include the systematic error taken as the difference $|a_\text{2-state}-a_\text{3-state}|$ that quantifies systematic uncertainties in the analysis of the excited states, are given by
\begin{equation}
\begin{aligned}
    \vec{a}_\text{ final} = [4.62(33),-2.2(2.5),-2.9(3.7),-1.2(2.4)]\\
    \text{corr}_\text{ final} = \begin{pmatrix}
1.0 & -0.804 & 0.435 & 0.14 \\
-0.804 & 1.0 & -0.825 & 0.217 \\
0.435 & -0.825 & 1.0 & -0.694 \\
0.14 & 0.217 & -0.694 & 1.0 \\
    \end{pmatrix}\,.
\end{aligned}
\end{equation}

\bibliography{main}

\end{document}